\pdfoutput=1

\documentclass[12pt,a4paper]{article}

\usepackage{ifthen} 
\usepackage{rotating}
\usepackage{dashrule}
\usepackage{array} 
\usepackage{dcolumn}
\usepackage{comment}
\usepackage{xcolor}

\usepackage{graphpap} 
\usepackage{makecell} 
\usepackage{rotating} 
\usepackage{tikz}
\usepackage{xcolor}
\usepackage{longtable} 

\usepackage{lscape}
\usepackage{multirow} 
\usepackage{mathrsfs}
\usepackage{mathtools} 
\usetikzlibrary{patterns}
\usepackage{nicefrac} 
\usepackage{transparent}
\usepackage{afterpage} 
\usepackage[normalem]{ulem}

\newboolean{pdflatex}
\setboolean{pdflatex}{true} 

\newboolean{articletitles}
\setboolean{articletitles}{true} 

\newboolean{uprightparticles}
\setboolean{uprightparticles}{true} 

\def\paperauthors{LHCb collaboration} 
\def\paperasciititle{Study of Bc+ decays to charmonia plus multihadron final states} 
\def\papertitle{Study of \Bc~meson decays to charmonia plus multihadron final states} 

\def\paperkeywords{{High Energy Physics}, {LHCb}} 
\def\papercopyright{\the\year\ CERN for the benefit of the LHCb collaboration} 
\def\paperlicence{CC BY 4.0 licence}
\def\paperlicenceurl{https://creativecommons.org/licenses/by/4.0/}

\usepackage{wasysym}

\DeclareMathOperator*{\bigplus}{\scalerel*{+}{\sum}}
\usepackage{scalerel}

\makeatletter
\g@addto@macro\bfseries{\boldmath}
\makeatother

\usepackage[top=1in, bottom=1.25in, left=1in, right=1in]{geometry}

\usepackage{microtype}
\usepackage{lineno}  
\usepackage{xspace} 
\usepackage{caption} 

\usepackage{graphicx}  
\usepackage{color}
\usepackage{colortbl}
\graphicspath{{./figs/}} 
\DeclareGraphicsExtensions{.pdf,.PDF,png,.PNG}

\usepackage{amsmath} 
\usepackage{amssymb}
\usepackage{amsfonts}
\usepackage{upgreek} 

\newcommand*\patchAmsMathEnvironmentForLineno[1]{%
\expandafter\let\csname old#1\expandafter\endcsname\csname #1\endcsname
\expandafter\let\csname oldend#1\expandafter\endcsname\csname
end#1\endcsname
 \renewenvironment{#1}%
   {\linenomath\csname old#1\endcsname}%
   {\csname oldend#1\endcsname\endlinenomath}%
}
\newcommand*\patchBothAmsMathEnvironmentsForLineno[1]{%
  \patchAmsMathEnvironmentForLineno{#1}%
  \patchAmsMathEnvironmentForLineno{#1*}%
}
\AtBeginDocument{%
\patchBothAmsMathEnvironmentsForLineno{equation}%
\patchBothAmsMathEnvironmentsForLineno{align}%
\patchBothAmsMathEnvironmentsForLineno{flalign}%
\patchBothAmsMathEnvironmentsForLineno{alignat}%
\patchBothAmsMathEnvironmentsForLineno{gather}%
\patchBothAmsMathEnvironmentsForLineno{multline}%
\patchBothAmsMathEnvironmentsForLineno{eqnarray}%
}

\usepackage{hyperxmp}

\usepackage[pdftex,
            pdfauthor={\paperauthors},
            pdftitle={\paperasciititle},
            pdfkeywords={\paperkeywords},
            pdfcopyright={Copyright (C) \papercopyright},
            pdflicenseurl={\paperlicenceurl}]{hyperref}

\usepackage[colorinlistoftodos,textsize=scriptsize]{todonotes}

\usepackage[all]{hypcap} 

\usepackage{xspace} 
\usepackage{upgreek}

\def\lhcb   {\mbox{LHCb}\xspace}

\def\cdf    {\mbox{CDF}\xspace}

\def\MagUp {\mbox{\em Mag\kern -0.05em Up}\xspace}

\ifthenelse{\boolean{uprightparticles}}%
{

 \def\Pmu         {\ensuremath{\upmu}\xspace}

 \def\Ppi         {\ensuremath{\uppi}\xspace}                 
                  
 \def\Prho        {\ensuremath{\uprho}\xspace}                 
                  
 \def\Ptau        {\ensuremath{\uptau}\xspace}                 
                  
 \def\Pphi        {\ensuremath{\upphi}\xspace}

 \def\Ppsi        {\ensuremath{\uppsi}\xspace}

 \def\PDelta      {\ensuremath{\Delta}\xspace}                 
 \def\PXi         {\ensuremath{\Xi}\xspace}                 
 \def\PLambda     {\ensuremath{\Lambda}\xspace}                 
 \def\PSigma      {\ensuremath{\Sigma}\xspace}                 
 \def\POmega      {\ensuremath{\Omega}\xspace}                 
 \def\PUpsilon    {\ensuremath{\Upsilon}\xspace}
 \let\oldPi\Pi
 \def\PPi         {\ensuremath{\oldPi}\xspace}

 \def\PB      {\ensuremath{\mathrm{B}}\xspace}                 
                  
 \def\PD      {\ensuremath{\mathrm{D}}\xspace}

 \def\PJ      {\ensuremath{\mathrm{J}}\xspace}                 
 \def\PK      {\ensuremath{\mathrm{K}}\xspace}

 \def\PW      {\ensuremath{\mathrm{W}}\xspace}                 
 \def\PX      {\ensuremath{\mathrm{X}}\xspace}                 
 \def\PY      {\ensuremath{\mathrm{Y}}\xspace}

 \def\Pb      {\ensuremath{\mathrm{b}}\xspace}                 
 \def\Pc      {\ensuremath{\mathrm{c}}\xspace}                 
                  
 \def\Pe      {\ensuremath{\mathrm{e}}\xspace}

 \def\Ph      {\ensuremath{\mathrm{h}}\xspace}                 
 \def\Pi      {\ensuremath{\mathrm{i}}\xspace}

 \def\Pp      {\ensuremath{\mathrm{p}}\xspace}

 \def\Ps      {\ensuremath{\mathrm{s}}\xspace}

 \def\thebaroffset{0.0em}
}
{

 \def\Pmu         {\ensuremath{\mu}\xspace}

 \def\Ppi         {\ensuremath{\pi}\xspace}                 
                  
 \def\Prho        {\ensuremath{\rho}\xspace}                 
                  
 \def\Ptau        {\ensuremath{\tau}\xspace}                 
                  
 \def\Pphi        {\ensuremath{\phi}\xspace}

 \def\Ppsi        {\ensuremath{\psi}\xspace}                 
                  
 \mathchardef\PDelta="7101
 \mathchardef\PXi="7104
 \mathchardef\PLambda="7103
 \mathchardef\PSigma="7106
 \mathchardef\POmega="710A
 \mathchardef\PUpsilon="7107
 \mathchardef\PPi="7105
                  
 \def\PB      {\ensuremath{B}\xspace}                 
                  
 \def\PD      {\ensuremath{D}\xspace}

 \def\PJ      {\ensuremath{J}\xspace}                 
 \def\PK      {\ensuremath{K}\xspace}

 \def\PW      {\ensuremath{W}\xspace}                 
 \def\PX      {\ensuremath{X}\xspace}                 
 \def\PY      {\ensuremath{Y}\xspace}

 \def\Pb      {\ensuremath{b}\xspace}                 
 \def\Pc      {\ensuremath{c}\xspace}                 
                  
 \def\Pe      {\ensuremath{e}\xspace}

 \def\Ph      {\ensuremath{h}\xspace}                 
 \def\Pi      {\ensuremath{i}\xspace}

 \def\Pp      {\ensuremath{p}\xspace}

 \def\Ps      {\ensuremath{s}\xspace}

 \def\thebaroffset{0.18em}
}
\newcommand{\offsetoverline}[2][\thebaroffset]{\kern #1\overline{\kern -#1 #2}}%

\makeatletter
\ifcase \@ptsize \relax
  \newcommand{\miniscule}{\@setfontsize\miniscule{4}{5}}
\or
  \newcommand{\miniscule}{\@setfontsize\miniscule{5}{6}}
\or
  \newcommand{\miniscule}{\@setfontsize\miniscule{5}{6}}
\fi
\makeatother

\DeclareRobustCommand{\optbar}[1]{\shortstack{{\miniscule (\rule[.5ex]{1.25em}{.18mm})}
  \\ [-.7ex] $#1$}}

\def\epem       {{\ensuremath{\Pe^+\Pe^-}}\xspace}

\def\mup        {{\ensuremath{\Pmu^+}}\xspace}
\def\mun        {{\ensuremath{\Pmu^-}}\xspace} 

\def\mumu       {{\ensuremath{\Pmu^+\Pmu^-}}\xspace}

\def\Wp     {{\ensuremath{\PW^+}}\xspace}

\def\squark    {{\ensuremath{\Ps}}\xspace}

\def\cquark    {{\ensuremath{\Pc}}\xspace}

\def\bquark    {{\ensuremath{\Pb}}\xspace}

\def\pion   {{\ensuremath{\Ppi}}\xspace}

\def\pip    {{\ensuremath{\pion^+}}\xspace}
\def\pim    {{\ensuremath{\pion^-}}\xspace}
\def\pipm   {{\ensuremath{\pion^\pm}}\xspace}
\def\pimp   {{\ensuremath{\pion^\mp}}\xspace}

\def\rhomeson {{\ensuremath{\Prho}}\xspace}
\def\rhoz     {{\ensuremath{\rhomeson^0}}\xspace}

\def\kaon    {{\ensuremath{\PK}}\xspace}
\def\Kbar    {{\ensuremath{\offsetoverline{\PK}}}\xspace}

\def\KorKbar {\kern \thebaroffset\optbar{\kern -\thebaroffset \PK}{}\xspace}

\def\Kp      {{\ensuremath{\kaon^+}}\xspace}
\def\Km      {{\ensuremath{\kaon^-}}\xspace}
\def\Kpm     {{\ensuremath{\kaon^\pm}}\xspace}

\def\KS      {{\ensuremath{\kaon^0_{\mathrm{S}}}}\xspace}

\def\Kstarz  {{\ensuremath{\kaon^{*0}}}\xspace}
\def\Kstarzb {{\ensuremath{\Kbar{}^{*0}}}\xspace}

\def\D       {{\ensuremath{\PD}}\xspace}

\def\DorDbar {\kern \thebaroffset\optbar{\kern -\thebaroffset \PD}\xspace}
\def\Dz      {{\ensuremath{\D^0}}\xspace}

\def\Dp      {{\ensuremath{\D^+}}\xspace}
\def\Dm      {{\ensuremath{\D^-}}\xspace}

\def\DpDm    {\ensuremath{\Dp {\kern -0.16em \Dm}}\xspace}

\def\Dstarp  {{\ensuremath{\D^{*+}}}\xspace}

\def\Ds      {{\ensuremath{\D^+_\squark}}\xspace}

\def\B       {{\ensuremath{\PB}}\xspace}

\def\BorBbar {\kern \thebaroffset\optbar{\kern -\thebaroffset \PB}\xspace}

\def\Bd      {{\ensuremath{\B^0}}\xspace}

\def\BdorBdbar {\kern \thebaroffset\optbar{\kern -\thebaroffset \Bd}\xspace}
\def\Bu      {{\ensuremath{\B^+}}\xspace}

\def\Bp      {{\ensuremath{\Bu}}\xspace}

\def\Bs      {{\ensuremath{\B^0_\squark}}\xspace}

\def\BsorBsbar {\kern \thebaroffset\optbar{\kern -\thebaroffset \Bs}\xspace}
\def\Bc      {{\ensuremath{\B_\cquark^+}}\xspace}

\def\jpsi     {{\ensuremath{{\PJ\mskip -3mu/\mskip -2mu\Ppsi}}}\xspace}
\def\psitwos  {{\ensuremath{\Ppsi{(2{\mathrm{S}})}}}\xspace}

\def\Y#1S{\ensuremath{\PUpsilon{(#1S)}}\xspace}

\def\proton      {{\ensuremath{\Pp}}\xspace}
\def\antiproton  {{\ensuremath{\overline \proton}}\xspace}

\def\LorLbar     {\kern \thebaroffset\optbar{\kern -\thebaroffset \PLambda}\xspace}

\def\BF         {{\ensuremath{\mathcal{B}}}\xspace}
\def\BR         {\BF}

\newcommand{\decay}[2]{\ensuremath{#1\!\to #2}\xspace} 

\def\to                 {\ensuremath{\rightarrow}\xspace}

\def\AT#1     {\ensuremath{A_{\mathrm{T}}^{#1}}\xspace}           

\def\C#1      {\ensuremath{\mathcal{C}_{#1}}\xspace}                       
\def\Cp#1     {\ensuremath{\mathcal{C}_{#1}^{'}}\xspace}                    
\def\Ceff#1   {\ensuremath{\mathcal{C}_{#1}^{\mathrm{(eff)}}}\xspace}        
\def\Cpeff#1  {\ensuremath{\mathcal{C}_{#1}^{'\mathrm{(eff)}}}\xspace}       
\def\Ope#1    {\ensuremath{\mathcal{O}_{#1}}\xspace}                       
\def\Opep#1   {\ensuremath{\mathcal{O}_{#1}^{'}}\xspace}                    

\newcommand{\nospaceunit}[1]{\ensuremath{\text{#1}}}       
\newcommand{\aunit}[1]{\ensuremath{\text{\,#1}}}       

\newcommand{\tev}{\aunit{Te\kern -0.1em V}\xspace}
\newcommand{\gev}{\aunit{Ge\kern -0.1em V}\xspace}
\newcommand{\mev}{\aunit{Me\kern -0.1em V}\xspace}
\newcommand{\kev}{\aunit{ke\kern -0.1em V}\xspace}
\newcommand{\ev}{\aunit{e\kern -0.1em V}\xspace}
 
\newcommand{\mevc}{\ensuremath{\aunit{Me\kern -0.1em V\!/}c}\xspace}
\newcommand{\gevc}{\ensuremath{\aunit{Ge\kern -0.1em V\!/}c}\xspace}
\newcommand{\mevcc}{\ensuremath{\aunit{Me\kern -0.1em V\!/}c^2}\xspace}
\newcommand{\gevcc}{\ensuremath{\aunit{Ge\kern -0.1em V\!/}c^2}\xspace}

\def\mum  {\ensuremath{\,\upmu\nospaceunit{m}}\xspace}

\def\fb   {\ensuremath{\aunit{fb}}\xspace}
\def\invfb   {\ensuremath{\fb^{-1}}\xspace}

\newcommand{\stat}{\aunit{(stat)}\xspace}
\newcommand{\syst}{\aunit{(syst)}\xspace}

\newcommand{\chisq}{\ensuremath{\chi^2}\xspace}

\newcommand{\chisqip}{\ensuremath{\chi^2_{\text{IP}}}\xspace}

\def\gsim{{~\raise.15em\hbox{$>$}\kern-.85em
          \lower.35em\hbox{$\sim$}~}\xspace}
\def\lsim{{~\raise.15em\hbox{$<$}\kern-.85em
          \lower.35em\hbox{$\sim$}~}\xspace}

\def\sPlot{\mbox{\em sPlot}\xspace}

\def\pt         {\ensuremath{p_{\mathrm{T}}}\xspace}

\def\evtgen     {\mbox{\textsc{EvtGen}}\xspace}

\def\geant      {\mbox{\textsc{Geant4}}\xspace}

\def\photos     {\mbox{\textsc{Photos}}\xspace}

\def\pythia     {\mbox{\textsc{Pythia}}\xspace}

\def\tell1  {TELL1\xspace}
\def\ukl1   {UKL1\xspace}

\newcommand{\lhcborcid}[1]{\href{https://orcid.org/#1}{\hspace*{0.1em}\raisebox{-0.45ex}{\includegraphics[width=1em]{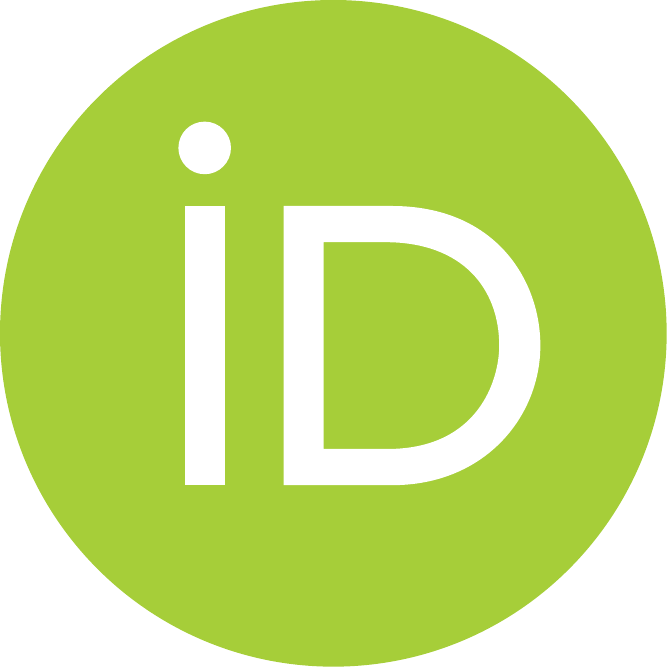}}}}

\def\BcTopsitwostripi {\decay{\Bc}{\psitwos\pip\pip\pim}}

\def\BcTojpsifivepi    {\decay{\Bc}{\jpsi3\pip2\pim}}
\def\BcTojpsisevenpi    {\decay{\Bc}{\jpsi4\pip3\pim}}
\def\BcTojpsikkpipipi    {\decay{\Bc}{\jpsi\Kp\Km\pip\pip\pim}}

\def\BcTopsitwosjpptripi {\decay{\Bc}{(\decay{\psitwos}{\jpsi\pip\pim})\pip\pip\pim}}

\def\Tojpsifivepi     {\jpsi 3\pip 2\pim}
\def\Tojpsisevenpi    {\jpsi 4\pip 3\pim}
\def\Tojpsikkpipipi   {{\jpsi\Kp\Km\pip\pip\pim}}

\def\JpsiPiPi{\jpsi\pip\pim}

\def\KorKbarzstar{\ensuremath{\KorKbar^{*0}}\xspace}

\def\Tojpsifivepi     {\jpsi 3\pip 2\pim}
\def\Tojpsisevenpi    {\jpsi 4\pip 3\pim}
\def\Tojpsikkpipipi   {{\jpsi\Kp\Km\pip\pip\pim}}

\def\BcTopsitwostripi {\decay{\Bc}{\psitwos\pip\pip\pim}}

\def\Topsitwostripi {\psitwos\pip\pip\pim}

\def\psitwosTojpsipipi {\mbox{\decay{\psitwos}{\jpsi\pip\pim}}}

\usepackage{cite} 
\usepackage{mciteplus}

\usepackage{longtable} 

\newcolumntype{d}[1]{D{,}{\,\pm\,}{#1} }
\newcolumntype{f}[1]{D{,}{.}{#1} }

\begin{document}

\renewcommand{\thefootnote}{\fnsymbol{footnote}}
\setcounter{footnote}{1}


\begin{titlepage}
\pagenumbering{roman}

\vspace*{-1.5cm}
\centerline{\large EUROPEAN ORGANIZATION FOR NUCLEAR RESEARCH (CERN)}
\vspace*{1.5cm}
\noindent
\begin{tabular*}{\linewidth}{lc@{\extracolsep{\fill}}r@{\extracolsep{0pt}}}
\ifthenelse{\boolean{pdflatex}}
{\vspace*{-1.5cm}\mbox{\!\!\!\includegraphics[width=.14\textwidth]{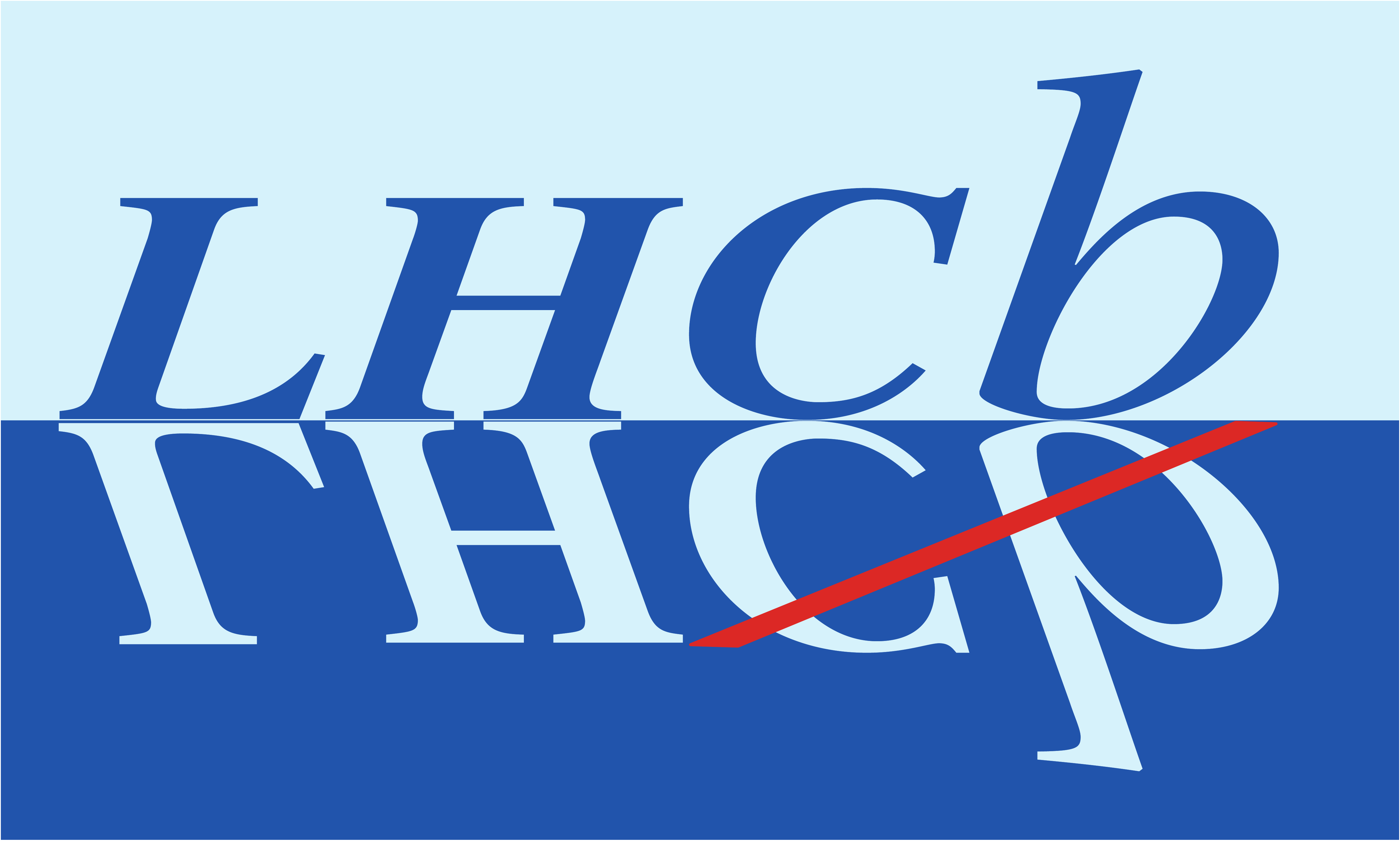}} & &}%
{\vspace*{-1.2cm}\mbox{\!\!\!\includegraphics[width=.12\textwidth]{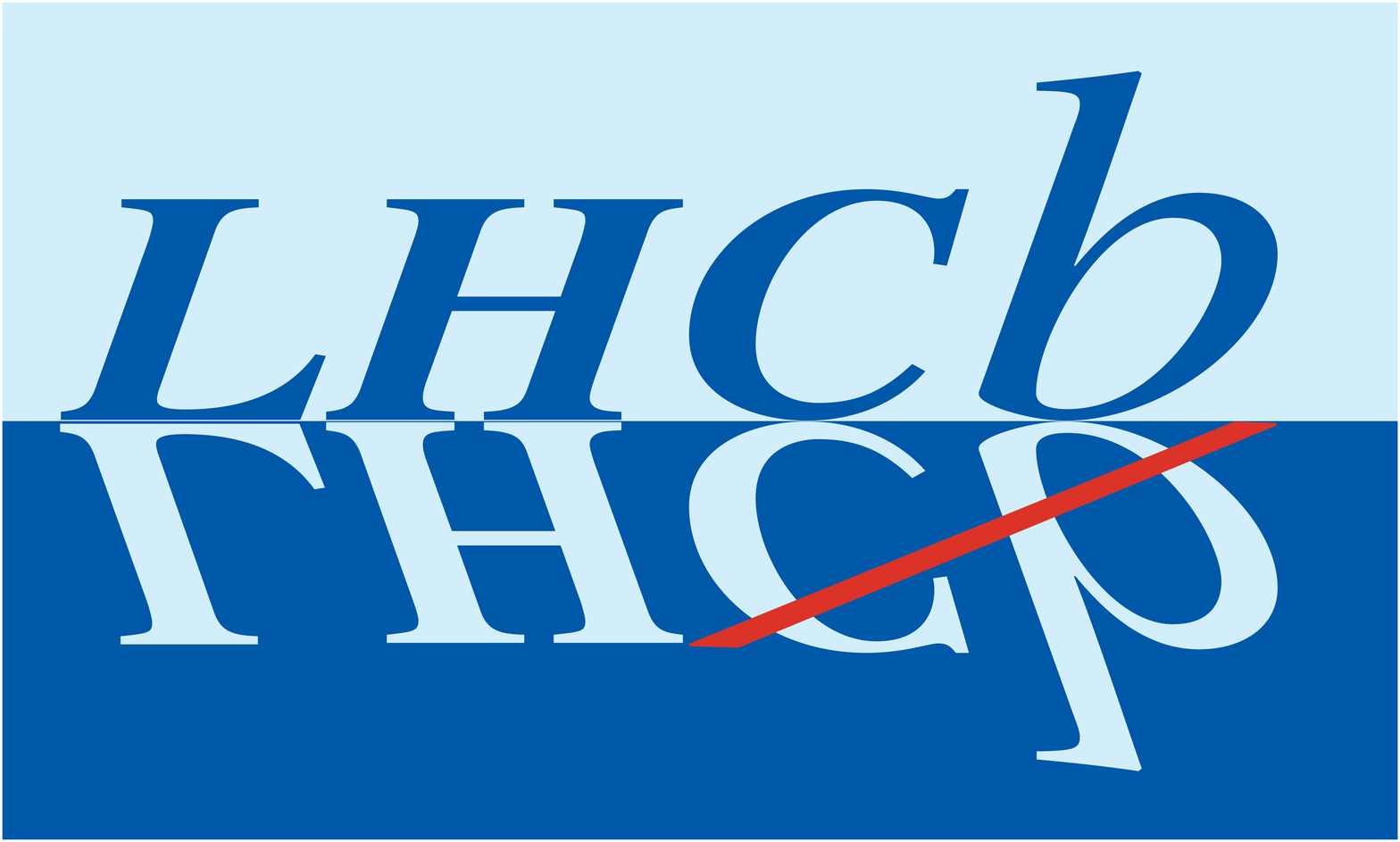}} & &}%
\\
 & & CERN-EP-2022-162 \\  
 & & LHCb-PAPER-2022-025 \\  
& & August 16, 2022\\ 
\end{tabular*}

\vspace*{3.5cm}

{\normalfont\bfseries\boldmath\huge
\begin{center}
  \papertitle 
\end{center}
}

\vspace*{0.2cm}

\begin{center}
\paperauthors\footnote{Authors are listed at the end of this paper.}
\end{center}


\begin{abstract}
\noindent 
Four~decay modes of the~\Bc~meson 
into a~\jpsi~meson 
and multiple charged 
kaons or pions are studied
using proton\nobreakdash-proton collision data, 
collected with the~\lhcb detector at 
centre\nobreakdash-of\nobreakdash-mass energies of 
7, 8, and 13\tev and 
corresponding~to~an~integrated~~luminosity~of~$9\invfb$.
The~decay 
\mbox{$\decay{\Bc}{\jpsi\Kp\Km\pip\pip\pim}$}
is observed for the~first time,
and evidence  for 
the~\mbox{$\decay{\Bc}{\jpsi4\pip3\pim}$}~decay
is found. 
The~decay 
\mbox{$\decay{\Bc}{\jpsi3\pip2\pim}$}
is observed
and the~previous observation 
of 
the~\mbox{$\decay{\Bc}{}\psitwos\pip\pip\pim$} decay
is confirmed
using 
the~\mbox{$\decay{\psitwos}{\jpsi\pip\pim}$}~decay mode.
Ratios of the~branching fractions 
of these 
four \Bc~decay channels
are measured.  
\end{abstract}

\vspace{\fill}

\begin{center}
  Published in \href{https://doi.org/10.1007/JHEP07(2023)198}{JHEP 07\,(2023) 198} 
\end{center}

\vspace*{0.5cm}

{\footnotesize 
\centerline{\copyright~\papercopyright. \href{\paperlicenceurl}{\paperlicence}.}}
\vspace*{2mm}

\end{titlepage}


\newpage
\setcounter{page}{2}
\mbox{~}
%

\cleardoublepage


\renewcommand{\thefootnote}{\arabic{footnote}}
\setcounter{footnote}{0}



\pagestyle{plain} 
\setcounter{page}{1}
\pagenumbering{arabic}


%

\section{Introduction}
\label{sec:Introduction}
The~\Bc meson,
discovered in 1998 by 
the~\cdf collaboration~\cite{PhysRevLett.81.2432, 
PhysRevD.58.112004}
at the~Tevatron $\proton\antiproton$~collider, 
is the~only known meson that contains
two different heavy\nobreakdash-flavour quarks, 
charm and beauty. 
The~high \bquark-quark production cross-section at 
the~Large Hadron
Collider\,(LHC)~\cite{LHCb-PAPER-2010-002,
LHCb-PAPER-2011-003,
LHCb-PAPER-2011-043,
LHCb-PAPER-2013-004,
LHCb-PAPER-2013-016,
LHCb-PAPER-2015-037} 
enables the~\lhcb, ATLAS and CMS experiments 
to study in detail the~production, 
decays and other properties 
of the~\Bc meson~\cite{LHCb-PAPER-2011-044, 
LHCb-PAPER-2012-054, 
LHCb-PAPER-2013-010, 
LHCb-PAPER-2013-021, 
LHCb-PAPER-2013-044, 
LHCb-PAPER-2013-047, 
LHCb-PAPER-2014-009, 
LHCb-PAPER-2014-039, 
LHCb-PAPER-2014-050, 
LHCb-PAPER-2014-060, 
CMS:2014oqy,
LHCb-PAPER-2015-024, 
ATLAS:2015jep,
LHCb-PAPER-2016-001, 
LHCb-PAPER-2016-022, 
LHCb-PAPER-2016-020, 
LHCb-PAPER-2016-055, 
LHCb-PAPER-2016-058, 
LHCb-PAPER-2017-035, 
LHCb-PAPER-2017-045, 
CMS:2017ygm,         
CMS:2019uhm,         
LHCb-PAPER-2019-007, 
LHCb-PAPER-2019-033, 
LHCb-PAPER-2020-003, 
LHCb-PAPER-2021-023,
LHCb-PAPER-2021-034}. 
The~\Bc~meson has 
a~rich set of 
decay modes 
since either of the~heavy quarks can 
decay while the~other
behaves as a~spectator quark, 
or both quarks can annihilate 
via a~virtual \Wp~boson. 

Decays of the~\Bc meson to charmonium and 
light hadrons can be  described
using the~quantum chromodynamics~(QCD) 
factorisation approach~\cite{Bauer:1986bm,
Wirbel:1988ft},
which relies on the~form factors of 
the~\mbox{\decay{\Bc}
{\jpsi\Wp}}~transition~\cite{Gershtein:1994jw,
Gershtein:1997qy,
Kiselev:1999sc,
Kiselev:2000pp,
Ebert:2002pp}
and on the~universal spectral function 
for the~virtual \Wp~boson fragmenting
into light hadrons~\cite{Likhoded:2009ib, 
Likhoded:2013iua, 
Berezhnoy:2011is}.
The~spectral function can be calculated 
or, alternatively, determined 
using the~multihadron decays of 
the~\Ptau~lepton 
or \epem~annihilation to light hadrons. 
The~phenomenological model proposed by 
Berezhnoy, Likhoded and 
Luchinsky\,(BLL model)~\cite{Likhoded:2009ib, 
Berezhnoy:2011is,
Luchinsky:2012rk,
Likhoded:2013iua,
Luchinsky:2013yla,
Luchinsky:2018lfj,
Luchinsky:2022pxu}, 
based on this approach, 
describes well 
the~measured branching fractions
for the~\mbox{$\decay{\Bc}
{\jpsi\pip\pip\pim}$},
\mbox{$\decay{\Bc}{\psitwos\pip\pip\pim}$},
\mbox{$\decay{\Bc}
{\jpsi\Kp\pip\pim}$},
\mbox{$\decay{\Bc}
{\jpsi\Kp\Km\pip}$},
and \mbox{$\decay{\Bc}
{\jpsi\Kp\Kp\Km}$}~decays~\cite{LHCb-PAPER-2012-054, LHCb-PAPER-2021-034, LHCb-PAPER-2013-047} 
as well as the~major characteristics of 
their light\nobreakdash-hadron 
systems and resonance 
structure.
Additional~measurements of the~branching 
fractions of 
various 
\Bc~decays into 
the~final states consisting of 
charmonium and 
multiple light hadrons 
would allow for more precise tests of
the~factorisation hypothesis.

Special 
interest in 
the~decays of 
the~\Bc meson 
to a~\jpsi meson and multiple light
hadrons
arises for the~case 
where both the~number of light hadrons 
and the~energy released 
in the decay~are large. 
In such~a scenario,  
one expects that 
the~statistical, 
or quasi\nobreakdash-classical,
approach~\cite{PhysRev.63.137,
PhysRev.66.149}
could be applied to 
describe the~multibody system of the~light hadrons
recoiling against the~\jpsi~meson. 
The~properties of such systems of light hadrons could be
comparable to those from  
models used for the~description 
of correlations
in multihadron production, in particular 
in heavy\nobreakdash-ion
collisions~\cite{PhysRevC.20.2267}.
Experimentally, evidence for 
$32\pm8$ decays
of \Bc~mesons into 
a~\jpsi~meson and five charged pions,
\mbox{$\decay{\Bc}{\jpsi3\pip2\pim}$},
was obtained by 
the~LHCb collaboration~\cite{LHCb-PAPER-2014-009}.
This study was done using data
collected in 
proton\nobreakdash-proton\,($\proton\proton$)~collisions 
at centre\nobreakdash-of\nobreakdash-mass energies 
of 7 and~8\tev, 
corresponding to an~integrated 
luminosity~of~3\invfb.
The~measured branching fraction, 
relative to 
the~\mbox{$\decay{\Bc}{\jpsi\pip}$}~decay mode,
and 
characteristics of the~multipion system, are 
consistent with expectations from 
the~BLL~model~\cite{Luchinsky:2012rk}.

This paper reports a~study of 
the~\Bc meson decaying into 
final states with charmonium and five 
light hadrons,\footnote{Inclusion of charge-conjugate
decays is implied throughout the paper.}
namely \mbox{$\decay{\Bc}{\jpsi3\pip2\pim}$},
\mbox{$\decay{\Bc}{\jpsi\Kp\Km\pip\pip\pim}$},
\mbox{$\decay{\Bc}{\left(\decay{\psitwos}
{\jpsi\pip\pim}\right)\pip\pip\pim}$},  
and the~final state with 
seven charged pions, 
\mbox{$\decay{\Bc}{\jpsi4\pip3\pim}$}. 
The~analysis is based 
on $\proton\proton$ collision data, 
corresponding to an~integrated 
luminosity of~9\,\invfb,
collected with the~\lhcb detector 
at~centre-of-mass energies of 7, 8, and 13\,\tev. 

\section{Detector and simulation}
\label{sec:Detector}

The \lhcb detector~\cite{Alves:2008zz,LHCb-DP-2014-002} is a single-arm forward
spectrometer covering the~pse\-udora\-pi\-dity range \mbox{$2<\eta <5$},
designed for the study of particles containing $\bquark$~or~$\cquark$~quarks. 
The~detector includes a high-precision tracking system consisting of a 
silicon-strip vertex detector surrounding the \proton\proton interaction
region~\cite{LHCb-DP-2014-001}, a large-area silicon-strip detector located
upstream of a dipole magnet with a bending power of about $4{\mathrm{\,Tm}}$,
and three stations of silicon-strip detectors and straw
drift tubes~\cite{LHCb-DP-2013-003,LHCb-DP-2017-001} placed downstream of the magnet. 
The tracking system provides a measurement of the momentum of charged particles
with a relative uncertainty that varies from $0.5\%$ at low momentum to $1.0\%$~at~$200 \gevc$. 
The~momentum scale is calibrated using samples of $\decay{\jpsi}{\mumu}$ 
and $\decay{\Bu}{\jpsi\Kp}$~decays collected concurrently
with the~data sample used for this analysis~\cite{LHCb-PAPER-2012-048,LHCb-PAPER-2013-011}. 
The~relative accuracy of this
procedure is estimated to be $3 \times 10^{-4}$ using samples of other
fully reconstructed $\bquark$~hadrons, 
$\PUpsilon$~and
$\KS$~mesons.
The~minimum distance between a track and 
a~primary 
$\proton\proton$\nobreakdash-collision vertex\,(PV)\cite{Bowen:2014tca,Dziurda:2115353}, 
the~impact parameter, 
is~measured with a~resolution of $(15+29/\pt)\mum$, where \pt is the component 
of the~momentum transverse to the beam, in\,\gevc. Different types of charged hadrons
are distinguished using information from 
two ring\nobreakdash-imaging Cherenkov 
detectors\,(RICH)~\cite{LHCb-DP-2012-003}. Photons,~electrons and hadrons are identified 
by a~calorimeter system consisting of scintillating\nobreakdash-pad 
and preshower detectors, 
an electromagnetic and 
a~hadronic calorimeter. Muons are~identified by a~system 
composed of alternating layers of iron and multiwire proportional chambers~\cite{LHCb-DP-2012-002}.

The online event selection is performed by a trigger~\cite{LHCb-DP-2012-004}, 
which consists of a hardware stage, based on information from the calorimeter and muon systems,
followed by a~software stage, which performs a full event reconstruction. 
The~hardware trigger selects muon candidates 
with high transverse momentum 
or dimuon candidates with a~high value of 
the~product
of the~transverse momenta of the~two muons.
In~the~software trigger, 
two 
oppositely-charged muons are required to form 
a~good\nobreakdash-quality
vertex that is significantly 
displaced from any~PV,
and the~mass of the~$\mumu$~pair 
is required to  
exceed~$2.7\gevcc$.

Simulated events are used 
to model the~signal mass shapes 
and to~compute the~efficiencies needed to determine 
the~branching fraction ratios.
In~the~simulation, \proton\proton collisions are generated 
using \pythia~\cite{Sjostrand:2007gs}  with a~specific \lhcb configuration~\cite{LHCb-PROC-2010-056}. 
Decays of unstable particles are described by 
the~\evtgen 
package~\cite{Lange:2001uf}, 
in which final-state radiation is generated using \photos~\cite{davidson2015photos}. 
The~decay channels in this study are simulated 
using the~BLL model~\cite{
Berezhnoy:2011nx,
Luchinsky:2022pxu}.
The~interaction of the~generated particles 
with the~detector,
and its response, are implemented using
the~\geant 
toolkit~\cite{Allison:2006ve,*Agostinelli:2002hh} 
as described in Ref.~\cite{LHCb-PROC-2011-006}.
To~account for imperfections in the~simulation of
charged\nobreakdash-particle reconstruction, 
the~track\nobreakdash-reconstruction efficiency
determined from simulation 
is corrected using 
calibration samples~\cite{LHCb-DP-2013-002}.

\section{Event selection}
\label{sec:Selection}
 
 The~\decay{\Bc}{\jpsi n\Ph^\pm}~candidates, where $n = 5,\,7$
 represents the number of light hadrons 
 in the~final 
 state and $\Ph^{\pm}$ 
 stands for a~charged kaon or pion,
are reconstructed using 
the~\mbox{$\decay{\jpsi}{\mumu}$}~decay mode.
The~selection criteria largely follow 
those described in 
Refs.~\cite{LHCb-PAPER-2013-047,
 LHCb-PAPER-2014-009, 
 LHCb-PAPER-2016-040,
 LHCb-PAPER-2021-034}.
The~selection starts from 
 reconstructed charged tracks
of good quality
and 
 muon, pion and kaon candidates 
 are identified 
 by combining information 
 from the~RICH, 
 calorimeter and muon
 detectors~\cite{LHCb-PROC-2011-008}. 
The~muon candidates 
are  required to have a~transverse 
momentum larger than 550\mevc.
Pairs of oppositely charged 
muons consistent with originating from a~common vertex 
are combined to form \decay{\jpsi}{\mumu} candidates. 
The~reconstructed mass of the~\mumu~pair 
is required to be
in the~range $3.0<m_{\mumu}<3.2\gevcc$, 
which approximately corresponds to  
a~$\pm7\sigma$ region
around the known \jpsi meson mass~\cite{PDG2021}, 
where $\sigma$ is 
the~\mup\mun mass resolution.

 To~form the~$\Bc$~candidates, 
 the~selected $\jpsi$~candidates 
 are combined with charged tracks identified as 
 kaons or pions, 
 requiring a~well reconstructed vertex.
Kaons and pions are required to have 
a~momentum between 3.2~and~150\gevc,
to ensure a~good performance 
of the~particle identification~\cite{LHCb-PROC-2011-008,
LHCb-DP-2012-003}.
To~reduce the~combinatorial background, 
only tracks that are inconsistent 
with originating from any 
reconstructed PV in the~event are considered, and
    the~scalar sum 
of the~transverse momenta 
of the~light\nobreakdash-hadron candidates
is required to be larger than a~minimum value.
%
Each~$\Bc$~candidate is associated with
the~PV that yields the~smallest~$\chisqip$, 
where \chisqip is defined as the~difference 
in the~vertex\nobreakdash-fit 
\chisq of a~given PV 
reconstructed with and without 
the~particle under consideration.
 To~improve the~mass resolution
 for the~$\Bc$\nobreakdash~candidates,
 a~kinematic fit  is performed~\cite{Hulsbergen:2005pu}. 
 This~fit constrains the~mass of the~$\mumu$~pair 
 to the~known mass of the~$\jpsi$  meson~\cite{PDG2021} 
 and  constrains the~\Bc~candidate 
 to originate from its associated PV.
 A~requirement on
the~quality of this fit is applied 
to further suppress combinatorial background.
Such~a~requirement also reduces 
contributions from 
the~\Bc~decays
proceeding through 
intermediate
$\Dp$, $\Ds$, $\Bu$ or $\Bd$~mesons. 
    The~proper decay time of the~\Bc~candidate, 
    calculated with respect to the~associated PV,
    is required to be lager than a~minimum value, 
    which suppresses 
random combinations of \jpsi candidates and charged tracks, 
which include tracks
originating from the~PV.
 The~mass of selected \Bc candidates 
 is required to be between 6.15~and~$6.45\gevcc$.

For the~selected \mbox{$\decay{\Bc}
{\jpsi\Kp\Km\pip\pip\pim}$}~candidates, 
an~excess of events is seen
in 
the~\mbox{$\jpsi\Kpm\pipm\pimp$}~mass
spectra at the~known mass of 
the~\mbox{\Bp~meson~\cite{PDG2021}}. Similarly, 
for the~selected 
\mbox{$\decay{\Bc}{\jpsi3\pip2\pim}$}~candidates 
a~slight excess of events
is seen in the~$\pip\pip\pim$~mass
spectrum close to the~known mass
of the~\Dp~meson~\cite{PDG2021}. 
Such \Bc~candidates are excluded 
from further 
analysis. No excess of candidates
is observed in the $\pip\pip\pim$~mass distribution 
near the mass of the ~\Ds~meson.
For~the~\mbox{$\decay{\Bc}{\jpsi3\pip2\pim}$}
and~\mbox{$\decay{\Bc}
{\jpsi\Kp\Km\pip\pip\pim}$}~decays,
the~contributions 
from the~\mbox{$\decay{\Bc}
{ \left( \decay{\psitwos}
{\jpsi\pip\pim}\right)\Ph^+\Ph^+\Ph^-} $}~decays
are removed by rejecting  candidates with 
any \mbox{$\jpsi\pip\pim$}~combination having mass within
the~range~\mbox{$3.68<m_{\jpsi\pip\pim}<3.69\gevcc$}.
The~\mbox{$\decay{\Bc}{\jpsi3\pip2\pim}$}~candidates 
with at least one $\jpsi\pip\pim$~mass within 
the~\mbox{$3.67<m_{\jpsi\pip\pim}<3.70\gevcc$}
range are considered as 
\mbox{$\decay{\Bc}
{ \left( \decay{\psitwos}
{\jpsi\pip\pim}\right)\pip\pip\pim}$}~candidates
in the~subsequent analysis. 

\begin{figure}[t]
	\setlength{\unitlength}{1mm}
	\centering
	\begin{picture}(150,120)
	\definecolor{root8}{rgb}{0.35, 0.83, 0.33}
	
    \put( 2, 62) {\includegraphics*[width=75mm]{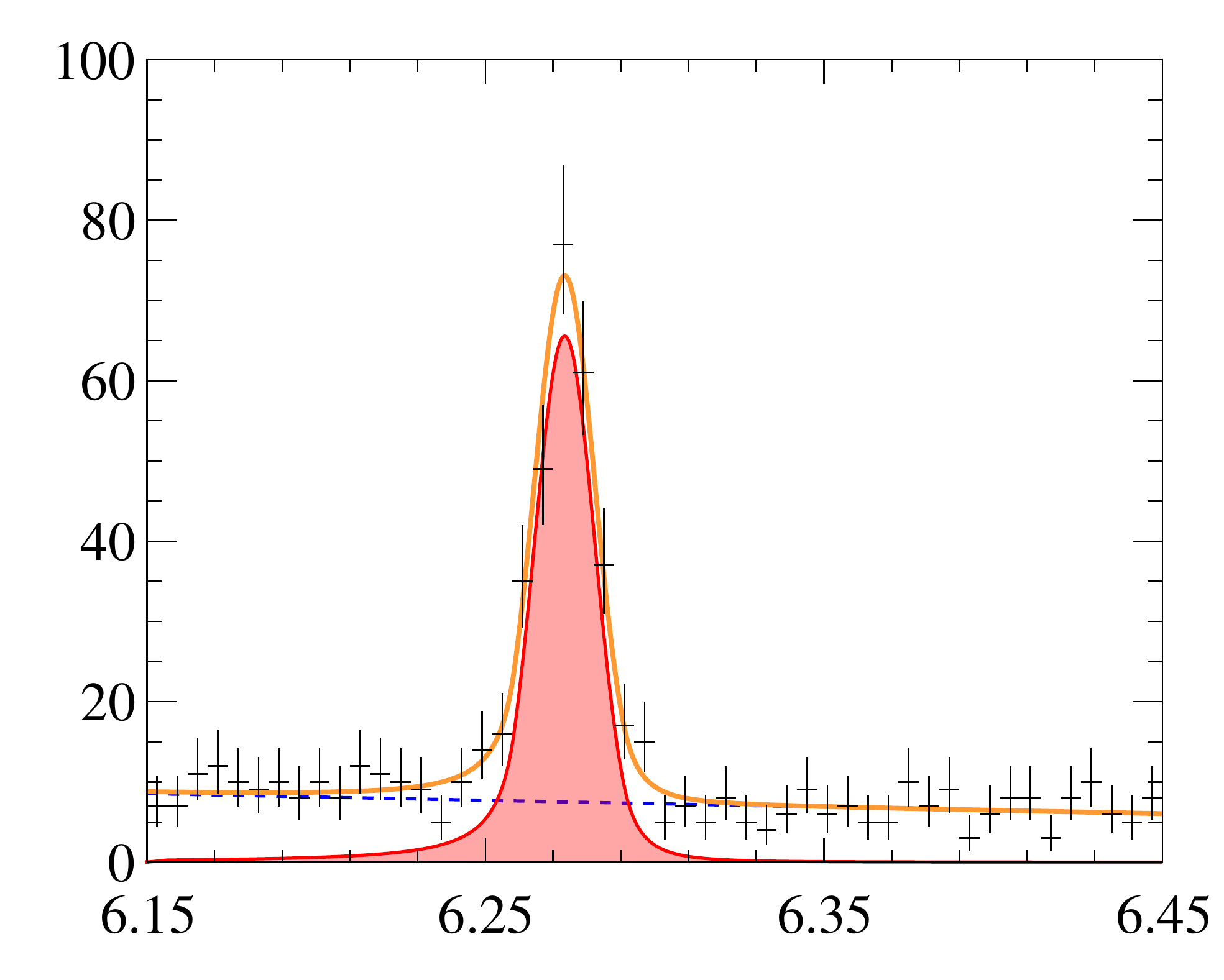}}
    \put(82, 62) {\includegraphics*[width=75mm]{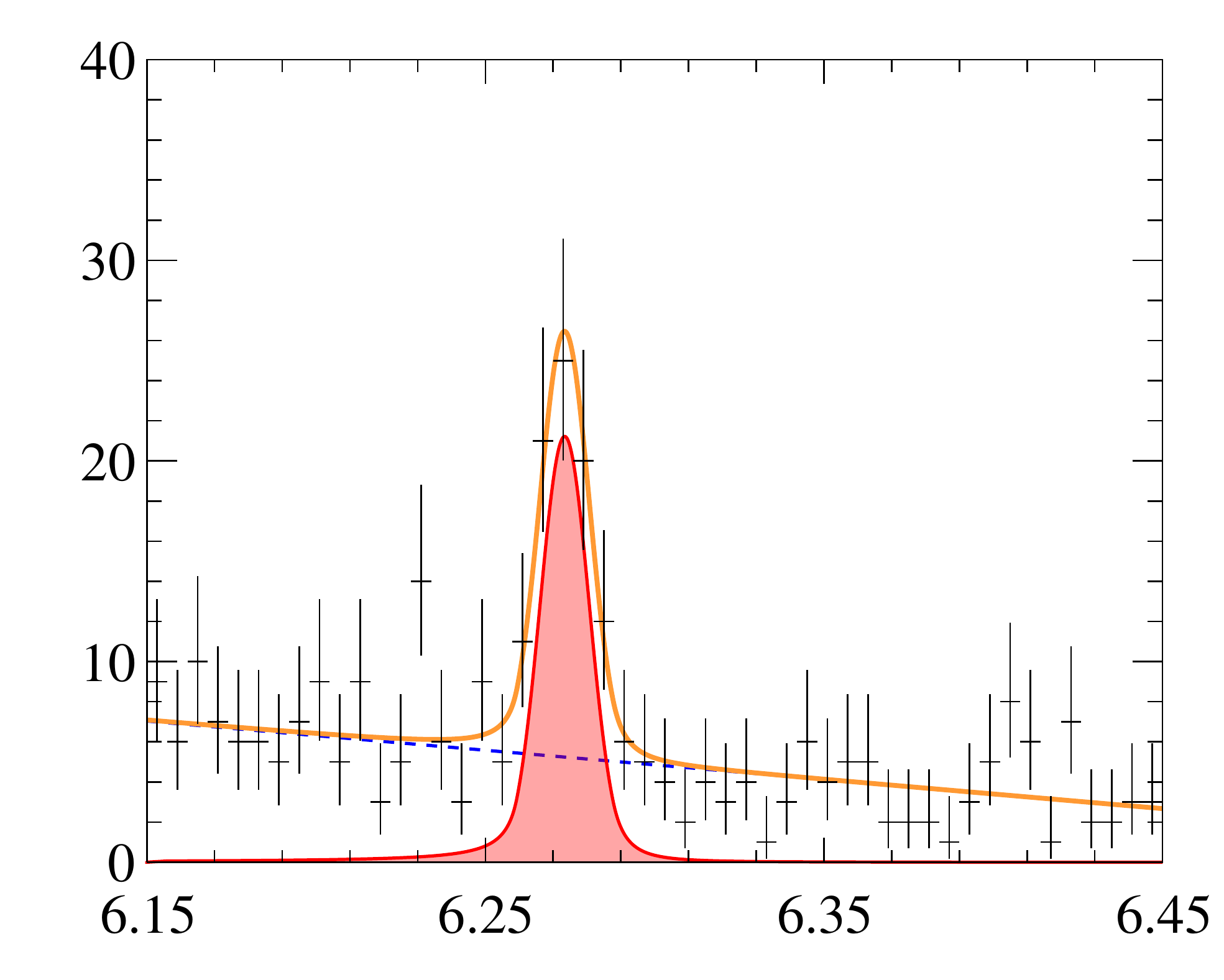}}
	\put( 2,  0) {\includegraphics*[width=75mm]{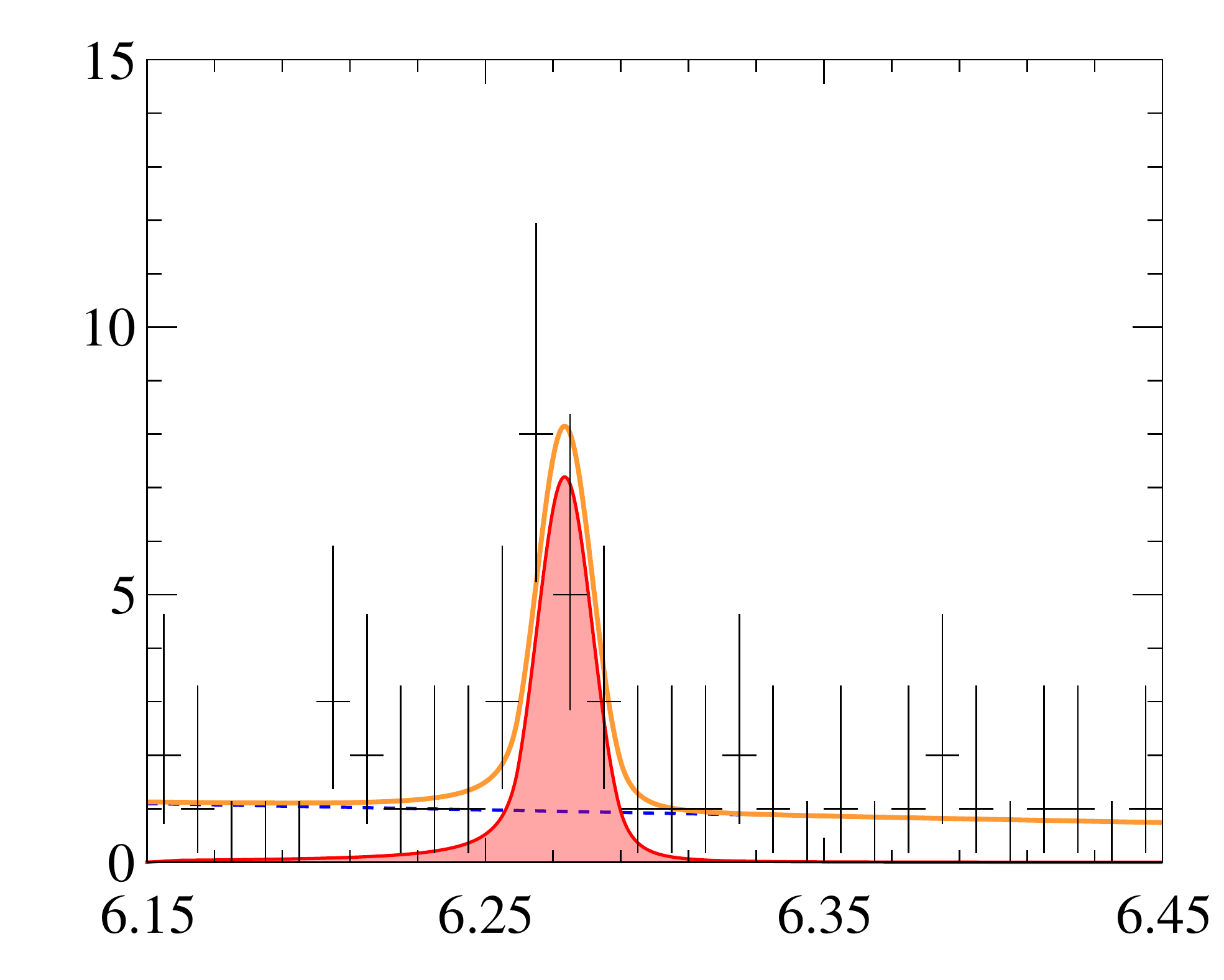}}

	\put(  0, 78){\begin{sideways}{Candidates/$(6\mevcc)$}\end{sideways}}
	\put( 80, 78){\begin{sideways}{Candidates/$(6\mevcc)$}\end{sideways}}
	\put(  0, 14){\begin{sideways}{Candidates/$(10\mevcc)$}\end{sideways}}

	\put( 33 , 60){$m_{\jpsi3\pip2\pim}$}
	\put(108 , 60){$m_{\jpsi\Kp\Km\pip\pip\pim}$}
	\put( 33 , -1){$m_{\jpsi4\pip3\pim}$}

	\put( 59, 60){$\left[\!\gevcc\right]$}
	\put( 59, -1){$\left[\!\gevcc\right]$}
	
	\put( 139, 60){$\left[\!\gevcc\right]$}

	\put( 58,111){$\begin{array}{l}\lhcb\\ 9\invfb \end{array}$}
    \put( 58, 49){$\begin{array}{l}\lhcb\\ 9\invfb \end{array}$}
	\put(138,111){$\begin{array}{l}\lhcb\\ 9\invfb \end{array}$}
	 \put(38,99){\footnotesize$\begin{array}{cl}
	 \!\bigplus\mkern-5mu&\mathrm{data} 
	 \\ 
	 \begin{tikzpicture}[x=1mm,y=1mm]\filldraw[fill=red!35!white,draw=red,thick]  (0,0) rectangle (6,3);\end{tikzpicture} & \decay{\Bc}{\jpsi n\Ph^{\pm}} 
	 \\
	 {\color[RGB]{0,0,255}{\hdashrule[0.0ex][x]{6mm}{1.0pt}{1.0mm 0.4mm}}} & \mathrm{background}
	 \\
	 {\color[RGB]{255,153,51} {\rule{6mm}{2.0pt}}} & \mathrm{total}
	 \end{array}$}

	\end{picture}
	\caption {\small 
	Mass distributions 
	for selected 
	(top left)~\mbox{$\decay{\Bc}{\jpsi3\pip2\pim}$},
	(top right)~\mbox{$\decay{\Bc}{\jpsi\Kp\Km\pip\pip\pim}$}
	and (bottom)~\mbox{$\decay{\Bc}{\jpsi4\pip3\pim}$}~candidates.
    Projections of the~fit, described in the~text, are overlaid.}
	\label{fig:signal_fit_1D}
\end{figure}

\begin{figure}[t]
	\setlength{\unitlength}{1mm}
	\centering
	\begin{picture}(150,60)
	\definecolor{root8}{rgb}{0.35, 0.83, 0.33}
	\put( 2,0) {\includegraphics*[width=75mm]{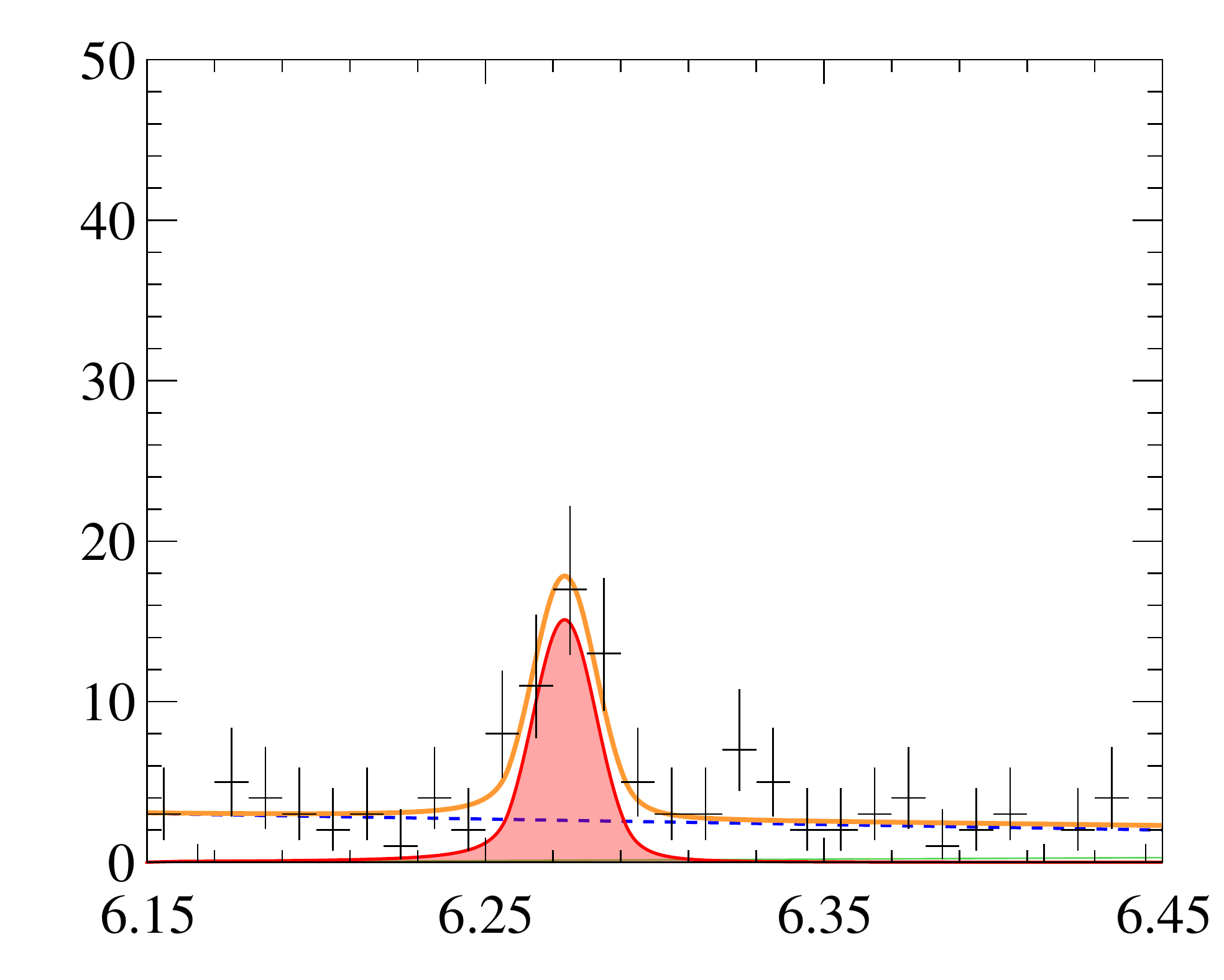}}
	\put(82,0) {\includegraphics*[width=75mm]{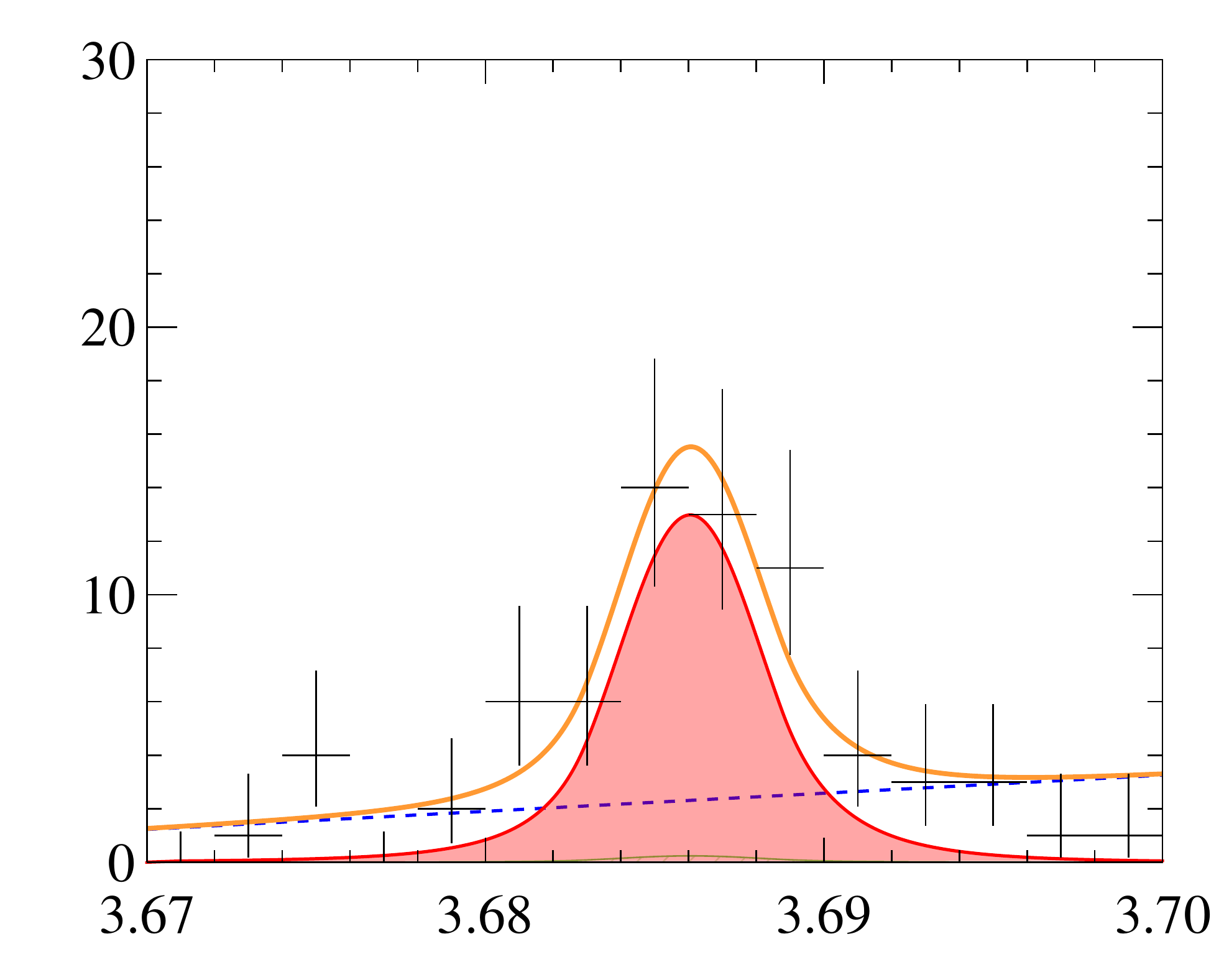}}
	\put( 0,14){\begin{sideways}{Candidates/$(10\mevcc)$}\end{sideways}}
	\put(80,16){\begin{sideways}{Candidates/$(2\mevcc)$}\end{sideways}}
	
	\put(32 ,-1){$m_{\jpsi3\pip2\pim}$}
	\put(116 ,-1){$m_{\jpsi\pip\pim}$}
	\put( 59, -1){$\left[\!\gevcc\right]$}
	\put( 139, -1){$\left[\!\gevcc\right]$}
    \put( 58, 49){$\begin{array}{l}\lhcb\\ 9\invfb \end{array}$}
    \put(138, 49){$\begin{array}{l}\lhcb\\ 9\invfb \end{array}$}

	 \put(12,40){\footnotesize$\begin{array}{cl}
	  \!\bigplus\mkern-10mu&\hspace{-1.5mm}\mathrm{data} 
	 \\ 
	  \begin{tikzpicture}[x=1mm,y=1mm]
	  \filldraw[fill=red!35!white,draw=red,thick]  (0,0) rectangle  (6,3);\end{tikzpicture} 
	   &\hspace{-1.5mm}\BcTopsitwostripi
	   \\
	   \begin{tikzpicture}[x=1mm,y=1mm]
	   \draw[thin,blue,pattern=north west lines, pattern color=blue]  (0,0) rectangle (6,3);\end{tikzpicture} 
       &\hspace{-1.5mm}  \mbox{\decay{\Bc}{\left(\jpsi\pip\pim\right)_{\mathrm{NR}}\pip\pip\pim}} 
       \\ 
	   \begin{tikzpicture}[x=1mm,y=1mm]\draw[thin,root8,pattern=north east lines, pattern color=root8]  (0,0) rectangle (6,3);\end{tikzpicture} 
 	    & \hspace{-1.5mm} \text{comb.}\ \psitwos\pip\pip\pim
 	   \\
 	   \color[RGB]{0,0,255}  {\hdashrule[0.0ex][x]{6mm}{1.0pt}{1.0mm 0.3mm} } 
	   &  \hspace{-1.5mm}\text{background}  
	   \\ 
      \color[RGB]{255,153,51} {\rule{6mm}{2.0pt}}
	   & \hspace{-1.5mm}\text{total} 
	 \end{array}$} 
	\end{picture}
	\caption {\small 
	(left)~Distribution of the~\mbox{$\jpsi3\pip2\pim$}~mass
	for selected 
    \mbox{$\decay{\Bc}{\psitwos\pip\pip\pim}$}~candidates
     with the~$\jpsi\pip\pim$~mass between 3.679 and 3.692\gevcc.
    (right)~Distribution of the~\mbox{$\jpsi\pip\pim$}~mass
    for selected 
    \mbox{$\decay{\Bc}{\psitwos\pip\pip\pim}$}~candidates
    with the~$\jpsi3\pip2\pim$~mass between 6.245 and 6.301\gevcc.
    Projections of the~fit, described in the~text, 
    are overlaid.}
	\label{fig:signal_fit_2D}
\end{figure}

For each decay channel,
when two or more \Bc~candidates 
are found in the~same event,
only one randomly chosen candidate 
is retained
for further analysis.
The~mass distributions for selected 
\BcTojpsifivepi,
\BcTojpsikkpipipi,
and \mbox{\BcTojpsisevenpi} candidates 
are shown in Fig.~\ref{fig:signal_fit_1D}.
Figure~\ref{fig:signal_fit_2D}
shows 
the~mass distributions for selected 
\mbox{\BcTopsitwosjpptripi}~candidates
and for \mbox{$\jpsi\pip\pim$}~combinations
for these candidates. 

\section{Signal yields}
\label{sec:Sig_eff}

The yields  for the~\mbox{\decay{\Bc}{\jpsi n\Ph^\pm}}~decays 
are determined 
using an~extended unbinned 
maximum\nobreakdash-likelihood fit. 
The~fit
is performed simultaneously to 
the~three mass distributions of selected 
\mbox{\BcTojpsifivepi}, 
\BcTojpsikkpipipi
and
\BcTojpsisevenpi~candidates; 
and to the~two\nobreakdash-dimensional distribution
of the~$\Tojpsifivepi$  mass, $m_{\jpsi3\pip2\pim}$, 
versus 
the~$\jpsi\pip\pim$~mass, $m_{\jpsi\pip\pim}$, for
the~\mbox{\BcTopsitwosjpptripi}~candidates. 
Following Refs.~\cite{LHCb-PAPER-2020-009,
LHCb-PAPER-2020-035},
to~improve the~resolution on 
the~$\jpsi\pip\pim$~mass
for the~\mbox{\BcTopsitwosjpptripi}~candidates 
and to eliminate a~small correlation between 
the~$m_{\jpsi3\pip2\pim}$ and 
$m_{\JpsiPiPi}$~variables,
the~$m_{\JpsiPiPi}$~variable 
is computed~\cite{Hulsbergen:2005pu} 
by constraining the~mass of the~\Bc~candidate
to its known value~\cite{LHCb-PAPER-2020-003}.

For~each \Bc~mass distribution, the~one-dimensional 
fit function consists of two components: 
\begin{enumerate}
	\item signal $\decay{\Bc}{\jpsi n\Ph^\pm}$ decays,
	parameterised by a~modified Gaussian function 
	with power\nobreakdash-law tails on both sides of the~distribution~\cite{LHCb-PAPER-2011-013,
	Skwarnicki:1986xj}.
	The~tail parameters are fixed to 
    the~values obtained from simulation; 
    \item random $\jpsi n\Ph^{\pm}$ combinations, 
    modelled by a~first\nobreakdash-order polynomial function.
\end{enumerate}
The two-dimensional fit function for 
the~\mbox{\BcTopsitwosjpptripi} channel 
is defined as the~sum of four components: 
\begin{enumerate}
	\item signal  \BcTopsitwosjpptripi decays,
	parameterised as the~product of 
	\Bc~and $\psitwos$~signal functions 
	each modelled  by a~modified Gaussian function 
	with power\nobreakdash-law tails on both sides 
	of the~distribution~\cite{LHCb-PAPER-2011-013,
	Skwarnicki:1986xj}. 
	The~tail parameters are fixed to 
    the~values obtained from simulation;
	\item 
	contributions 
	from non\nobreakdash-resonant 
	\mbox{\decay{\Bc}
	{\left(\jpsi\pip\pim
	\right)_{\mathrm{NR}}\pip\pip\pim}}~decays, 
	not proceeding through 
	the~intermediate $\psitwos$~state,
            but falling into
            the~\mbox{$3.67<m_{\jpsi\pip\pim}<3.70\gevcc$}~region,
	parameterised as the~product of 
	the~\Bc~signal function and 
	a~phase\nobreakdash-space function describing
	a~three-body out of the~six-body 
	final state~\cite{Byckling}, 
	modified by
	a~positive 
	linear function of 
	the~$\jpsi\pip\pim$~mass; 
	\item random combinations for  
	$\psitwos$ and $\pip\pip\pim$~candidates,
     parameterised as the~product 
     of the~$\psitwos$~signal function
     and a~positive linear function 
     of the~mass of the~$\jpsi3\pip2\pim$~system; 
	\item random $\jpsi3\pip2\pim$ combinations, 
	described by a~two\nobreakdash-dimensional 
	positive\nobreakdash-definite  
    second\nobreakdash-order polynomial function.
\end{enumerate}
\begin{table}[tb]
	\centering
	\caption{\small 
	Signal yields obtained 
	from the~simultaneous  
	unbinned extended 
	maximum\protect\nobreakdash-likelihood fit.
	The uncertainties are statistical only.
	The~last column shows 
	the~statistical significance 
	estimated using Wilks' theorem, 
	in units of standard deviations. 
	}
	\begin{tabular}{lr@{$\,\pm\,$}lc}
	Decay & \multicolumn{2}{c}{Yield}  &  $\mathcal{S}~\left[\upsigma\right]$ 
    \\[1.5mm]
    \hline     \\[-3mm]
  \BcTojpsifivepi
  &  268 & 20  &   $21.0\phantom{0}$
  \\
  \BcTojpsikkpipipi
  &  69 & 11  &  9.1 
  \\
   \BcTojpsisevenpi
   &  16 & 5  &   4.9  
   \\
     \BcTopsitwosjpptripi
  &  40 & 8  &   6.4  
	\end{tabular}
	\label{tab:sim_fit_res}
\end{table}
For all \Bc~signal~functions, 
the~peak\nobreakdash-position parameter 
is shared by all decays and allowed 
to vary in the~fit. 
The~ratio of the~mass resolutions  
of the~\Bc~decays in data and simulation,
$s_{\Bc}$, is shared by all 
decay modes and is allowed 
to vary in the~fit, 
to account for a~discrepancy  
in the~mass resolution  
between data and
simulation~\cite{LHCb-PAPER-2020-008,
LHCb-PAPER-2020-009,
LHCb-PAPER-2020-035}.
The~ratio of the~mass resolution  
of the~\mbox{$\decay{\psitwos}
{\jpsi\pip\pim}$}~decays in data and simulation,
\mbox{$s_{\psitwos}=1.048\pm0.004$},
and the~peak\nobreakdash-position 
parameter 
for the~\psitwos~signal component  
are Gaussian constrained to the~values
obtained from a~previous
\lhcb~study~\cite{LHCb-PAPER-2020-009}.
The~projections of the~fit
are overlaid 
in Fig.\,\ref{fig:signal_fit_1D} for
\mbox{\BcTojpsifivepi}, 
\mbox{\BcTojpsikkpipipi},  
and
\mbox{\BcTojpsisevenpi}~candidates
and in Fig.~\ref{fig:signal_fit_2D} for 
the~\mbox{\BcTopsitwosjpptripi}~candidates.
The~signal yields obtained  from the~fit 
are listed in Table~\ref{tab:sim_fit_res},
along with the~statistical significance 
estimated
using Wilks' theorem~\cite{Wilks:1938dza}. 
The~resolution correction factors 
are found to be 
\mbox{$s_{\Bc} =  1.00 \pm 0.06$} 
and  
\mbox{$s_{\psitwos} = 1.048 \pm 0.004$}.
For all previously unobserved modes, 
the~significance is confirmed by simulating 
a~large number of pseudoexperiments according 
to the~background 
distribution observed in data.

\begin{figure}[t]
  \setlength{\unitlength}{1mm}
  \centering
  \begin{picture}(150,120)
  
 \put( 2,62) {\includegraphics*[width=75mm]{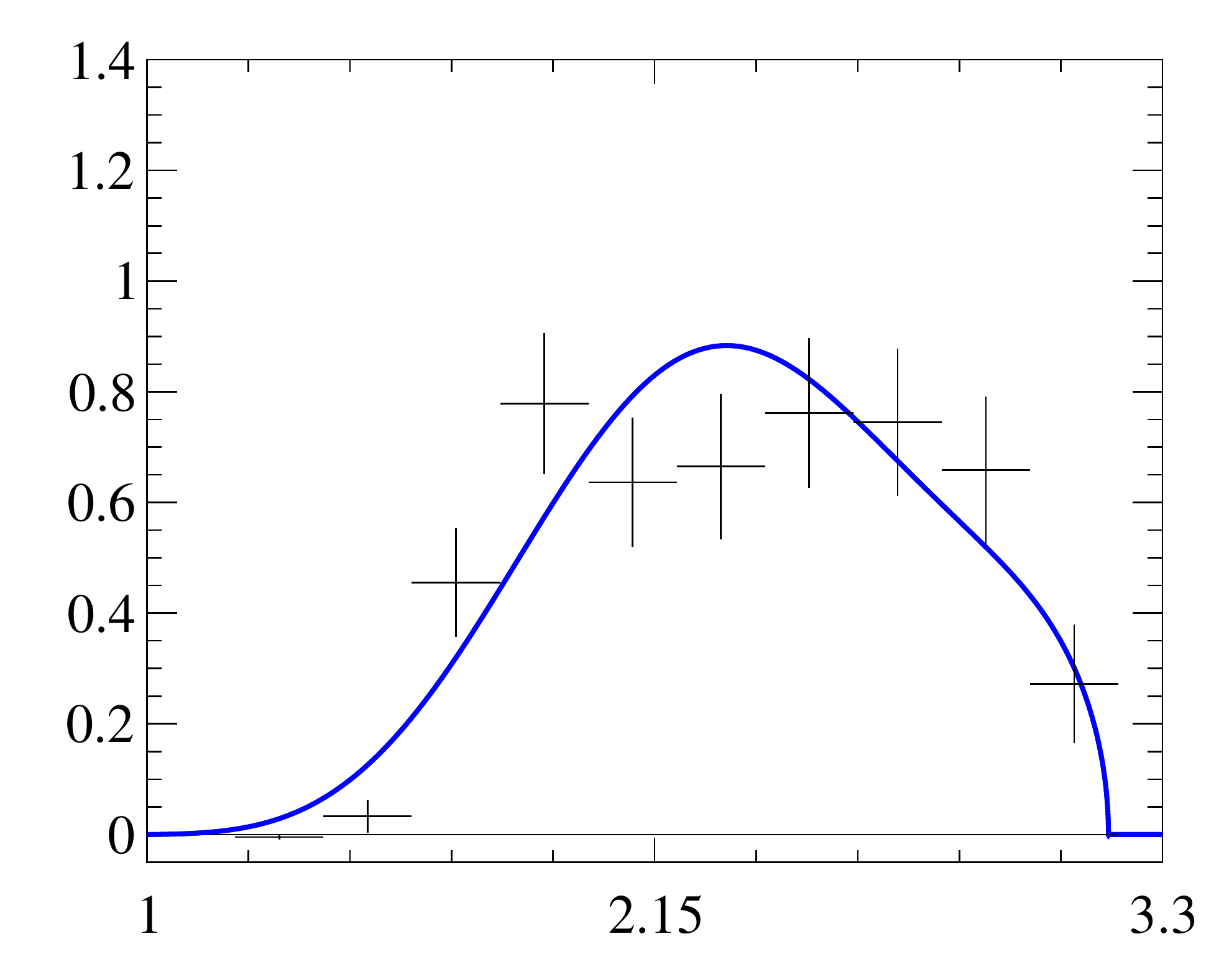}}
 \put(82,62) {\includegraphics*[width=75mm]{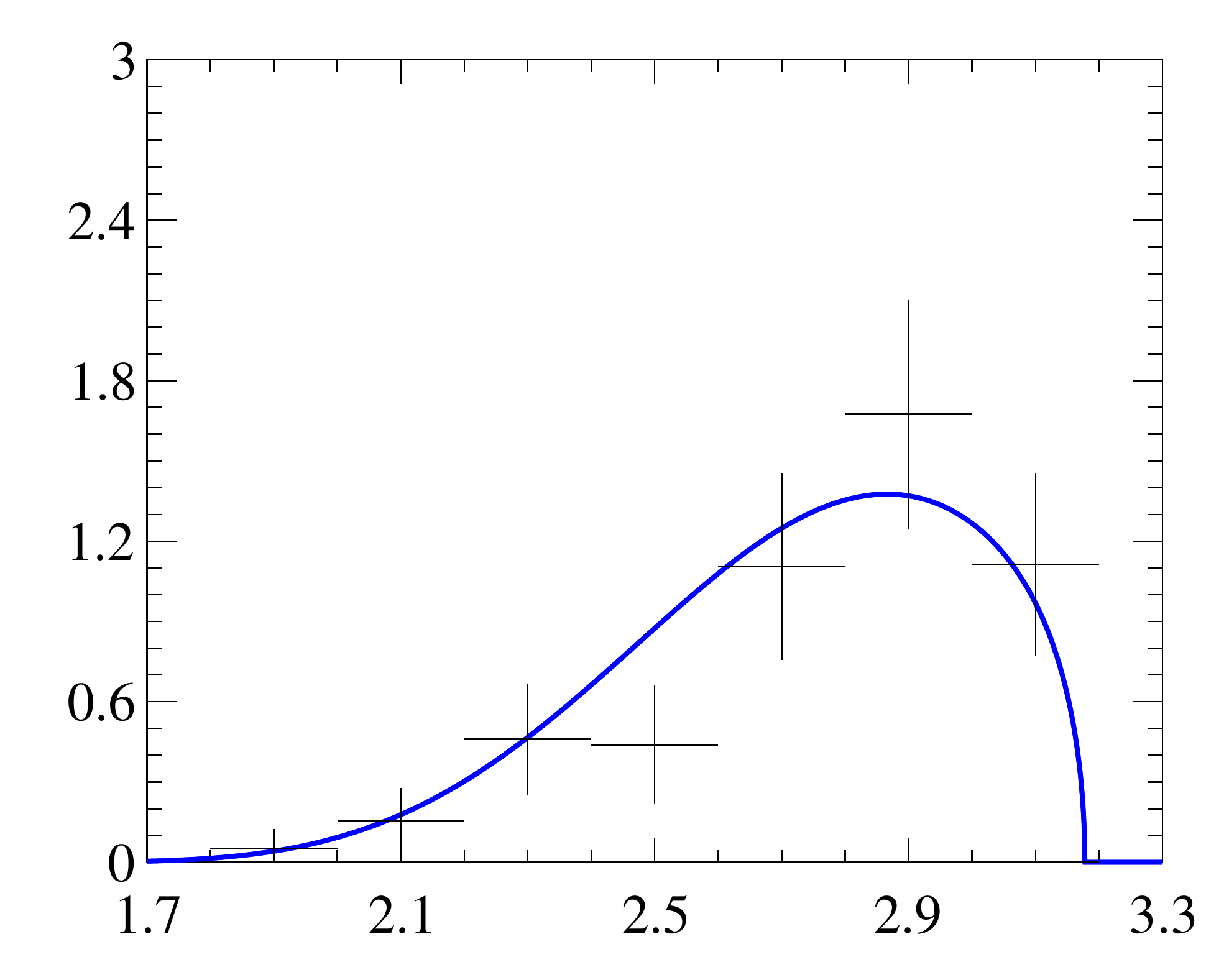}}
 \put( 2, 0) {\includegraphics*[width=75mm]{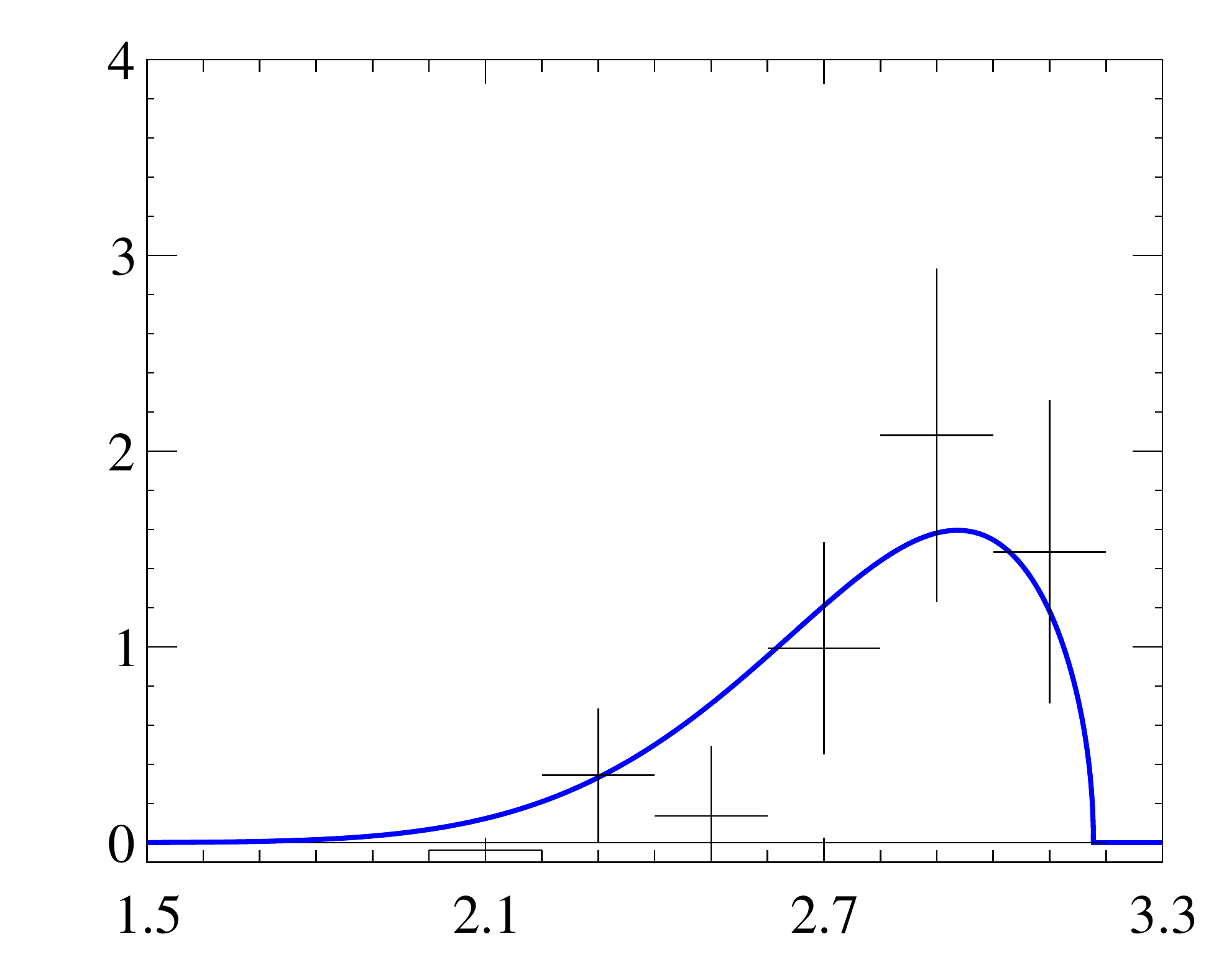}}
	\put(92,105){\small$\begin{array}{cl}
		 \color{black}{\bigplus\mkern-18mu\phantom{\bullet}}\  & \mathrm{data}  \\ \,	
		 \color[RGB]{0,0,255} {\rule{6mm}{2.0pt}}  & \mathrm{simulation\,(BLL)}
	\end{array}$}
	\put(-4, 90) {\begin{sideways}\large 
	$\frac{1}{N}\frac{\mathrm{d}N}
	{\mathrm{d}m}~\left[\frac{1}{\!\gevcc}\right]$
	\end{sideways}}
	\put(77, 90) {\begin{sideways}\large 
	 $\frac{1}{N}\frac{\mathrm{d}N}
	 {\mathrm{d}m}~\left[\frac{1}{\!\gevcc}\right]$
	\end{sideways}}
	\put(35 ,60){$m_{3\pip2\pim}$}
	\put(110,60){$m_{\Kp\Km\pip\pip\pim}$}
	\put( 59,60){$\left[\!\gevcc\right]$}
	\put(139,60){$\left[\!\gevcc\right]$}
	\put( 58,111){$\begin{array}{l} \lhcb \\ 9\invfb \end{array}$} 
	\put(138,111){$\begin{array}{l} \lhcb \\ 9\invfb \end{array}$} 
    \put(13,113){\BcTojpsifivepi}    
    \put(93,113){\decay{\Bc}{\jpsi\Kp\Km\pip\pip\pim}}    

    \put( 32,-1){$m_{4\pip3\pim}$}
	\put( 59,-1){$\left[\!\gevcc\right]$}
	\put( 58,49){$\begin{array}{l} \lhcb \\ 9\invfb \end{array}$} 
    \put( 13,51){\decay{\Bc}{\jpsi4\pip3\pim}}  
    \put(-3, 28) {\begin{sideways}\large 
	 $\frac{1}{N}\frac{\mathrm{d}N}
	 {\mathrm{d}m}~\left[\frac{1}{\!\gevcc}\right]$
	\end{sideways}}
	\end{picture}
	\caption {\small 
     Mass spectra for 
     the~light\protect\nobreakdash-hadron
     system for 
     the~(top left)~\mbox{$\decay{\Bc}{\jpsi3\pip2\pim}$},
     (top right)~\mbox{$\decay{\Bc}{\jpsi\Kp\Km\pip\pip\pim}$}
     and (bottom)~~\mbox{$\decay{\Bc}{\jpsi4\pip3\pim}$}~decays.
     Expectations from the~BLL model are overlaid. 
     }
	\label{fig:5h_mass}
\end{figure}

\begin{figure}[t]
  \setlength{\unitlength}{1mm}
  \centering
  \begin{picture}(150,60)
  \definecolor{root8}{rgb}{0.35, 0.83, 0.33}

  \put( 2,0) {\includegraphics*[width=75mm]{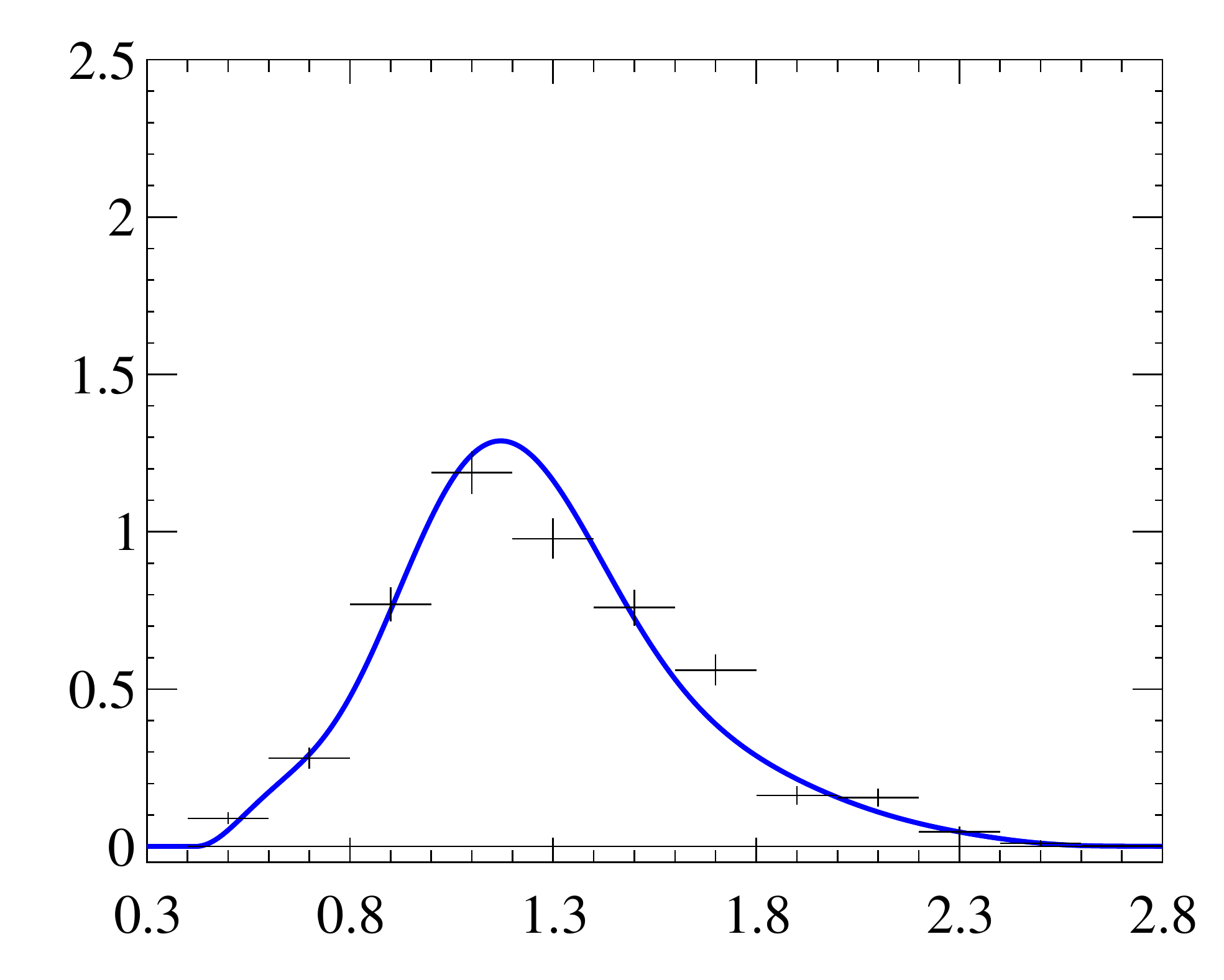}}
  \put(82,0) {\includegraphics*[width=75mm]{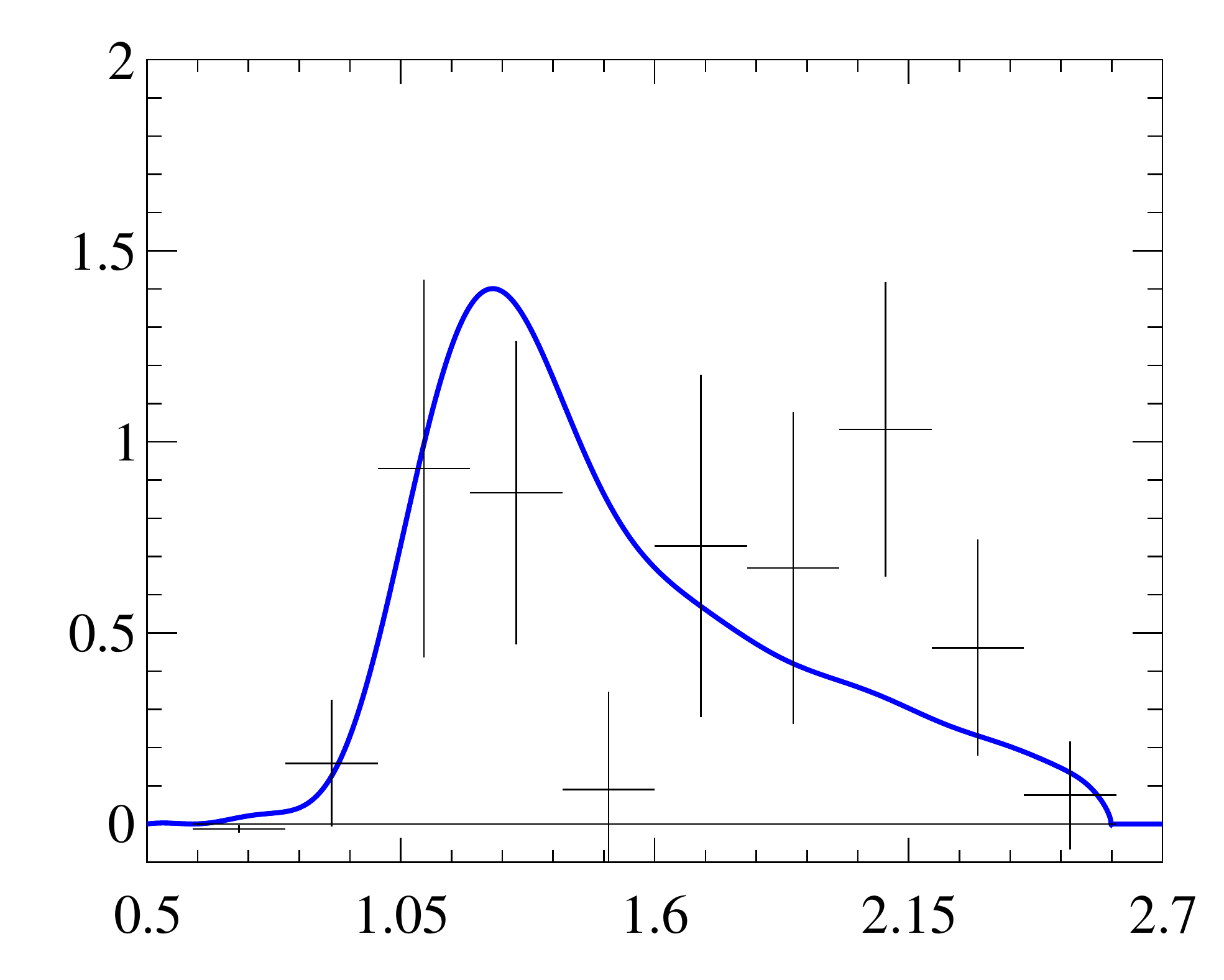}}
  
	\put(-3, 28) {\begin{sideways}\large 
	$\frac{1}{N}\frac{\mathrm{d}N}
	{\mathrm{d}m}~\left[\frac{1}{\!\gevcc}\right]$
	\end{sideways}}	
   \put(77, 28) {\begin{sideways}\large 
   $\frac{1}{N}\frac{\mathrm{d}N}
   {\mathrm{d}m}~\left[\frac{1}{\!\gevcc}\right]$
   \end{sideways}}	

   \put(13,51){\BcTojpsifivepi} 
   \put(93,51){$\decay{\Bc}{\psitwos\pip\pip\pim}$} 

	\put(12,40){\small$\begin{array}{cl}
		 \color{black}{\bigplus\mkern-18mu\phantom{\bullet}}\  & \mathrm{data}  \\ \,	
		 \color[RGB]{0,0,255} {\rule{6mm}{2.0pt}}  & \mathrm{simulation\,(BLL)}
	\end{array}$}
	\put(35  ,-1){$m_{\pip\pip\pim}$}
	\put(59  ,-1){$\left[\!\gevcc\right]$}
 	\put(115 ,-1){$m_{\pip\pip\pim}$}
	\put(139 ,-1){$\left[\!\gevcc\right]$}
    \put( 58, 49){$\begin{array}{l}
    \lhcb\\ 9\invfb \end{array}$}
    \put(138, 49){$\begin{array}{l}
    \lhcb\\ 9\invfb \end{array}$}

  \end{picture}
  \caption{\small
  Mass spectra for 
  the~\mbox{$\pip\pip\pim$} combinations
  from 
  the~(left)~\mbox{$\BcTojpsifivepi$} 
  (6~entries per \Bc~candidate)
  and (right)~\mbox{$\decay{\Bc}
  {\psitwos\pip\pip\pim}$} decays.
  The expectations from the~BLL~model
  are overlaid.
  }
  \label{fig:m_3pi} 
\end{figure}

The~background\nobreakdash-subtracted 
mass spectra for the~light\nobreakdash-hadron 
system
for the~observed decays of the~\Bc~mesons
are obtained using 
the~\sPlot technique~\cite{Pivk:2004ty},
based on the~results of the~fit described above. 
The~distributions are shown in 
Figs.~\ref{fig:5h_mass} 
and~\ref{fig:m_3pi}\,(right) 
together
with the~expectations from the~BLL~model.
For all cases,
good agreement with the~BLL~model is observed.
No~\mbox{$\decay{\Ds}{3\pip2\pim}$},  
\mbox{$\decay{\Ds}{4\pip3\pim}$}
or \mbox{$\decay{\Ds}{\pip\pip\pim}$}~signals are observed in the studied spectra. 

\begin{figure}[t]
  \setlength{\unitlength}{1mm}
  \centering
  \begin{picture}(150,60)
  
  	\definecolor{root8}{rgb}{0.35, 0.83, 0.33}
	
  	\put( 2,0) {\includegraphics*[width=75mm]{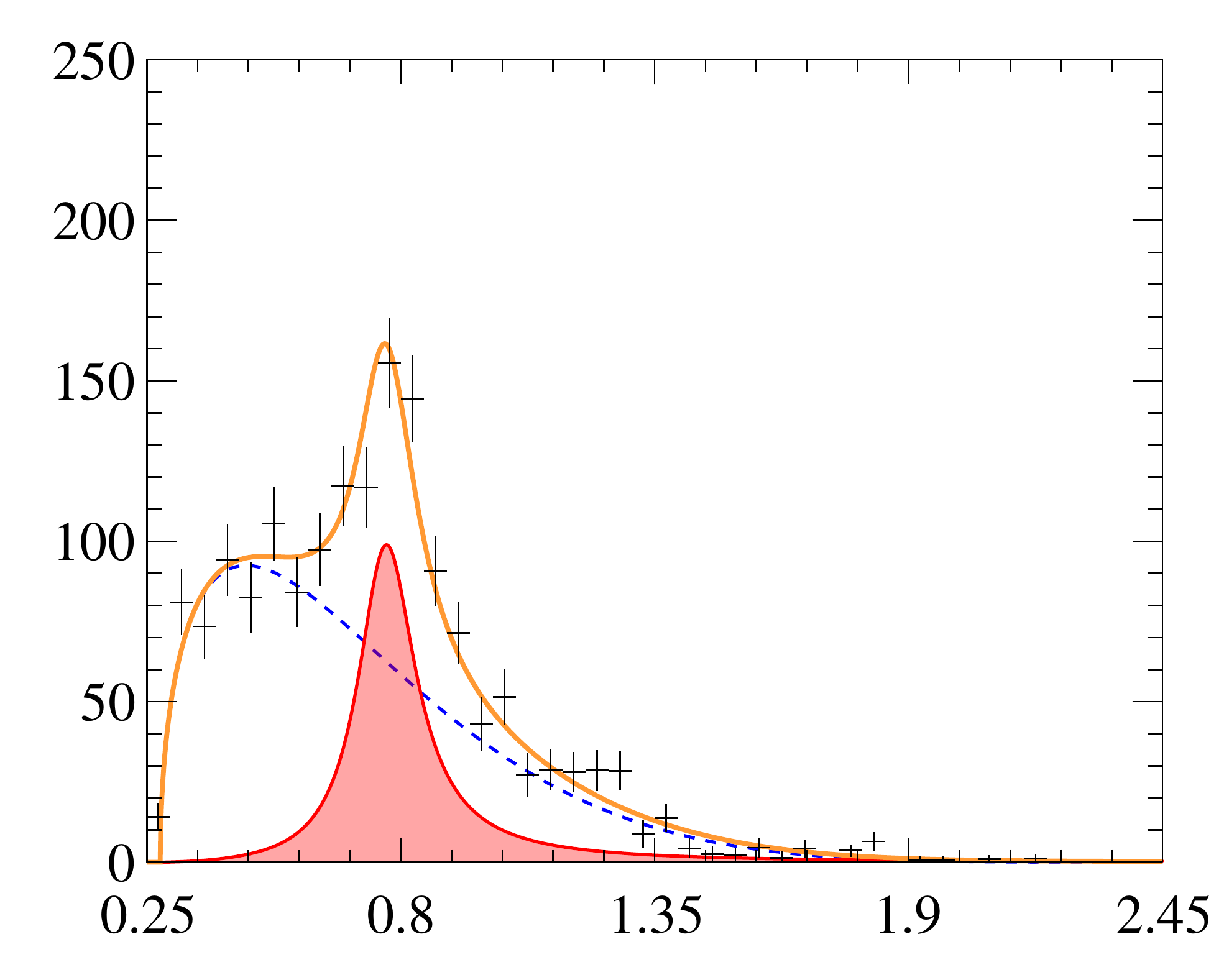}}
	\put(82,0) {\includegraphics*[width=75mm]{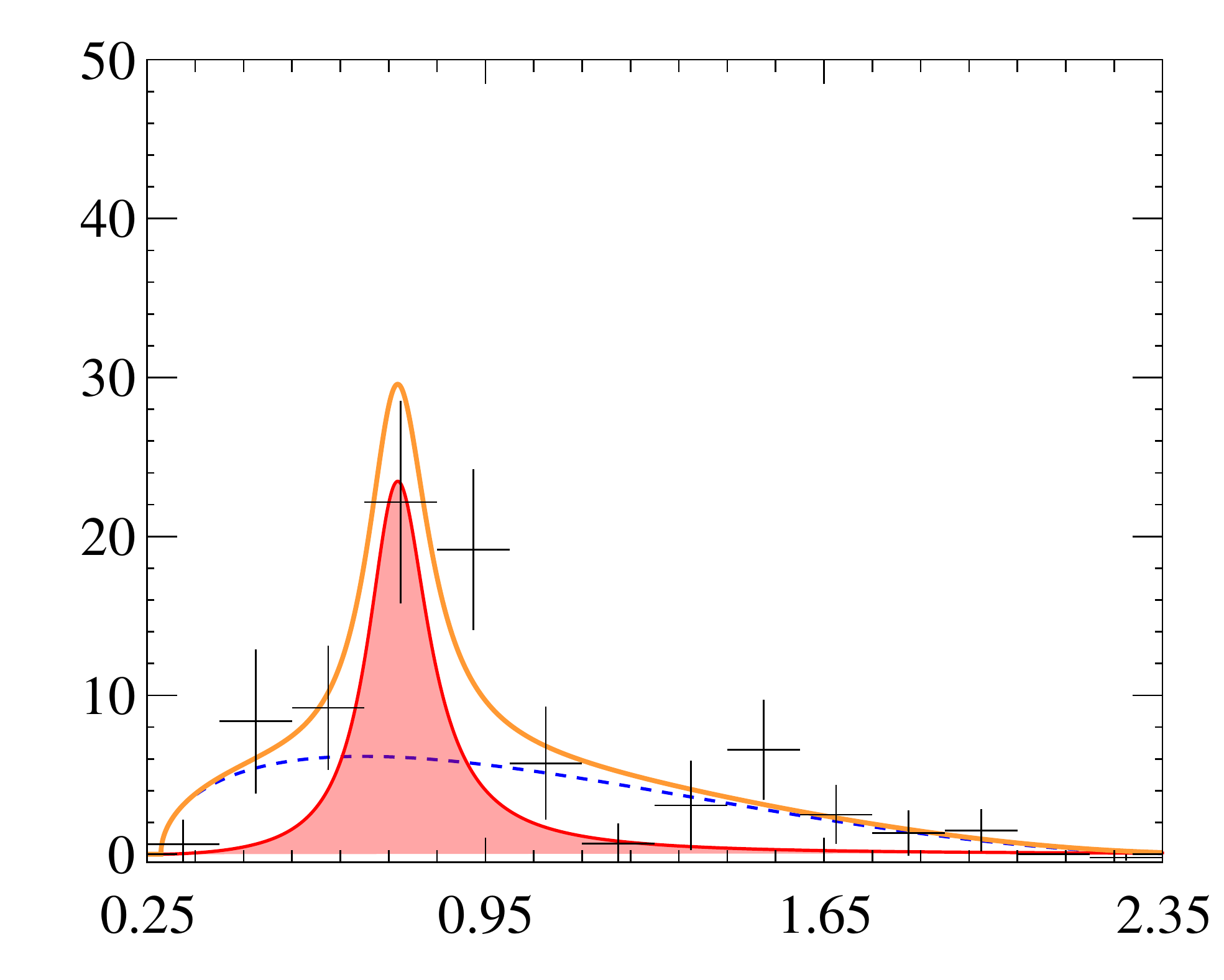}}
	\put(0,25){\begin{sideways}{Yield/$(50\mevcc)$}\end{sideways}}
	\put(79,22){\begin{sideways}{Yield/$(150\mevcc)$}\end{sideways}}
	
	\put(37 ,-1){$m_{\pip\pim}$}
	\put(117,-1){$m_{\pip\pim}$}
	\put(59 ,-1){$\left[\!\gevcc\right]$}
	\put(139,-1){$\left[\!\gevcc\right]$}

   \put( 58, 49){$\begin{array}{l}\lhcb\\ 9\invfb \end{array}$}
   \put( 138, 49){$\begin{array}{l}\lhcb\\ 9\invfb \end{array}$}
 
    \put(13,51){\decay{\Bc}{\jpsi3\pip2\pim}} 
    \put(93,51){\decay{\Bc}{\psitwos\pip\pip\pim}} 
     
	\put(33,35){\footnotesize$\begin{array}{cl}
	\!\bigplus\mkern-10mu&\hspace{-1.5mm}\mathrm{data} 
	\\ 
	\begin{tikzpicture}[x=1mm,y=1mm]
	\filldraw[fill=red!35!white,draw=red,thick]  (0,0) rectangle (6,3);
	\end{tikzpicture} 
	&\hspace{-1.5mm}\decay{\Prho^0}{\pip\pim}
    \\ 
 	\color[RGB]{0,0,255}     {\hdashrule[0.0ex][x]{6mm}{1.0pt}{1.0mm 0.3mm} } 
	&  \hspace{-1.5mm}\mathrm{\text{non-resonant}}
	\\ 
    \color[RGB]{255,153,51} {\rule{6mm}{2.0pt}}
	& \hspace{-1.5mm}\text{total} 
	\end{array}$} 

	\put(115,35){\footnotesize$\begin{array}{cl}
	\!\bigplus\mkern-10mu&\hspace{-1.5mm}\mathrm{data} 
	\\ 
	\begin{tikzpicture}[x=1mm,y=1mm]
	\filldraw[fill=red!35!white,draw=red,thick]  (0,0) rectangle (6,3);
	\end{tikzpicture} 
	&\hspace{-1.5mm}\decay{\Prho^0}{\pip\pim}
    \\ 
 	\color[RGB]{0,0,255}     {\hdashrule[0.0ex][x]{6mm}{1.0pt}{1.0mm 0.3mm} } 
	&  \hspace{-1.5mm}\mathrm{\text{non-resonant}}
	\\ 
    \color[RGB]{255,153,51} {\rule{6mm}{2.0pt}}
	& \hspace{-1.5mm}\text{total} 
	\end{array}$} 

  \end{picture}
  \caption{\small
  Background-subtracted 
  $\pip\pim$~mass distributions
  from (left)~\mbox{$\decay{\Bc}
  {\jpsi3\pip2\pim}$}\,(6~entries per \Bc~candidate)
  and (right)~\mbox{$\decay{\Bc}{\psitwos\pip\pip\pim}$}\,(2~entries per \Bc~candidate)~decays.
  The~results of the~fits described in the~text are overlaid.
  }
  \label{fig:mpipi} 
\end{figure}
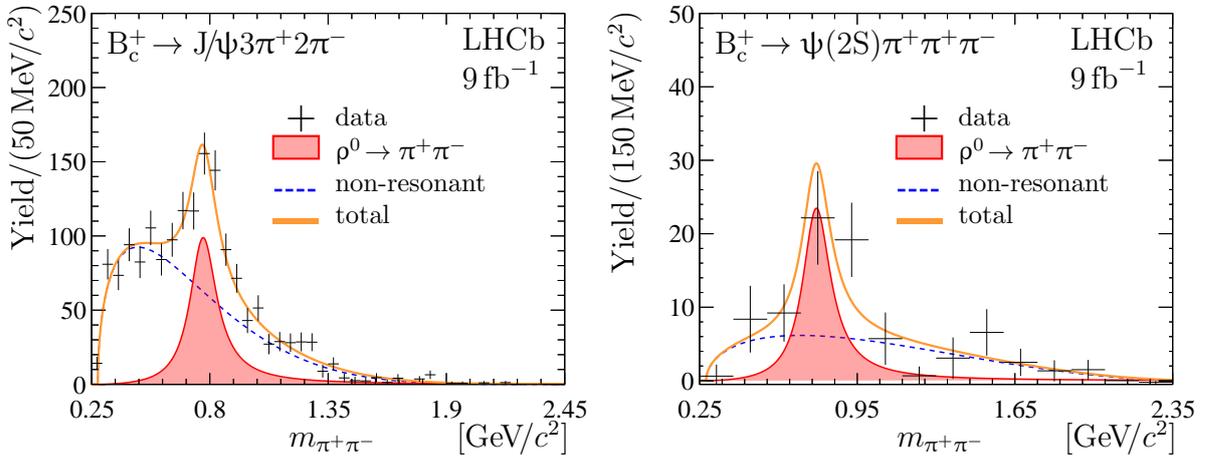

The~background\nobreakdash-subtracted
$\pip\pip\pim$~mass distribution from 
the~\mbox{$\decay{\Bc}{\jpsi3\pip2\pim}$}~decays
is shown in Fig.~\ref{fig:m_3pi}\,(left). 
The~observed spectrum is in good agreement with 
the~expectations from 
the~BLL~model. 
The~background\nobreakdash-subtracted
$\pip\pim$~mass spectra 
from  the~\mbox{$\decay{\Bc}{\jpsi3\pip2\pim}$}
and~\mbox{$\decay{\Bc}
{\psitwos\pip\pip\pim}$}~decays are shown in 
Fig.~\ref{fig:mpipi}. 
Figures~\ref{fig:m_3pi} and~\ref{fig:mpipi}
contain all possible 
$\pip\pip\pim$ 
and $\pip\pim$  
combinations from 
a~single \Bc~candidate.
The~fits
to the~$\pip\pim$~mass
distributions are performed using  
a~function that contains two~terms:
a~component corresponding to decays 
via the~intermediate
\mbox{$\decay{\rhoz}{\pip\pim}$}~resonance
and a~smooth function describing
the~$\pip\pim$~mass spectrum
without 
a~\mbox{$\decay{\rhoz}{\pip\pim}$}~signal,
labelled as ``non\nobreakdash-resonant'' in Fig.~\ref{fig:mpipi}. 
The~resonance component is parameterised with
a~relativistic P\nobreakdash-wave
Breit\nobreakdash--Wigner function  
with a~Blatt\nobreakdash--Weisskopf form factor 
with a~meson radius
of~$3.5\gev^{-1}$~\cite{Blatt:1952ije}.
The~non\nobreakdash-resonant component is parameterised 
with the~product of the~phase\nobreakdash-space function 
describing
a~two\nobreakdash-body
combination
from a~six\nobreakdash-body combination
in the~$\decay{\Bc}{\jpsi3\pip2\pim}$~case
and
    a~two\nobreakdash-body
    combination from
a~four\nobreakdash-body combination in the~\mbox{$\decay{\Bc}
{\psitwos\pip\pip\pim}$}~case~\cite{Byckling},
and a~positive first\nobreakdash-order 
polynomial function 
that accounts for the~unknown decay dynamics. 
The~results of the~fits, overlaid in
Fig.~\ref{fig:mpipi},
are consistent
with a~large fraction of the~decays 
proceeding 
via an~intermediate
\mbox{$\decay{\rhoz}{\pip\pim}$}~resonance, 
as expected within the~BLL model. 
Making a~more quantitative statement 
would require a~more complicated treatment 
of the~multihadron system, 
which is beyond the~scope of this~paper.

\begin{figure}[t]
  \setlength{\unitlength}{1mm}
  \centering
  \begin{picture}(150,60)
  \definecolor{root8}{rgb}{0.35, 0.83, 0.33}
	
  	\put( 2,0) {\includegraphics*[width=75mm]{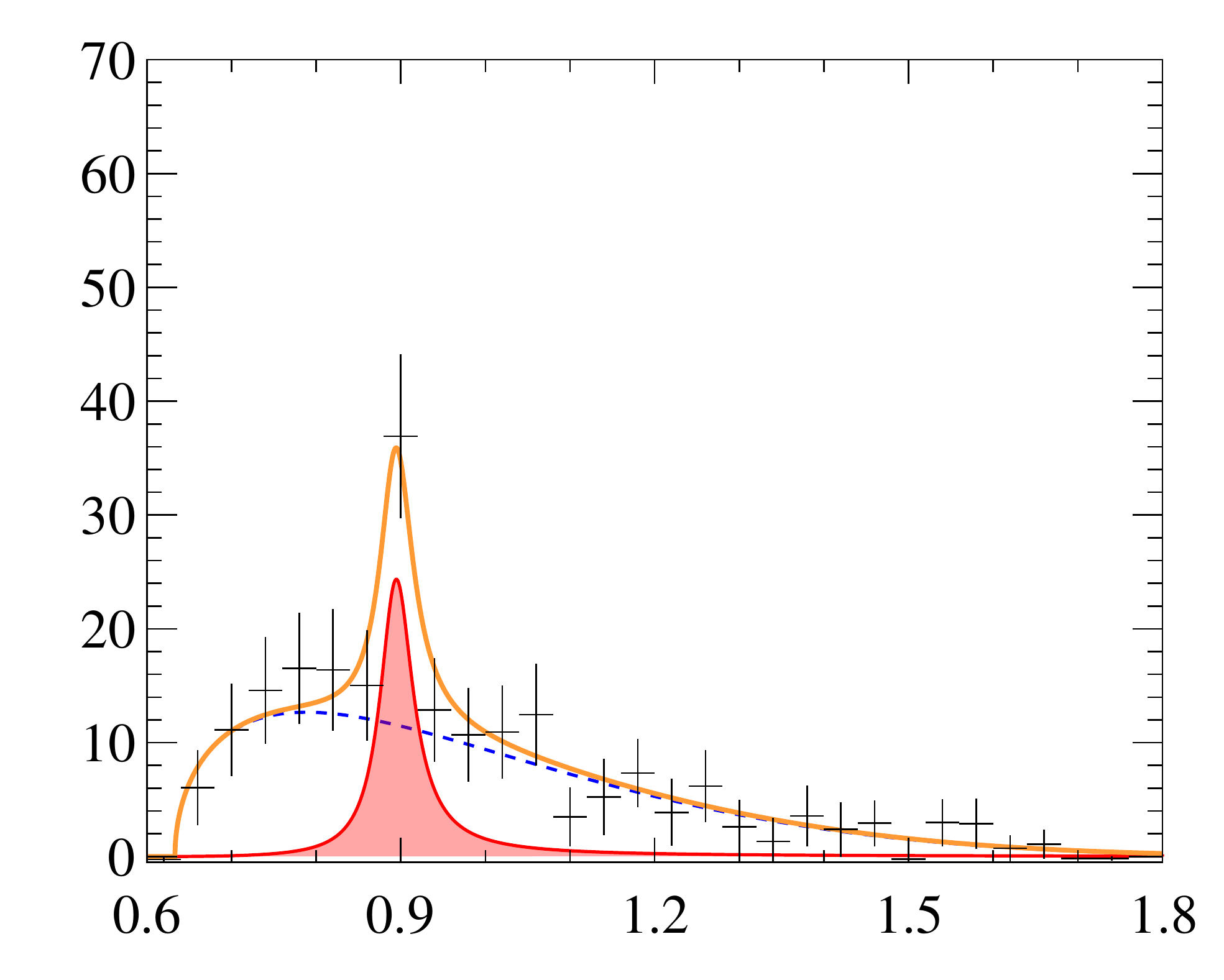}}
	\put(82,0) {\includegraphics*[width=75mm]{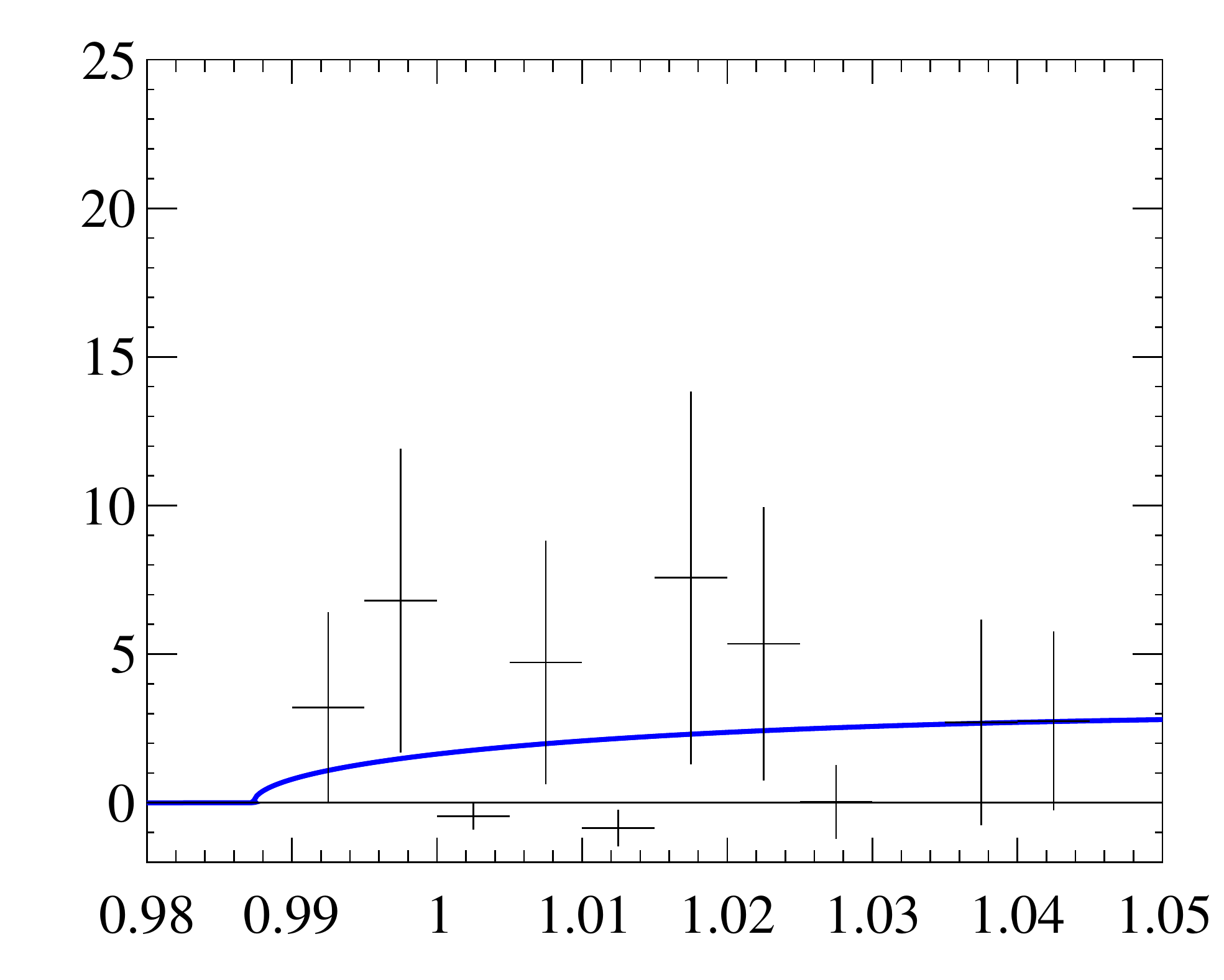}}
	\put(0,25){\begin{sideways}{Yield/$(40\mevcc)$}\end{sideways}}
	\put(77, 28) {\begin{sideways}\large 
	 $\frac{1}{N}\frac{\mathrm{d}N}
	 {\mathrm{d}m}~\left[\frac{1}{\!\gevcc}\right]$
	\end{sideways}}	
   \put(13,50){\decay{\Bc}{\jpsi\Kp\Km\pip\pip\pim}}  \put(93,50){\decay{\Bc}{\jpsi\Kp\Km\pip\pip\pim}} 
	\put(32,35){\small$\begin{array}{cl}
	\!\bigplus\mkern-10mu&\hspace{-1.5mm}\mathrm{data} 
	\\ 
	\begin{tikzpicture}[x=1mm,y=1mm]
	\filldraw[fill=red!35!white,draw=red,thick]  (0,0) rectangle  (6,3);\end{tikzpicture} 
	&\hspace{-1.5mm}\decay{\KorKbarzstar}{\kaon^{\pm}\pion^{\mp}}
    \\ 
 	\color[RGB]{0,0,255}     {\hdashrule[0.0ex][x]{6mm}{1.0pt}{1.0mm 0.3mm} } 
	&  \hspace{-1.5mm}\mathrm{\text{non-resonant}} 
	\\ 
    \color[RGB]{255,153,51} {\rule{6mm}{2.0pt}}
	& \hspace{-1.5mm}\text{total} 
	\end{array}$} 
	\put(94,40){\small$\begin{array}{cl}
		 \color{black}{\bigplus\mkern-18mu\phantom{\bullet}}\  & \mathrm{data}  \\ \,	
		 \color[RGB]{0,0,255} {\rule{6mm}{2.0pt}}  & \mathrm{simulation\,(BLL)}
	\end{array}$}
	\put(35  ,-1){$m_{\Kpm\pimp}$}
	\put(113 ,-1){$m_{\Kp\Km}$}
	\put( 59 ,-1){$\left[\!\gevcc\right]$}
	\put( 139,-1){$\left[\!\gevcc\right]$}
    \put( 58, 49){$\begin{array}{l}\lhcb\\ 9\invfb \end{array}$}
    \put( 138, 49){$\begin{array}{l}\lhcb\\ 9\invfb \end{array}$}
  \end{picture}
  \caption{\small
  Background-subtracted 
  (left)~$\kaon^{\pm}\pion^{\mp}$\,(3~entries per \Bc~candidate)~mass
  and (right)~low\protect\nobreakdash-mass
  part of the~$\Kp\Km$~mass distribution 
  from the~\mbox{$\decay{\Bc}{\jpsi\Kp\Km\pip\pip\pim}$}~decays. 
  The~results of the~fit described in the~text are overlaid
  on the~left plot, while expectations from the~BLL~model
  are overlaid on the~right plot.
  }
  \label{fig:m_kk3pi} 
\end{figure}
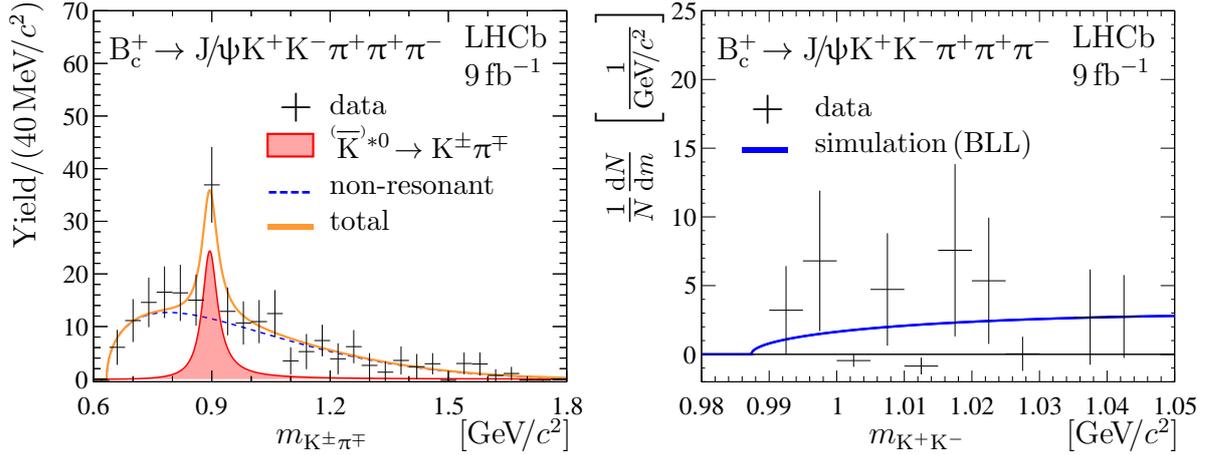

The~background\nobreakdash-subtracted $\Kp\pim$
and $\Km\pip$~mass spectra 
and the~low\nobreakdash-mass part of  
the~$\Kp\Km$~mass spectrum 
from 
the~\mbox{$\decay{\Bc}{\jpsi\Kp\Km\pip\pip\pim}$}~decays
are shown 
in Fig.~\ref{fig:m_kk3pi}.
A~fit to
the~$\kaon^{\pm}\pion^{\mp}$~mass spectrum 
is performed using 
a~two\nobreakdash-component function, 
similar to the~function described above,  
and 
consisting of 
a~component corresponding to decays 
via the~intermediate $\Kstarz$ or \Kstarzb~resonance
and a~smooth function describing 
decays without a~$\Kstarz$ or $\Kstarzb$~resonance.
The~resonance component is parameterised with
a~relativistic P\nobreakdash-wave
Breit\nobreakdash--Wigner function.
Fit~results are overlaid in 
Fig.~\ref{fig:m_kk3pi}\,(left) and 
indicate a~presence of 
decays via intermediate 
$\Kstarz$ and $\Kstarzb$~mesons.
The~$\Kp\Km$~mass spectrum, 
shown in Fig.~\ref{fig:m_kk3pi}\,(right),  
exhibits no sign of
the~\mbox{$\Pphi$}~resonance, 
in agreement both with the~expected suppression 
of the~\Pphi~meson production due to
the~Okubo\nobreakdash--Zweig\nobreakdash--Iizuka~rule~\cite{Okubo:1963fa, 
Zweig2, 
Iizuka:1966fk,
Singh:1977sa,
Singh:1978mi} and with 
expectations from the~BLL~model.
A~similar suppression has been observed 
for the~\mbox{$\decay{\Bc}
{\jpsi\Kp\Km\pip}$}~decays~\cite{LHCb-PAPER-2013-047,
LHCb-PAPER-2021-034}.

 \section{Ratios of branching fractions}
 \label{sec:ratios}

Three ratios of branching fractions 
are reported in this paper,  
\begin{subequations}
\label{eq:ratios}
\begingroup
\allowdisplaybreaks
\begin{eqnarray}
\mathcal{R}^{\Tojpsikkpipipi}_{\Tojpsifivepi} 
& \equiv & 
\dfrac{   \BR({\BcTojpsikkpipipi})  } 
{   \BR({\BcTojpsifivepi})  }     \,, 
\\
\mathcal{R}^{\Tojpsisevenpi}_{\Tojpsifivepi} 
& \equiv & 
\dfrac{\BR({\BcTojpsisevenpi})} 
      {\BR({\BcTojpsifivepi})}     \,, 
\\
\mathcal{R}^{\Topsitwostripi}_{\Tojpsifivepi} 
& \equiv & 
\dfrac{   \BR({\BcTopsitwostripi}) 
\times    \BR({\decay{\psitwos}{\JpsiPiPi}}) } 
{         \BR({\BcTojpsifivepi})  }  \,. 
\end{eqnarray}
\endgroup
\end{subequations}
Each ratio of branching fractions for 
the~decays of \Bc~mesons
into the~final states $\PX$ and $\PY$
is calculated~as  
\begin{equation}
     \mathcal{R}^{\mathrm{X}}_{\mathrm{Y}} =
     \dfrac{ N_{\mathrm{X}}}
           { N_{\mathrm{Y}}} \times 
      \dfrac{ \upvarepsilon_{\mathrm{Y}}}
           { \upvarepsilon_{\mathrm{X}}}\,, \label{eq:br_rat}
\end{equation}
where $N$ is the~signal yield 
reported in Table~\ref{tab:sim_fit_res} 
and $\upvarepsilon$~denotes 
the~corresponding efficiency.
The~efficiency is defined as the~product 
of geometric acceptance and of 
reconstruction, selection, hadron-identification 
and trigger efficiencies. 
All of these~contributions, 
except that 
of the~hadron\nobreakdash-identification
efficiency, 
are determined using simulated samples, 
corrected as described in Sec.~\ref{sec:Detector}.
The~hadron\nobreakdash-identification 
efficiency is  calculated 
separately for each hadron track~\cite{LHCb-DP-2012-003},
determined from large calibration 
samples of
\mbox{$\decay{\Dstarp}
{\left(\decay{\Dz}{\Km\pip}\right)\pip}$}, 
\mbox{$\decay{\KS}{\pip\pim}$}
and 
\mbox{$\decay{\Ds}{\left(\decay{\Pphi}
{\Kp\Km}\right)\pip}$}~decays~\cite{LHCb-DP-2018-001}. 
The~measured ratios of branching fractions are
\begin{subequations}
\begingroup
\allowdisplaybreaks
\begin{eqnarray*}
\mathcal{R}^{\Tojpsikkpipipi}_{\Tojpsifivepi}   
&=& 
\left(33.7 \pm 5.7 \right)\times10^{-2}  \,, 
\\
\mathcal{R}^{\Tojpsisevenpi}_{\Tojpsifivepi}     
&=& 
\left(28.5 \pm 8.7 \right)\times10^{-2}  \,,
\\
\mathcal{R}^{\Topsitwostripi}_{\Tojpsifivepi} 
&=&\left(17.6\pm3.6\right)\times10^{-2} \,, 
\end{eqnarray*}
\endgroup
\end{subequations}
where uncertainties are statistical only and 
correlation coefficients are listed 
in Table~\ref{tab:corr_stat}.

\section{Systematic uncertainties}
\label{sec:Systematics}

The~decay channels under study 
have similar kinematics
and topologies, therefore,
many sources of systematic uncertainty
cancel 
in~the~branching fraction
ratios,~$\mathcal{R}^{\PX}_{\PY}$. 
The~remaining contributions 
to the systematic uncertainty are 
summarised in Table~\ref{tab:systematics} 
and are discussed below.

An important source of systematic uncertainty on the~ratios
is the imperfect knowledge 
of the~shapes of signal and background 
components used in the~fits. 
To~estimate this
uncertainty, several alternative models 
are tested.  
For~the~\Bc and \psitwos
signal shapes, a~generalized Student's
$t$\nobreakdash-distribution~\cite{Student,
Jackman}
and~a~modified 
Apollonios function~\cite{Santos:2013gra} 
are employed as an alternative
model. 
For~the~background components, the~degree
of the~polynomials used in the~fits is 
increased by one.
Also, the~product of an~exponential function 
and a~first\nobreakdash-order 
polynomial function is considered as 
an~alternative background shape.
The~systematic uncertainty related to 
the~fit model is estimated 
with large ensembles 
of pseudoexperiments.
For~each alternative model 
an~ensemble of pseudoexperiments 
is generated and 
each pseudoexperiment
is fitted with the~baseline model.   
The~maximal deviations 
in the~ratios of 
the~mean values of 
signal yields over the~ensemble 
with  respect to the~baseline model
do not exceed 2.5\% for the~variations 
of the~signal model
and 1.0\% for the~variations of 
background model, and   
are taken as systematic uncertainties.
The~sample of \mbox{\BcTojpsifivepi}~decays 
is used to~assess the~systematic uncertainty 
due to the~procedure  of multiple 
candidate exclusion, if two or 
more \Bc~candidates are found from 
the~same \proton\proton collision.
A~large set of pseudoexperiments 
is performed with a~random rejection of 
multiple candidates. 
The~variation of the~signal yield 
for the~\BcTojpsifivepi channel
between the~pseudoexperiments
is found to be of 1.3\%
and this value is assigned as 
the~corresponding systematic uncertainty.

\begin{table}[t]
	\centering
	\caption{\small
	Ranges of relative systematic uncertainties for 
    the~various ratios of branching fractions, $\mathcal{R}^{\PX}_{\PY}$.
    The~total systematic uncertainty is 
    the~quadratic sum of individual contributions.} 
	\label{tab:systematics}
	\vspace{2mm}
	\begin{tabular*}{0.70\textwidth}
	{@{\hspace{5mm}}l@{\extracolsep{\fill}}c@{\hspace{5mm}}}
	Source  &   Uncertainty~$\left[\%\right]$ 
   \\[1.5mm]
  \hline 
  \\[-1.5mm]
  Fit model                  &   
  \\
  ~~~Signal shape            & $   0.1 - 2.5$ 
  \\
  ~~~Background shape        & $   0.4 - 1.0$ 
  \\
  Multiple candidates exclusion& $1.3$  
  \\
  \Bc~decay model            & $   2.2 - 5.1$  
  \\
  Efficiency corrections     & $  0.1  - 1.1$
  \\
  Hadron interactions 
    & $0.0 - 2.8$ 
  \\ 
  Trigger efficiency         & $  1.1$ 
  \\
  Data-simulation difference & $  2.3$ 
  \\
  Size of simulated sample   &  $ 1.5 - 2.4 $ 
    \\[1.5mm]
  \hline 
  \\[-1.5mm]
  Total 
  &   $4.4 - 7.1 $ 
 
	\end{tabular*}
	\vspace{3mm}
\end{table}

To~assess the~systematic uncertainty 
related  to~the~\Bc~decay model used in 
the~simulation~\cite{Berezhnoy:2011nx,
Luchinsky:2022pxu}, 
the~reconstructed 
mass distributions of
the~light\nobreakdash-hadron 
systems in simulation
are adjusted to reproduce 
the~distributions observed in data. 
The~uncertainty associated with 
the~low yield of the~target data distributions 
is accounted for 
by varying  them
within their uncertainties.
The~changes in the~ratios $\mathcal{R}^{\PX}_{\PY}$
do not exceed~$5.1\%$ and are taken
as systematic uncertainties 
related to the~\Bc~decay model.

An~additional uncertainty arises from 
the~difference between  data and  
simulation 
in the~reconstruction efficiency 
of charged\nobreakdash-particle tracks.
The~track\nobreakdash-finding
efficiencies obtained from simulation 
are corrected using data calibration
samples~\cite{LHCb-DP-2013-002}.
The~uncertainties related to 
the~correction factors, 
together with the~uncertainty in 
the~hadron\nobreakdash-identification 
efficiency  due to the~finite size of 
the~calibration 
samples~\cite{LHCb-DP-2012-003,
LHCb-DP-2018-001},
are propagated to the~ratio of 
total efficiencies using 
pseudoexperiments. 
The~obtained 
systematic uncertainty for 
the~$\mathcal{R}^{\PX}_{\PY}$~ratios 
does not exceed 1.1\%. 
The~hadronic interaction 
length of the~detector is known 
with 10\%~uncertainty~\cite{LHCb-PAPER-2010-001}.
It~corresponds to an~additional 
uncertainty for 
the~track\nobreakdash-finding efficiency 
of~1.1\%\,(1.4\%) 
per~charged kaon\,(pion)
track~\cite{LHCb-PAPER-2010-001,
LHCb-DP-2013-002,
LHCb-PAPER-2015-041}.
This~uncertainty is 
assumed to be totally correlated
and partly cancels for the~ratios.
The~systematic uncertainty of 1.1\% related 
to the~trigger efficiency is estimated 
by comparing 
the~ratios of trigger
efficiencies in data and simulation
using large samples of~\mbox{$\decay{\Bp}{\jpsi\Kp}$} and 
\mbox{$\decay{\Bp}{\psitwos\Kp}$}~decays~\cite{LHCb-PAPER-2012-010}. 
Another source of uncertainty 
is a~potential disagreement
between data and simulation 
in the~estimation of efficiencies, 
due to possible effects not explicitly 
considered above. 
This is studied by varying 
the~selection criteria
of the~high
yield \mbox{\BcTojpsifivepi} data sample
in ranges that lead 
up to a~$\pm 20\%$ change in the~measured 
signal yields. 
The~resulting difference between 
the~efficiencies estimated using data 
and simulation does not exceed $2.3\%$,
which is taken as a~systematic uncertainty 
for the~ratios $\mathcal{R}^{\PX}_{\PY}$.
The~last systematic uncertainty 
considered is 
due to the~finite size of the~simulated samples,
and it varies between 1.5\% and 2.4\%.
The~total systematic uncertainty is 
estimated as the~quadratic sum of 
individual contributions.
For~each choice of the~alternative 
fit model the~statistical significance 
for the~channels under study
is recalculated from data using Wilks' 
theorem~\cite{Wilks:1938dza}. 
The~smallest significances found are 
$9.0$,
$5.2$ and $4.7$~standard deviations 
for 
the~\mbox{$\decay{\Bc}{\jpsi\Kp\Km\pip\pip\pim}$},
\mbox{$\decay{\Bc}
{\left(\decay{\psitwos}
{\jpsi\pip\pim} \right) \pip\pip\pim}$}
and~\mbox{$\decay{\Bc}
{\jpsi4\pip3\pim}$}~decays, respectively.
These values are 
taken as 
the~significance 
including systematic
uncertainty.

\section{Summary}
\label{sec:Results}

Several~\decay{\Bc}{\jpsi n\Ph^\pm} decays  are studied
using proton\nobreakdash-proton collision data, 
corresponding to an~integrated luminosity of 
$9\invfb$,
collected with the~\lhcb detector at 
centre\nobreakdash-of\nobreakdash-mass energies of 7, 8, and 13\tev.
The~first observation of   
the~decay
\mbox{$\BcTojpsikkpipipi$} is reported. 
The~decays 
\mbox{\BcTojpsifivepi} and
\mbox{\BcTopsitwostripi}, with $\psitwosTojpsipipi$,
are confirmed 
and  the~first evidence for 
the \mbox{\BcTojpsisevenpi}~decay 
is obtained with 
a~significance
of~$4.7$~standard deviations.

Three ratios of branching fractions, 
defined in Eqs.~\eqref{eq:ratios},
are measured as  
\begingroup
\allowdisplaybreaks
\begin{eqnarray*}
\mathcal{R}_{\Tojpsifivepi}^{\Tojpsikkpipipi} 
&=&
\left( 33.7\pm 5.7 \pm 1.6 \right) \times 10^{-2} \,, 
\\
\mathcal{R}^{\Tojpsisevenpi}_{\Tojpsifivepi} 
& = &  
\left( 28.5 \pm 8.7 \pm 2.0 \right) \times 10^{-2}   \,, 
\\
\mathcal{R}^{\Topsitwostripi}_{\Tojpsifivepi} 
& = & 
\left( 17.6 \pm 3.6 \pm 0.8 \right) \times 10^{-2} \,,  
\end{eqnarray*}
\endgroup
 where the~first uncertainty 
 is statistical and the~second
 systematic. Correlation coefficients for 
 statistical and systematic 
 uncertainties for the measured ratios
 of branching fractions are given in
 Appendix~\ref{sec:correlations}.
The~mass spectra for the~light\nobreakdash-hadron 
system, 
as well as 
the~mass~spectra for 
the~intermediate combinations of light hadrons
agree with the~phenomenological 
model by Berezhnoy, 
Likhoded and 
Luchinsky
based on QCD~factorisation~\cite{Likhoded:2009ib, 
Berezhnoy:2011is,
Luchinsky:2012rk,
Likhoded:2013iua,
Luchinsky:2013yla,
Luchinsky:2018lfj,
Luchinsky:2022pxu}.
 The~ratio 
 $\mathcal{R}_{\Tojpsifivepi}^{\Tojpsikkpipipi}$
 is found to be higher than the~analogous 
 ratio of the~branching  fractions of the~\mbox{$\decay{\Bc}{\jpsi\Kp\Km\pip}$}
 to \mbox{$\decay{\Bc}{\jpsi\pip\pip\pim}$}~decays,
 which was measured to be equal to \mbox{$\left(18.5 \pm 1.3 \pm 0.6\right)
 \times 10^{-2}$}~\cite{LHCb-PAPER-2021-034}.
 

The~majority of branching fractions 
 for the~\Bc~mesons 
 are known relative to
 the~\mbox{$\decay{\Bc}{\jpsi\pip}$} 
 mode.
 All measurements presented here 
 can be related 
 to the~reference  
 \mbox{$\decay{\Bc}{\jpsi\pip}$}~decay mode
 through
 the~\mbox{$\decay{\Bc}
 {\left(\decay{\psitwos}
   {\jpsi\pip\pim}\right)\pip\pip\pim}$}~decay mode.
     The~most precise determinaiton
     can be achieved using a~combination of 
 the~measurements of
 the~ratio of branching
 fractions for
 the~\mbox{$\decay{\Bc}{\psitwos\pip\pip\pim}$}
 and \mbox{$\decay{\Bc}{\psitwos\pip}$}~decays in Ref.~\cite{LHCb-PAPER-2021-034},
 and the~ratios
 of the~branching fractions
 for the~\mbox{$\decay{\Bc}{\psitwos\pip}$},
 \mbox{$\decay{\Bc}{\jpsi\pip\pip\pim}$}
 and
\mbox{$\decay{\Bc}{\jpsi\pip}$}~decays
from
 Refs.~\cite{LHCb-PAPER-2011-044,
 CMS:2014oqy,
 LHCb-PAPER-2015-024}.

\section*{Acknowledgements}
%
%
\noindent We express our gratitude to our colleagues in
the~CERN
accelerator departments for the~excellent performance of the~LHC.
We~thank the technical and administrative staff at the~LHCb
institutes.
We acknowledge support from CERN and from the national agencies:
CAPES, CNPq, FAPERJ and FINEP\,(Brazil); 
MOST and NSFC\,(China); 
CNRS/IN2P3\,(France); 
BMBF, DFG and MPG\,(Germany); 
INFN\,(Italy); 
NWO\,(Netherlands); 
MNiSW and NCN\,(Poland); 
MEN/IFA\,(Romania); 
MICINN\,(Spain); 
SNSF and SER\,(Switzerland); 
NASU\,(Ukraine); 
STFC\,(United Kingdom); 
DOE NP and NSF\,(USA).
We acknowledge the computing resources that are provided by CERN,
IN2P3\.(France),
KIT and DESY\.(Germany),
INFN\,(Italy),
SURF\,(Netherlands),
PIC\,(Spain),
GridPP\,(United Kingdom), 
CSCS\,(Switzerland),
IFIN-HH\,(Romania),
CBPF\,(Brazil),
Polish WLCG\,(Poland)
and NERSC\,(USA).
We~are indebted to the~communities behind the~multiple
open\nobreakdash-source
software packages on which we depend.
Individual groups or members have received support from
ARC and ARDC\,(Australia);
Minciencias\,(Colombia);
AvH Foundation\,(Germany);
EPLANET, Marie Sk\l{}odowska\nobreakdash-Curie Actions and
ERC\,(European Union);
A*MIDEX, ANR, IPhU and Labex P2IO, and R\'{e}gion
Auvergne\nobreakdash-Rh\^{o}ne\nobreakdash-Alpes\,(France);
Key Research Program of Frontier Sciences of CAS, CAS PIFI, CAS CCEPP, 
Fundamental Research Funds for the Central Universities, 
and Sci. \& Tech. Program of Guangzhou\,(China);
GVA, XuntaGal, GENCAT and Prog.~Atracci\'on Talento, CM\,(Spain);
SRC\,(Sweden);
the~Leverhulme Trust, the~Royal Society
 and UKRI\,(United Kingdom).

\clearpage 
\appendix

 \renewcommand{\thetable}{A.\arabic{table}}
 \renewcommand{\thefigure}{A.\arabic{figure}}
 \renewcommand{\theequation}{A.\arabic{equation}}
 \setcounter{figure}{0}
 \setcounter{table}{0}
 \setcounter{equation}{0}

\section{Correlation matrices}\label{sec:correlations}

The correlation coefficients for 
the~statistical and systematic
uncertainties of the~measured ratios,
$\mathcal{R}^{\PX}_{\PY}$,
are shown in Table~\ref{tab:corr_stat}.

\begin{table}[htb]
	\centering
	\caption{\small
	Off-diagonal 
	correlation 
	coefficients (in percent) 
	for statistical and systematic 
	uncertainties of 
	the~measured ratios $\mathcal{R}^{\PX}_{\PY}$.
   } 
	\label{tab:corr_stat}
	\vspace{2mm}
\begin{tabular*}{0.70\textwidth}
        {@{\hspace{5mm}}l@{\extracolsep{\fill}}cccc@{\hspace{5mm}}}
&  
\multicolumn{2}{c}{$\mathcal{R}_{\Tojpsifivepi}^{\Tojpsisevenpi}$}
&
\multicolumn{2}{c}{$\mathcal{R}^{\Topsitwostripi}_{\Tojpsifivepi}$}

\\
 &  \stat & \syst 
 & \stat  & \syst 
   \\[1.5mm]
  \hline 
  \\[-1.5mm]
  $\mathcal{R}_{\Tojpsifivepi}^{\Tojpsikkpipipi}$   
  & $+10$ & $+19$  
  & $+15$ & $+33$
  \\
  $\mathcal{R}_{\Tojpsifivepi}^{\Tojpsisevenpi}$          
  &        &     
  & $+8 \phantom{0}$ & $+20$ 

  \\
  \end{tabular*}
  \vspace{3mm}
\end{table}





\usetikzlibrary{patterns}

\clearpage
\addcontentsline{toc}{section}{References}
\bibliographystyle{LHCb}
\bibliography{main,standard,LHCb-PAPER,LHCb-CONF,LHCb-DP,LHCb-TDR}

\ifx\mcitethebibliography\mciteundefinedmacro
\PackageError{LHCb.bst}{mciteplus.sty has not been loaded}
{This bibstyle requires the use of the mciteplus package.}\fi
\providecommand{\href}[2]{#2}
\begin{mcitethebibliography}{10}
\mciteSetBstSublistMode{n}
\mciteSetBstMaxWidthForm{subitem}{\alph{mcitesubitemcount})}
\mciteSetBstSublistLabelBeginEnd{\mcitemaxwidthsubitemform\space}
{\relax}{\relax}

\bibitem{PhysRevLett.81.2432}
\cdf collaboration, F.~Abe {\em et~al.},
  \ifthenelse{\boolean{articletitles}}{\emph{Observation of the ${\PB}_{\Pc}$
  meson in $\proton\antiproton$ collisions at $\sqrt{s}=1.8\,\tev$},
  }{}\href{https://doi.org/10.1103/PhysRevLett.81.2432}{Phys.\ Rev.\ Lett.\
  \textbf{81} (1998) 2432},
  \href{http://arxiv.org/abs/hep-ex/9805034}{{\normalfont\ttfamily
  arXiv:hep-ex/9805034}}\relax
\mciteBstWouldAddEndPuncttrue
\mciteSetBstMidEndSepPunct{\mcitedefaultmidpunct}
{\mcitedefaultendpunct}{\mcitedefaultseppunct}\relax
\EndOfBibitem
\bibitem{PhysRevD.58.112004}
CDF collaboration, F.~Abe {\em et~al.},
  \ifthenelse{\boolean{articletitles}}{\emph{Observation of ${\PB}_{\Pc}$
  mesons in $\proton\antiproton$ collisions at $\sqrt{s}=1.8\,\tev$},
  }{}\href{https://doi.org/10.1103/PhysRevD.58.112004}{Phys.\ Rev.\
  \textbf{D58} (1998) 112004},
  \href{http://arxiv.org/abs/hep-ex/9804014}{{\normalfont\ttfamily
  arXiv:hep-ex/9804014}}\relax
\mciteBstWouldAddEndPuncttrue
\mciteSetBstMidEndSepPunct{\mcitedefaultmidpunct}
{\mcitedefaultendpunct}{\mcitedefaultseppunct}\relax
\EndOfBibitem
\bibitem{LHCb-PAPER-2010-002}
LHCb collaboration, R.~Aaij {\em et~al.},
  \ifthenelse{\boolean{articletitles}}{\emph{{Measurement of
  \mbox{$\sigma(\decay{\proton\proton}{\bbbar \PX})$} at \mbox{$\sqs=7\tev$} in
  the~forward region}},
  }{}\href{https://doi.org/10.1016/j.physletb.2010.10.010}{Phys.\ Lett.\
  \textbf{B694} (2010) 209},
  \href{http://arxiv.org/abs/1009.2731}{{\normalfont\ttfamily
  arXiv:1009.2731}}\relax
\mciteBstWouldAddEndPuncttrue
\mciteSetBstMidEndSepPunct{\mcitedefaultmidpunct}
{\mcitedefaultendpunct}{\mcitedefaultseppunct}\relax
\EndOfBibitem
\bibitem{LHCb-PAPER-2011-003}
LHCb collaboration, R.~Aaij {\em et~al.},
  \ifthenelse{\boolean{articletitles}}{\emph{{Measurement of \jpsi~production
  in \proton\proton~collisions at \mbox{$\sqs=7\tev$}}},
  }{}\href{https://doi.org/10.1140/epjc/s10052-011-1645-y}{Eur.\ Phys.\ J.\
  \textbf{C71} (2011) 1645},
  \href{http://arxiv.org/abs/1103.0423}{{\normalfont\ttfamily
  arXiv:1103.0423}}\relax
\mciteBstWouldAddEndPuncttrue
\mciteSetBstMidEndSepPunct{\mcitedefaultmidpunct}
{\mcitedefaultendpunct}{\mcitedefaultseppunct}\relax
\EndOfBibitem
\bibitem{LHCb-PAPER-2011-043}
LHCb collaboration, R.~Aaij {\em et~al.},
  \ifthenelse{\boolean{articletitles}}{\emph{{Measurement of
  the~\Bpm~production cross-section in \proton\proton~collisions at
  \mbox{$\sqs=7\tev$}}}, }{}\href{https://doi.org/10.1007/JHEP04(2012)093}{JHEP
  \textbf{04} (2012) 093},
  \href{http://arxiv.org/abs/1202.4812}{{\normalfont\ttfamily
  arXiv:1202.4812}}\relax
\mciteBstWouldAddEndPuncttrue
\mciteSetBstMidEndSepPunct{\mcitedefaultmidpunct}
{\mcitedefaultendpunct}{\mcitedefaultseppunct}\relax
\EndOfBibitem
\bibitem{LHCb-PAPER-2013-004}
LHCb collaboration, R.~Aaij {\em et~al.},
  \ifthenelse{\boolean{articletitles}}{\emph{{Measurement of \B~meson
  production cross-sections in proton-proton collisions
  at~\mbox{$\sqs=7\tev$}}},
  }{}\href{https://doi.org/10.1007/JHEP08(2013)117}{JHEP \textbf{08} (2013)
  117}, \href{http://arxiv.org/abs/1306.3663}{{\normalfont\ttfamily
  arXiv:1306.3663}}\relax
\mciteBstWouldAddEndPuncttrue
\mciteSetBstMidEndSepPunct{\mcitedefaultmidpunct}
{\mcitedefaultendpunct}{\mcitedefaultseppunct}\relax
\EndOfBibitem
\bibitem{LHCb-PAPER-2013-016}
LHCb collaboration, R.~Aaij {\em et~al.},
  \ifthenelse{\boolean{articletitles}}{\emph{{Production of \jpsi~and
  \Upsilonres~mesons in \proton\proton~collisions at \mbox{$\sqs=8\tev$}}},
  }{}\href{https://doi.org/10.1007/JHEP06(2013)064}{JHEP \textbf{06} (2013)
  064}, \href{http://arxiv.org/abs/1304.6977}{{\normalfont\ttfamily
  arXiv:1304.6977}}\relax
\mciteBstWouldAddEndPuncttrue
\mciteSetBstMidEndSepPunct{\mcitedefaultmidpunct}
{\mcitedefaultendpunct}{\mcitedefaultseppunct}\relax
\EndOfBibitem
\bibitem{LHCb-PAPER-2015-037}
LHCb collaboration, R.~Aaij {\em et~al.},
  \ifthenelse{\boolean{articletitles}}{\emph{{Measurement~\,of\,~forward\,
  \jpsi~production cross\protect\nobreakdash-sections in \proton\proton
  collisions at \mbox{$\sqs=13\tev$}}},
  }{}\href{https://doi.org/10.1007/JHEP10(2015)172}{JHEP \textbf{10} (2015)
  172}, Erratum \href{https://doi.org/10.1007/JHEP05(2017)063}{ibid.\
  \textbf{05} (2017) 063},
  \href{http://arxiv.org/abs/1509.00771}{{\normalfont\ttfamily
  arXiv:1509.00771}}\relax
\mciteBstWouldAddEndPuncttrue
\mciteSetBstMidEndSepPunct{\mcitedefaultmidpunct}
{\mcitedefaultendpunct}{\mcitedefaultseppunct}\relax
\EndOfBibitem
\bibitem{LHCb-PAPER-2011-044}
LHCb collaboration, R.~Aaij {\em et~al.},
  \ifthenelse{\boolean{articletitles}}{\emph{{First observation of the decay
  \mbox{\decay{\Bc}{\jpsi\pip\pim\pip}}}},
  }{}\href{https://doi.org/10.1103/PhysRevLett.108.251802}{Phys.\ Rev.\ Lett.\
  \textbf{108} (2012) 251802},
  \href{http://arxiv.org/abs/1204.0079}{{\normalfont\ttfamily
  arXiv:1204.0079}}\relax
\mciteBstWouldAddEndPuncttrue
\mciteSetBstMidEndSepPunct{\mcitedefaultmidpunct}
{\mcitedefaultendpunct}{\mcitedefaultseppunct}\relax
\EndOfBibitem
\bibitem{LHCb-PAPER-2012-054}
LHCb collaboration, R.~Aaij {\em et~al.},
  \ifthenelse{\boolean{articletitles}}{\emph{{Observation of the decay
  \mbox{\decay{\Bc}{\psitwos\pip}}}},
  }{}\href{https://doi.org/10.1103/PhysRevD.87.071103}{Phys.\ Rev.\
  \textbf{D87} (2013) 071103(R)},
  \href{http://arxiv.org/abs/1303.1737}{{\normalfont\ttfamily
  arXiv:1303.1737}}\relax
\mciteBstWouldAddEndPuncttrue
\mciteSetBstMidEndSepPunct{\mcitedefaultmidpunct}
{\mcitedefaultendpunct}{\mcitedefaultseppunct}\relax
\EndOfBibitem
\bibitem{LHCb-PAPER-2013-010}
LHCb collaboration, R.~Aaij {\em et~al.},
  \ifthenelse{\boolean{articletitles}}{\emph{{Observation of
  \mbox{\decay{\Bc}{\jpsi\Dsp}} and \mbox{\decay{\Bc}{\jpsi\Dssp}} decays}},
  }{}\href{https://doi.org/10.1103/PhysRevD.87.112012}{Phys.\ Rev.\
  \textbf{D87} (2013) 112012}, Erratum
  \href{https://doi.org/10.1103/PhysRevD.89.019901}{ibid.\   \textbf{D89}
  (2014) 019901(E)},
  \href{http://arxiv.org/abs/1304.4530}{{\normalfont\ttfamily
  arXiv:1304.4530}}\relax
\mciteBstWouldAddEndPuncttrue
\mciteSetBstMidEndSepPunct{\mcitedefaultmidpunct}
{\mcitedefaultendpunct}{\mcitedefaultseppunct}\relax
\EndOfBibitem
\bibitem{LHCb-PAPER-2013-021}
LHCb collaboration, R.~Aaij {\em et~al.},
  \ifthenelse{\boolean{articletitles}}{\emph{{First observation of the decay
  \mbox{\decay{\Bc}{\jpsi\Kp}}}},
  }{}\href{https://doi.org/10.1007/JHEP09(2013)075}{JHEP \textbf{09} (2013)
  075}, \href{http://arxiv.org/abs/1306.6723}{{\normalfont\ttfamily
  arXiv:1306.6723}}\relax
\mciteBstWouldAddEndPuncttrue
\mciteSetBstMidEndSepPunct{\mcitedefaultmidpunct}
{\mcitedefaultendpunct}{\mcitedefaultseppunct}\relax
\EndOfBibitem
\bibitem{LHCb-PAPER-2013-044}
LHCb collaboration, R.~Aaij {\em et~al.},
  \ifthenelse{\boolean{articletitles}}{\emph{{Observation of the decay
  \mbox{\decay{\Bc}{\Bs\pip}}}},
  }{}\href{https://doi.org/10.1103/PhysRevLett.111.181801}{Phys.\ Rev.\ Lett.\
  \textbf{111} (2013) 181801},
  \href{http://arxiv.org/abs/1308.4544}{{\normalfont\ttfamily
  arXiv:1308.4544}}\relax
\mciteBstWouldAddEndPuncttrue
\mciteSetBstMidEndSepPunct{\mcitedefaultmidpunct}
{\mcitedefaultendpunct}{\mcitedefaultseppunct}\relax
\EndOfBibitem
\bibitem{LHCb-PAPER-2013-047}
LHCb collaboration, R.~Aaij {\em et~al.},
  \ifthenelse{\boolean{articletitles}}{\emph{{Observation of the decay
  \mbox{\decay{\Bc}{\jpsi\Kp\Km\pip}}}},
  }{}\href{https://doi.org/10.1007/JHEP11(2013)094}{JHEP \textbf{11} (2013)
  094}, \href{http://arxiv.org/abs/1309.0587}{{\normalfont\ttfamily
  arXiv:1309.0587}}\relax
\mciteBstWouldAddEndPuncttrue
\mciteSetBstMidEndSepPunct{\mcitedefaultmidpunct}
{\mcitedefaultendpunct}{\mcitedefaultseppunct}\relax
\EndOfBibitem
\bibitem{LHCb-PAPER-2014-009}
LHCb collaboration, R.~Aaij {\em et~al.},
  \ifthenelse{\boolean{articletitles}}{\emph{{Evidence for the decay
  \mbox{\decay{\Bc}{\jpsi 3\pip 2\pim}}}},
  }{}\href{https://doi.org/10.1007/JHEP05(2014)148}{JHEP \textbf{05} (2014)
  148}, \href{http://arxiv.org/abs/1404.0287}{{\normalfont\ttfamily
  arXiv:1404.0287}}\relax
\mciteBstWouldAddEndPuncttrue
\mciteSetBstMidEndSepPunct{\mcitedefaultmidpunct}
{\mcitedefaultendpunct}{\mcitedefaultseppunct}\relax
\EndOfBibitem
\bibitem{LHCb-PAPER-2014-039}
LHCb collaboration, R.~Aaij {\em et~al.},
  \ifthenelse{\boolean{articletitles}}{\emph{{First observation of a baryonic
  \Bcp decay}}, }{}\href{https://doi.org/10.1103/PhysRevLett.113.152003}{Phys.\
  Rev.\ Lett.\  \textbf{113} (2014) 152003},
  \href{http://arxiv.org/abs/1408.0971}{{\normalfont\ttfamily
  arXiv:1408.0971}}\relax
\mciteBstWouldAddEndPuncttrue
\mciteSetBstMidEndSepPunct{\mcitedefaultmidpunct}
{\mcitedefaultendpunct}{\mcitedefaultseppunct}\relax
\EndOfBibitem
\bibitem{LHCb-PAPER-2014-050}
LHCb collaboration, R.~Aaij {\em et~al.},
  \ifthenelse{\boolean{articletitles}}{\emph{{Measurement of \Bcp production in
  proton-proton collisions at \mbox{$\sqs=8\tev$}}},
  }{}\href{https://doi.org/10.1103/PhysRevLett.114.132001}{Phys.\ Rev.\ Lett.\
  \textbf{114} (2015) 132001},
  \href{http://arxiv.org/abs/1411.2943}{{\normalfont\ttfamily
  arXiv:1411.2943}}\relax
\mciteBstWouldAddEndPuncttrue
\mciteSetBstMidEndSepPunct{\mcitedefaultmidpunct}
{\mcitedefaultendpunct}{\mcitedefaultseppunct}\relax
\EndOfBibitem
\bibitem{LHCb-PAPER-2014-060}
LHCb collaboration, R.~Aaij {\em et~al.},
  \ifthenelse{\boolean{articletitles}}{\emph{{Measurement of the lifetime of
  the \Bcp meson using the \mbox{\decay{\Bc}{\jpsi\pip}} decay mode}},
  }{}\href{https://doi.org/10.1016/j.physletb.2015.01.010}{Phys.\ Lett.\
  \textbf{B742} (2015) 29},
  \href{http://arxiv.org/abs/1411.6899}{{\normalfont\ttfamily
  arXiv:1411.6899}}\relax
\mciteBstWouldAddEndPuncttrue
\mciteSetBstMidEndSepPunct{\mcitedefaultmidpunct}
{\mcitedefaultendpunct}{\mcitedefaultseppunct}\relax
\EndOfBibitem
\bibitem{CMS:2014oqy}
CMS collaboration, V.~Khachatryan {\em et~al.},
  \ifthenelse{\boolean{articletitles}}{\emph{{Measurement of the~ratio of
  the~production cross sections times branching fractions of
  \mbox{$\decay{\Bcpm}{\jpsi\pipm}$} and \mbox{$\decay{\Bpm}{\jpsi\Kpm}$} and
  \mbox{$\BR(\decay{\Bcpm}{\jpsi\pipm\pipm\pimp})
  /\BR(\decay{\Bcpm}{\jpsi\pipm})$} in $\proton\proton$~collisions at
  $\sqs=7\tev$}}, }{}\href{https://doi.org/10.1007/JHEP01(2015)063}{JHEP
  \textbf{01} (2015) 063},
  \href{http://arxiv.org/abs/1410.5729}{{\normalfont\ttfamily
  arXiv:1410.5729}}\relax
\mciteBstWouldAddEndPuncttrue
\mciteSetBstMidEndSepPunct{\mcitedefaultmidpunct}
{\mcitedefaultendpunct}{\mcitedefaultseppunct}\relax
\EndOfBibitem
\bibitem{LHCb-PAPER-2015-024}
LHCb collaboration, R.~Aaij {\em et~al.},
  \ifthenelse{\boolean{articletitles}}{\emph{{Measurement of the branching
  fraction ratio
  \mbox{$\BF(\decay{\Bcp}{\psitwos\pip})/\BF(\decay{\Bcp}{\jpsi\pip})$}}},
  }{}\href{https://doi.org/10.1103/PhysRevD.92.072007}{Phys.\ Rev.\
  \textbf{D92} (2015) 072007},
  \href{http://arxiv.org/abs/1507.03516}{{\normalfont\ttfamily
  arXiv:1507.03516}}\relax
\mciteBstWouldAddEndPuncttrue
\mciteSetBstMidEndSepPunct{\mcitedefaultmidpunct}
{\mcitedefaultendpunct}{\mcitedefaultseppunct}\relax
\EndOfBibitem
\bibitem{ATLAS:2015jep}
ATLAS collaboration, G.~Aad {\em et~al.},
  \ifthenelse{\boolean{articletitles}}{\emph{{Study of
  the~\mbox{$\decay{\Bc}{\jpsi\Ds}$} and~\mbox{$\decay{\Bc}{\jpsi\Dss}$}~decays
  with the~\atlas detector}},
  }{}\href{https://doi.org/10.1140/epjc/s10052-015-3743-8}{Eur.\ Phys.\ J.\
  \textbf{C76} (2016) 4},
  \href{http://arxiv.org/abs/1507.07099}{{\normalfont\ttfamily
  arXiv:1507.07099}}\relax
\mciteBstWouldAddEndPuncttrue
\mciteSetBstMidEndSepPunct{\mcitedefaultmidpunct}
{\mcitedefaultendpunct}{\mcitedefaultseppunct}\relax
\EndOfBibitem
\bibitem{LHCb-PAPER-2016-001}
LHCb collaboration, R.~Aaij {\em et~al.},
  \ifthenelse{\boolean{articletitles}}{\emph{{Search for \Bcp decays to the
  $\proton\antiproton\pip$ final state}},
  }{}\href{https://doi.org/10.1016/j.physletb.2016.05.074}{Phys.\ Lett.\
  \textbf{B759} (2016) 313},
  \href{http://arxiv.org/abs/1603.07037}{{\normalfont\ttfamily
  arXiv:1603.07037}}\relax
\mciteBstWouldAddEndPuncttrue
\mciteSetBstMidEndSepPunct{\mcitedefaultmidpunct}
{\mcitedefaultendpunct}{\mcitedefaultseppunct}\relax
\EndOfBibitem
\bibitem{LHCb-PAPER-2016-022}
LHCb collaboration, R.~Aaij {\em et~al.},
  \ifthenelse{\boolean{articletitles}}{\emph{{Study of \Bcp decays to the
  $\Kp\Km\pip$ final state and evidence for the decay
  \mbox{\decay{\Bc}{\chiczero\pip}}}},
  }{}\href{https://doi.org/10.1103/PhysRevD.94.091102}{Phys.\ Rev.\
  \textbf{D94} (2016) 091102(R)},
  \href{http://arxiv.org/abs/1607.06134}{{\normalfont\ttfamily
  arXiv:1607.06134}}\relax
\mciteBstWouldAddEndPuncttrue
\mciteSetBstMidEndSepPunct{\mcitedefaultmidpunct}
{\mcitedefaultendpunct}{\mcitedefaultseppunct}\relax
\EndOfBibitem
\bibitem{LHCb-PAPER-2016-020}
LHCb collaboration, R.~Aaij {\em et~al.},
  \ifthenelse{\boolean{articletitles}}{\emph{{Measurement of the ratio of
  branching fractions
  \mbox{$\BF(\decay{\Bcp}{\jpsi\Kp})/\BF(\decay{\Bcp}{\jpsi\pip})$}}},
  }{}\href{https://doi.org/10.1007/JHEP09(2016)153}{JHEP \textbf{09} (2016)
  153}, \href{http://arxiv.org/abs/1607.06823}{{\normalfont\ttfamily
  arXiv:1607.06823}}\relax
\mciteBstWouldAddEndPuncttrue
\mciteSetBstMidEndSepPunct{\mcitedefaultmidpunct}
{\mcitedefaultendpunct}{\mcitedefaultseppunct}\relax
\EndOfBibitem
\bibitem{LHCb-PAPER-2016-055}
LHCb collaboration, R.~Aaij {\em et~al.},
  \ifthenelse{\boolean{articletitles}}{\emph{{Observation of
  \mbox{\decay{\Bc}{\jpsi \PD^{(\ast)} \PK^{(\ast)}}} decays}},
  }{}\href{https://doi.org/10.1103/PhysRevD.95.032005}{Phys.\ Rev.\
  \textbf{D95} (2017) 032005},
  \href{http://arxiv.org/abs/1612.07421}{{\normalfont\ttfamily
  arXiv:1612.07421}}\relax
\mciteBstWouldAddEndPuncttrue
\mciteSetBstMidEndSepPunct{\mcitedefaultmidpunct}
{\mcitedefaultendpunct}{\mcitedefaultseppunct}\relax
\EndOfBibitem
\bibitem{LHCb-PAPER-2016-058}
LHCb collaboration, R.~Aaij {\em et~al.},
  \ifthenelse{\boolean{articletitles}}{\emph{{Observation of
  \mbox{\decay{\Bc}{\Dz\Kp}} decays}},
  }{}\href{https://doi.org/10.1103/PhysRevLett.118.111803}{Phys.\ Rev.\ Lett.\
  \textbf{118} (2017) 111803},
  \href{http://arxiv.org/abs/1701.01856}{{\normalfont\ttfamily
  arXiv:1701.01856}}\relax
\mciteBstWouldAddEndPuncttrue
\mciteSetBstMidEndSepPunct{\mcitedefaultmidpunct}
{\mcitedefaultendpunct}{\mcitedefaultseppunct}\relax
\EndOfBibitem
\bibitem{LHCb-PAPER-2017-035}
LHCb collaboration, R.~Aaij {\em et~al.},
  \ifthenelse{\boolean{articletitles}}{\emph{{Measurement of the ratio of
  branching fractions
  \mbox{$\mathcal{B}(\decay{\Bc}{\jpsi\taup\Pnu_{\Ptau}})/\mathcal{B}(\decay{\Bc}{\jpsi\mup\Pnu_{\Pmu}})$}}},
  }{}\href{https://doi.org/10.1103/PhysRevLett.120.121801}{Phys.\ Rev.\ Lett.\
  \textbf{120} (2018) 121801},
  \href{http://arxiv.org/abs/1711.05623}{{\normalfont\ttfamily
  arXiv:1711.05623}}\relax
\mciteBstWouldAddEndPuncttrue
\mciteSetBstMidEndSepPunct{\mcitedefaultmidpunct}
{\mcitedefaultendpunct}{\mcitedefaultseppunct}\relax
\EndOfBibitem
\bibitem{LHCb-PAPER-2017-045}
LHCb collaboration, R.~Aaij {\em et~al.},
  \ifthenelse{\boolean{articletitles}}{\emph{{Search for \Bc decays to two
  charm mesons}},
  }{}\href{https://doi.org/10.1016/j.nuclphysb.2018.03.015}{Nucl.\ Phys.\
  \textbf{B930} (2018) 563},
  \href{http://arxiv.org/abs/1712.04702}{{\normalfont\ttfamily
  arXiv:1712.04702}}\relax
\mciteBstWouldAddEndPuncttrue
\mciteSetBstMidEndSepPunct{\mcitedefaultmidpunct}
{\mcitedefaultendpunct}{\mcitedefaultseppunct}\relax
\EndOfBibitem
\bibitem{CMS:2017ygm}
CMS collaboration, A.~M. Sirunyan {\em et~al.},
  \ifthenelse{\boolean{articletitles}}{\emph{{Measurement of \bquark~hadron
  lifetimes in $\proton\proton$~collisions at $\sqs=8\tev$}},
  }{}\href{https://doi.org/10.1140/epjc/s10052-018-5929-3}{Eur.\ Phys.\ J.\
  \textbf{C78} (2018) 457}, Erratum
  \href{https://doi.org/10.1140/epjc/s10052-018-6014-7}{ibid.\   \textbf{C78}
  (2018) 561}, \href{http://arxiv.org/abs/1710.08949}{{\normalfont\ttfamily
  arXiv:1710.08949}}\relax
\mciteBstWouldAddEndPuncttrue
\mciteSetBstMidEndSepPunct{\mcitedefaultmidpunct}
{\mcitedefaultendpunct}{\mcitedefaultseppunct}\relax
\EndOfBibitem
\bibitem{CMS:2019uhm}
CMS collaboration, A.~M. Sirunyan {\em et~al.},
  \ifthenelse{\boolean{articletitles}}{\emph{{Observation of two~excited
  \Bc~states and measurement of the $\B^{+}_{\cquark}(2\PS)$~mass in
  $\proton\proton$~collisions at~$\sqs=13\tev$}},
  }{}\href{https://doi.org/10.1103/PhysRevLett.122.132001}{Phys.\ Rev.\ Lett.\
  \textbf{122} (2019) 132001},
  \href{http://arxiv.org/abs/1902.00571}{{\normalfont\ttfamily
  arXiv:1902.00571}}\relax
\mciteBstWouldAddEndPuncttrue
\mciteSetBstMidEndSepPunct{\mcitedefaultmidpunct}
{\mcitedefaultendpunct}{\mcitedefaultseppunct}\relax
\EndOfBibitem
\bibitem{LHCb-PAPER-2019-007}
LHCb collaboration, R.~Aaij {\em et~al.},
  \ifthenelse{\boolean{articletitles}}{\emph{{Observation of an excited \Bc
  state}}, }{}\href{https://doi.org/10.1103/PhysRevLett.122.232001}{Phys.\
  Rev.\ Lett.\  \textbf{122} (2019) 232001},
  \href{http://arxiv.org/abs/1904.00081}{{\normalfont\ttfamily
  arXiv:1904.00081}}\relax
\mciteBstWouldAddEndPuncttrue
\mciteSetBstMidEndSepPunct{\mcitedefaultmidpunct}
{\mcitedefaultendpunct}{\mcitedefaultseppunct}\relax
\EndOfBibitem
\bibitem{LHCb-PAPER-2019-033}
LHCb collaboration, R.~Aaij {\em et~al.},
  \ifthenelse{\boolean{articletitles}}{\emph{{Measurement of the~\Bcm~meson
  production fraction and asymmetry in $7$ and $13\tev$
  $\proton\proton$~collisions}},
  }{}\href{https://doi.org/10.1103/PhysRevD.100.112006}{Phys.\ Rev.\
  \textbf{D100} (2019) 112006},
  \href{http://arxiv.org/abs/1910.13404}{{\normalfont\ttfamily
  arXiv:1910.13404}}\relax
\mciteBstWouldAddEndPuncttrue
\mciteSetBstMidEndSepPunct{\mcitedefaultmidpunct}
{\mcitedefaultendpunct}{\mcitedefaultseppunct}\relax
\EndOfBibitem
\bibitem{LHCb-PAPER-2020-003}
LHCb collaboration, R.~Aaij {\em et~al.},
  \ifthenelse{\boolean{articletitles}}{\emph{{Precision measurement of the \Bc
  meson mass}}, }{}\href{https://doi.org/10.1007/JHEP07(2020)123}{JHEP
  \textbf{07} (2020) 123},
  \href{http://arxiv.org/abs/2004.08163}{{\normalfont\ttfamily
  arXiv:2004.08163}}\relax
\mciteBstWouldAddEndPuncttrue
\mciteSetBstMidEndSepPunct{\mcitedefaultmidpunct}
{\mcitedefaultendpunct}{\mcitedefaultseppunct}\relax
\EndOfBibitem
\bibitem{LHCb-PAPER-2021-023}
LHCb collaboration, R.~Aaij {\em et~al.},
  \ifthenelse{\boolean{articletitles}}{\emph{{Updated search for $\Bc$ decays
  to two charm mesons}}, }{}\href{https://doi.org/10.1007/JHEP12(2021)117}{JHEP
  \textbf{12} (2021) 117},
  \href{http://arxiv.org/abs/2109.00488}{{\normalfont\ttfamily
  arXiv:2109.00488}}\relax
\mciteBstWouldAddEndPuncttrue
\mciteSetBstMidEndSepPunct{\mcitedefaultmidpunct}
{\mcitedefaultendpunct}{\mcitedefaultseppunct}\relax
\EndOfBibitem
\bibitem{LHCb-PAPER-2021-034}
LHCb collaboration, R.~Aaij {\em et~al.},
  \ifthenelse{\boolean{articletitles}}{\emph{{Study of \Bc decays to charmonia
  and three light hadrons}},
  }{}\href{https://doi.org/10.1007/JHEP01(2022)065}{JHEP \textbf{01} (2022)
  065}, \href{http://arxiv.org/abs/2111.03001}{{\normalfont\ttfamily
  arXiv:2111.03001}}\relax
\mciteBstWouldAddEndPuncttrue
\mciteSetBstMidEndSepPunct{\mcitedefaultmidpunct}
{\mcitedefaultendpunct}{\mcitedefaultseppunct}\relax
\EndOfBibitem
\bibitem{Bauer:1986bm}
M.~Bauer, B.~Stech, and M.~Wirbel,
  \ifthenelse{\boolean{articletitles}}{\emph{{Exclusive non-leptonic decays of
  \PD-, $\PD_{\Ps}$-, and \PB-mesons}},
  }{}\href{https://doi.org/10.1007/BF01561122}{Z.\ Phys.\  \textbf{C34} (1987)
  103}\relax
\mciteBstWouldAddEndPuncttrue
\mciteSetBstMidEndSepPunct{\mcitedefaultmidpunct}
{\mcitedefaultendpunct}{\mcitedefaultseppunct}\relax
\EndOfBibitem
\bibitem{Wirbel:1988ft}
M.~Wirbel, \ifthenelse{\boolean{articletitles}}{\emph{{Description of weak
  decays of $\D$~and $\B$~mesons}},
  }{}\href{https://doi.org/10.1016/0146-6410(88)90031-2}{Prog.\ Part.\ Nucl.\
  Phys.\  \textbf{21} (1988) 33}\relax
\mciteBstWouldAddEndPuncttrue
\mciteSetBstMidEndSepPunct{\mcitedefaultmidpunct}
{\mcitedefaultendpunct}{\mcitedefaultseppunct}\relax
\EndOfBibitem
\bibitem{Gershtein:1994jw}
S.~S. Gershtein, V.~V. Kiselev, A.~K. Likhoded, and A.~V. Tkabladze,
  \ifthenelse{\boolean{articletitles}}{\emph{{Physics of $\B_\cquark$-mesons}},
  }{}\href{https://doi.org/10.1070/PU1995v038n01ABEH000063}{Phys.\ Usp.\
  \textbf{38} (1995) 1},
  \href{http://arxiv.org/abs/hep-ph/9504319}{{\normalfont\ttfamily
  arXiv:hep-ph/9504319}}\relax
\mciteBstWouldAddEndPuncttrue
\mciteSetBstMidEndSepPunct{\mcitedefaultmidpunct}
{\mcitedefaultendpunct}{\mcitedefaultseppunct}\relax
\EndOfBibitem
\bibitem{Gershtein:1997qy}
S.~S. Gershtein {\em et~al.},
  \ifthenelse{\boolean{articletitles}}{\emph{{Theoretical status of
  the~$\B_\cquark$~meson, International Workshop on Heavy Quark Physics}}, }{}
  1997, \href{http://arxiv.org/abs/hep-ph/9803433}{{\normalfont\ttfamily
  arXiv:hep-ph/9803433}}\relax
\mciteBstWouldAddEndPuncttrue
\mciteSetBstMidEndSepPunct{\mcitedefaultmidpunct}
{\mcitedefaultendpunct}{\mcitedefaultseppunct}\relax
\EndOfBibitem
\bibitem{Kiselev:1999sc}
V.~V. Kiselev, A.~K. Likhoded, and A.~I. Onishchenko,
  \ifthenelse{\boolean{articletitles}}{\emph{{Semileptonic $\B_\cquark$-meson
  decays in sum rules of QCD and NRQCD}},
  }{}\href{https://doi.org/10.1016/S0550-3213(99)00505-2}{Nucl.\ Phys.\
  \textbf{B569} (2000) 473},
  \href{http://arxiv.org/abs/hep-ph/9905359}{{\normalfont\ttfamily
  arXiv:hep-ph/9905359}}\relax
\mciteBstWouldAddEndPuncttrue
\mciteSetBstMidEndSepPunct{\mcitedefaultmidpunct}
{\mcitedefaultendpunct}{\mcitedefaultseppunct}\relax
\EndOfBibitem
\bibitem{Kiselev:2000pp}
V.~V. Kiselev, A.~E. Kovalsky, and A.~K. Likhoded,
  \ifthenelse{\boolean{articletitles}}{\emph{{$\B_\cquark$~decays and lifetime
  in QCD sum rules}},
  }{}\href{https://doi.org/10.1016/S0550-3213(00)00386-2}{Nucl.\ Phys.\
  \textbf{B585} (2000) 353},
  \href{http://arxiv.org/abs/hep-ph/0002127}{{\normalfont\ttfamily
  arXiv:hep-ph/0002127}}\relax
\mciteBstWouldAddEndPuncttrue
\mciteSetBstMidEndSepPunct{\mcitedefaultmidpunct}
{\mcitedefaultendpunct}{\mcitedefaultseppunct}\relax
\EndOfBibitem
\bibitem{Ebert:2002pp}
D.~Ebert, R.~N. Faustov, and V.~O. Galkin,
  \ifthenelse{\boolean{articletitles}}{\emph{{Properties of heavy quarkonia and
  $\B_{\cquark}$~mesons in the~relativistic quark model}},
  }{}\href{https://doi.org/10.1103/PhysRevD.67.014027}{Phys.\ Rev.\
  \textbf{D67} (2003) 014027},
  \href{http://arxiv.org/abs/hep-ph/0210381}{{\normalfont\ttfamily
  arXiv:hep-ph/0210381}}\relax
\mciteBstWouldAddEndPuncttrue
\mciteSetBstMidEndSepPunct{\mcitedefaultmidpunct}
{\mcitedefaultendpunct}{\mcitedefaultseppunct}\relax
\EndOfBibitem
\bibitem{Likhoded:2009ib}
A.~K. Likhoded and A.~V. Luchinsky,
  \ifthenelse{\boolean{articletitles}}{\emph{{Light hadron production in
  \decay{\PB_{\Pc}}{\jpsi + \PX} decays}},
  }{}\href{https://doi.org/10.1103/PhysRevD.81.014015}{Phys.\ Rev.\
  \textbf{D81} (2010) 014015},
  \href{http://arxiv.org/abs/0910.3089}{{\normalfont\ttfamily
  arXiv:0910.3089}}\relax
\mciteBstWouldAddEndPuncttrue
\mciteSetBstMidEndSepPunct{\mcitedefaultmidpunct}
{\mcitedefaultendpunct}{\mcitedefaultseppunct}\relax
\EndOfBibitem
\bibitem{Likhoded:2013iua}
A.~K. Likhoded and A.~V. Luchinsky,
  \ifthenelse{\boolean{articletitles}}{\emph{{Production of a pion system in
  exclusive $\decay{\PB_{\Pc}}{\PV(\PP) + n\pion}$ decays}},
  }{}\href{https://doi.org/10.1134/S1063778813050062}{Phys.\ Atom.\ Nucl.\
  \textbf{76} (2013) 787}\relax
\mciteBstWouldAddEndPuncttrue
\mciteSetBstMidEndSepPunct{\mcitedefaultmidpunct}
{\mcitedefaultendpunct}{\mcitedefaultseppunct}\relax
\EndOfBibitem
\bibitem{Berezhnoy:2011is}
A.~V. Berezhnoy, A.~K. Likhoded, and A.~V. Luchinsky,
  \ifthenelse{\boolean{articletitles}}{\emph{{$\PB_\Pc \to \jpsi (\PB^{}_\Ps,
  \PB^{*}_\Ps) + n\Ppi$ decays}},
  }{}\href{https://doi.org/10.22323/1.138.0076}{PoS \textbf{QFTHEP2011} (2012)
  076}, \href{http://arxiv.org/abs/1111.5952}{{\normalfont\ttfamily
  arXiv:1111.5952}}\relax
\mciteBstWouldAddEndPuncttrue
\mciteSetBstMidEndSepPunct{\mcitedefaultmidpunct}
{\mcitedefaultendpunct}{\mcitedefaultseppunct}\relax
\EndOfBibitem
\bibitem{Luchinsky:2012rk}
A.~V. Luchinsky, \ifthenelse{\boolean{articletitles}}{\emph{{Production of
  charged $\pion$ mesons in exclusive
  \mbox{$\decay{\B_\cquark}{\PV(\PP)+n\pion}$} decays}},
  }{}\href{https://doi.org/10.1103/PhysRevD.86.074024}{Phys.\ Rev.\
  \textbf{D86} (2012) 074024},
  \href{http://arxiv.org/abs/1208.1398}{{\normalfont\ttfamily
  arXiv:1208.1398}}\relax
\mciteBstWouldAddEndPuncttrue
\mciteSetBstMidEndSepPunct{\mcitedefaultmidpunct}
{\mcitedefaultendpunct}{\mcitedefaultseppunct}\relax
\EndOfBibitem
\bibitem{Luchinsky:2013yla}
A.~V. Luchinsky, \ifthenelse{\boolean{articletitles}}{\emph{{Production of
  \PK~mesons in exclusive $\B_\cquark$~decays}},
  }{}\href{http://arxiv.org/abs/1307.0953}{{\normalfont\ttfamily
  arXiv:1307.0953}}\relax
\mciteBstWouldAddEndPuncttrue
\mciteSetBstMidEndSepPunct{\mcitedefaultmidpunct}
{\mcitedefaultendpunct}{\mcitedefaultseppunct}\relax
\EndOfBibitem
\bibitem{Luchinsky:2018lfj}
A.~V. Luchinsky, \ifthenelse{\boolean{articletitles}}{\emph{{Excited
  $\Prho$~mesons in
  \mbox{$\decay{\B_\cquark}{\Ppsi^{(')}\kaon\PK_\PS}$}~decays}},
  }{}\href{https://doi.org/10.1103/PhysRevD.99.036019}{Phys.\ Rev.\
  \textbf{D99} (2019) 036019},
  \href{http://arxiv.org/abs/1812.09783}{{\normalfont\ttfamily
  arXiv:1812.09783}}\relax
\mciteBstWouldAddEndPuncttrue
\mciteSetBstMidEndSepPunct{\mcitedefaultmidpunct}
{\mcitedefaultendpunct}{\mcitedefaultseppunct}\relax
\EndOfBibitem
\bibitem{Luchinsky:2022pxu}
A.~V. Luchinsky, \ifthenelse{\boolean{articletitles}}{\emph{{Multiple charged
  meson production in exclusive $\B_\cquark$~decays: \mbox{$\PK+4\Ppi$},
  \mbox{$\PK\PK+3\Ppi$}, \mbox{$7\Ppi$}~cases}},
  }{}\href{https://doi.org/10.1016/j.physletb.2022.137269}{Phys.\ Lett.\
  \textbf{B832} (2022) 137269},
  \href{http://arxiv.org/abs/2204.01136}{{\normalfont\ttfamily
  arXiv:2204.01136}}\relax
\mciteBstWouldAddEndPuncttrue
\mciteSetBstMidEndSepPunct{\mcitedefaultmidpunct}
{\mcitedefaultendpunct}{\mcitedefaultseppunct}\relax
\EndOfBibitem
\bibitem{PhysRev.63.137}
G.~Wataghin, \ifthenelse{\boolean{articletitles}}{\emph{Thermal equilibrium
  between elementary particles},
  }{}\href{https://doi.org/10.1103/PhysRev.63.137}{Phys.\ Rev.\  \textbf{63}
  (1943) 137}\relax
\mciteBstWouldAddEndPuncttrue
\mciteSetBstMidEndSepPunct{\mcitedefaultmidpunct}
{\mcitedefaultendpunct}{\mcitedefaultseppunct}\relax
\EndOfBibitem
\bibitem{PhysRev.66.149}
G.~Wataghin, \ifthenelse{\boolean{articletitles}}{\emph{Statistical mechanics
  at extremely high temperatures},
  }{}\href{https://doi.org/10.1103/PhysRev.66.149}{Phys.\ Rev.\  \textbf{66}
  (1944) 149}\relax
\mciteBstWouldAddEndPuncttrue
\mciteSetBstMidEndSepPunct{\mcitedefaultmidpunct}
{\mcitedefaultendpunct}{\mcitedefaultseppunct}\relax
\EndOfBibitem
\bibitem{PhysRevC.20.2267}
M.~Gyulassy, S.~K. Kauffmann, and L.~W. Wilson,
  \ifthenelse{\boolean{articletitles}}{\emph{Pion interferometry of nuclear
  collisions. {I}. {Theory}},
  }{}\href{https://doi.org/10.1103/PhysRevC.20.2267}{Phys.\ Rev.\  \textbf{C20}
  (1979) 2267}\relax
\mciteBstWouldAddEndPuncttrue
\mciteSetBstMidEndSepPunct{\mcitedefaultmidpunct}
{\mcitedefaultendpunct}{\mcitedefaultseppunct}\relax
\EndOfBibitem
\bibitem{Alves:2008zz}
\lhcb collaboration, A.~A. Alves~Jr.\ {\em et~al.},
  \ifthenelse{\boolean{articletitles}}{\emph{{The \lhcb detector at the LHC}},
  }{}\href{https://doi.org/10.1088/1748-0221/3/08/S08005}{JINST \textbf{3}
  (2008) S08005}\relax
\mciteBstWouldAddEndPuncttrue
\mciteSetBstMidEndSepPunct{\mcitedefaultmidpunct}
{\mcitedefaultendpunct}{\mcitedefaultseppunct}\relax
\EndOfBibitem
\bibitem{LHCb-DP-2014-002}
LHCb collaboration, R.~Aaij {\em et~al.},
  \ifthenelse{\boolean{articletitles}}{\emph{{\lhcb detector performance}},
  }{}\href{https://doi.org/10.1142/S0217751X15300227}{Int.\ J.\ Mod.\ Phys.\
  \textbf{A30} (2015) 1530022},
  \href{http://arxiv.org/abs/1412.6352}{{\normalfont\ttfamily
  arXiv:1412.6352}}\relax
\mciteBstWouldAddEndPuncttrue
\mciteSetBstMidEndSepPunct{\mcitedefaultmidpunct}
{\mcitedefaultendpunct}{\mcitedefaultseppunct}\relax
\EndOfBibitem
\bibitem{LHCb-DP-2014-001}
R.~Aaij {\em et~al.}, \ifthenelse{\boolean{articletitles}}{\emph{{Performance
  of the LHCb Vertex Locator}},
  }{}\href{https://doi.org/10.1088/1748-0221/9/09/P09007}{JINST \textbf{9}
  (2014) P09007}, \href{http://arxiv.org/abs/1405.7808}{{\normalfont\ttfamily
  arXiv:1405.7808}}\relax
\mciteBstWouldAddEndPuncttrue
\mciteSetBstMidEndSepPunct{\mcitedefaultmidpunct}
{\mcitedefaultendpunct}{\mcitedefaultseppunct}\relax
\EndOfBibitem
\bibitem{LHCb-DP-2013-003}
R.~Arink {\em et~al.}, \ifthenelse{\boolean{articletitles}}{\emph{{Performance
  of the LHCb Outer Tracker}},
  }{}\href{https://doi.org/10.1088/1748-0221/9/01/P01002}{JINST \textbf{9}
  (2014) P01002}, \href{http://arxiv.org/abs/1311.3893}{{\normalfont\ttfamily
  arXiv:1311.3893}}\relax
\mciteBstWouldAddEndPuncttrue
\mciteSetBstMidEndSepPunct{\mcitedefaultmidpunct}
{\mcitedefaultendpunct}{\mcitedefaultseppunct}\relax
\EndOfBibitem
\bibitem{LHCb-DP-2017-001}
P.~d'Argent {\em et~al.}, \ifthenelse{\boolean{articletitles}}{\emph{{Improved
  performance of the LHCb Outer Tracker in LHC Run~2}},
  }{}\href{https://doi.org/10.1088/1748-0221/12/11/P11016}{JINST \textbf{12}
  (2017) P11016}, \href{http://arxiv.org/abs/1708.00819}{{\normalfont\ttfamily
  arXiv:1708.00819}}\relax
\mciteBstWouldAddEndPuncttrue
\mciteSetBstMidEndSepPunct{\mcitedefaultmidpunct}
{\mcitedefaultendpunct}{\mcitedefaultseppunct}\relax
\EndOfBibitem
\bibitem{LHCb-PAPER-2012-048}
LHCb collaboration, R.~Aaij {\em et~al.},
  \ifthenelse{\boolean{articletitles}}{\emph{{Measurement of the \Lb, \Xibm,
  and \Omegab baryon masses}},
  }{}\href{https://doi.org/10.1103/PhysRevLett.110.182001}{Phys.\ Rev.\ Lett.\
  \textbf{110} (2013) 182001},
  \href{http://arxiv.org/abs/1302.1072}{{\normalfont\ttfamily
  arXiv:1302.1072}}\relax
\mciteBstWouldAddEndPuncttrue
\mciteSetBstMidEndSepPunct{\mcitedefaultmidpunct}
{\mcitedefaultendpunct}{\mcitedefaultseppunct}\relax
\EndOfBibitem
\bibitem{LHCb-PAPER-2013-011}
LHCb collaboration, R.~Aaij {\em et~al.},
  \ifthenelse{\boolean{articletitles}}{\emph{{Precision measurement of \D meson
  mass differences}}, }{}\href{https://doi.org/10.1007/JHEP06(2013)065}{JHEP
  \textbf{06} (2013) 065},
  \href{http://arxiv.org/abs/1304.6865}{{\normalfont\ttfamily
  arXiv:1304.6865}}\relax
\mciteBstWouldAddEndPuncttrue
\mciteSetBstMidEndSepPunct{\mcitedefaultmidpunct}
{\mcitedefaultendpunct}{\mcitedefaultseppunct}\relax
\EndOfBibitem
\bibitem{Bowen:2014tca}
E.~E. Bowen, \ifthenelse{\boolean{articletitles}}{\emph{{Vertexing and tracking
  software at LHCb}}, }{}\href{https://doi.org/10.22323/1.227.0038}{PoS
  \textbf{Vertex2014} (2014) 038}\relax
\mciteBstWouldAddEndPuncttrue
\mciteSetBstMidEndSepPunct{\mcitedefaultmidpunct}
{\mcitedefaultendpunct}{\mcitedefaultseppunct}\relax
\EndOfBibitem
\bibitem{Dziurda:2115353}
A.~Dziurda, {\em {Studies of time\nobreakdash-dependent $\CP$~violation in
  charm decays of $\Bs$~mesons}}, PhD thesis, {Institute of Nuclear Physics,
  Kwakow}, 2015,
  {\href{http://cds.cern.ch/record/2115353}{CERN-THESES-2015-246}}\relax
\mciteBstWouldAddEndPuncttrue
\mciteSetBstMidEndSepPunct{\mcitedefaultmidpunct}
{\mcitedefaultendpunct}{\mcitedefaultseppunct}\relax
\EndOfBibitem
\bibitem{LHCb-DP-2012-003}
M.~Adinolfi {\em et~al.},
  \ifthenelse{\boolean{articletitles}}{\emph{{Performance of the \lhcb \rich
  detector at the LHC}},
  }{}\href{https://doi.org/10.1140/epjc/s10052-013-2431-9}{Eur.\ Phys.\ J.\
  \textbf{C73} (2013) 2431},
  \href{http://arxiv.org/abs/1211.6759}{{\normalfont\ttfamily
  arXiv:1211.6759}}\relax
\mciteBstWouldAddEndPuncttrue
\mciteSetBstMidEndSepPunct{\mcitedefaultmidpunct}
{\mcitedefaultendpunct}{\mcitedefaultseppunct}\relax
\EndOfBibitem
\bibitem{LHCb-DP-2012-002}
A.~A. Alves~Jr.\ {\em et~al.},
  \ifthenelse{\boolean{articletitles}}{\emph{{Performance of the LHCb muon
  system}}, }{}\href{https://doi.org/10.1088/1748-0221/8/02/P02022}{JINST
  \textbf{8} (2013) P02022},
  \href{http://arxiv.org/abs/1211.1346}{{\normalfont\ttfamily
  arXiv:1211.1346}}\relax
\mciteBstWouldAddEndPuncttrue
\mciteSetBstMidEndSepPunct{\mcitedefaultmidpunct}
{\mcitedefaultendpunct}{\mcitedefaultseppunct}\relax
\EndOfBibitem
\bibitem{LHCb-DP-2012-004}
R.~Aaij {\em et~al.}, \ifthenelse{\boolean{articletitles}}{\emph{{The \lhcb
  trigger and its performance in 2011}},
  }{}\href{https://doi.org/10.1088/1748-0221/8/04/P04022}{JINST \textbf{8}
  (2013) P04022}, \href{http://arxiv.org/abs/1211.3055}{{\normalfont\ttfamily
  arXiv:1211.3055}}\relax
\mciteBstWouldAddEndPuncttrue
\mciteSetBstMidEndSepPunct{\mcitedefaultmidpunct}
{\mcitedefaultendpunct}{\mcitedefaultseppunct}\relax
\EndOfBibitem
\bibitem{Sjostrand:2007gs}
T.~Sj\"{o}strand, S.~Mrenna, and P.~Skands,
  \ifthenelse{\boolean{articletitles}}{\emph{{A brief introduction to
  $\pythia\,8.1$}},
  }{}\href{https://doi.org/10.1016/j.cpc.2008.01.036}{Comput.\ Phys.\ Commun.\
  \textbf{178} (2008) 852},
  \href{http://arxiv.org/abs/0710.3820}{{\normalfont\ttfamily
  arXiv:0710.3820}}\relax
\mciteBstWouldAddEndPuncttrue
\mciteSetBstMidEndSepPunct{\mcitedefaultmidpunct}
{\mcitedefaultendpunct}{\mcitedefaultseppunct}\relax
\EndOfBibitem
\bibitem{LHCb-PROC-2010-056}
I.~Belyaev {\em et~al.}, \ifthenelse{\boolean{articletitles}}{\emph{{Handling
  of the generation of primary events in \gauss, the \lhcb simulation
  framework}}, }{}\href{https://doi.org/10.1088/1742-6596/331/3/032047}{J.\
  Phys.\ Conf.\ Ser.\  \textbf{331} (2011) 032047}\relax
\mciteBstWouldAddEndPuncttrue
\mciteSetBstMidEndSepPunct{\mcitedefaultmidpunct}
{\mcitedefaultendpunct}{\mcitedefaultseppunct}\relax
\EndOfBibitem
\bibitem{Lange:2001uf}
D.~J. Lange, \ifthenelse{\boolean{articletitles}}{\emph{{The \evtgen particle
  decay simulation package}},
  }{}\href{https://doi.org/10.1016/S0168-9002(01)00089-4}{Nucl.\ Instrum.\
  Meth.\  \textbf{A462} (2001) 152}\relax
\mciteBstWouldAddEndPuncttrue
\mciteSetBstMidEndSepPunct{\mcitedefaultmidpunct}
{\mcitedefaultendpunct}{\mcitedefaultseppunct}\relax
\EndOfBibitem
\bibitem{davidson2015photos}
N.~Davidson, T.~Przedzinski, and Z.~Was,
  \ifthenelse{\boolean{articletitles}}{\emph{{\photos interface in C++
  technical and physics documentation}},
  }{}\href{https://doi.org/https://doi.org/10.1016/j.cpc.2015.09.013}{Comput.\
  Phys.\ Commun.\  \textbf{199} (2016) 86},
  \href{http://arxiv.org/abs/1011.0937}{{\normalfont\ttfamily
  arXiv:1011.0937}}\relax
\mciteBstWouldAddEndPuncttrue
\mciteSetBstMidEndSepPunct{\mcitedefaultmidpunct}
{\mcitedefaultendpunct}{\mcitedefaultseppunct}\relax
\EndOfBibitem
\bibitem{Berezhnoy:2011nx}
A.~V. Berezhnoy, A.~K. Likhoded, and A.~V. Luchinsky,
  \ifthenelse{\boolean{articletitles}}{\emph{{ {\tt{BC\_NPI}} module for
  the~analysis of $\decay{\PB_{\Pc}}{\jpsi+n\pion}$ and
  $\decay{\PB_{\Pc}}{\PB_\Ps+n\pion}$~decays within the
  {\sc{EvtGen}}~package}},
  }{}\href{http://arxiv.org/abs/1104.0808}{{\normalfont\ttfamily
  arXiv:1104.0808}}\relax
\mciteBstWouldAddEndPuncttrue
\mciteSetBstMidEndSepPunct{\mcitedefaultmidpunct}
{\mcitedefaultendpunct}{\mcitedefaultseppunct}\relax
\EndOfBibitem
\bibitem{Allison:2006ve}
Geant4 collaboration, J.~Allison {\em et~al.},
  \ifthenelse{\boolean{articletitles}}{\emph{{\geant developments and
  applications}}, }{}\href{https://doi.org/10.1109/TNS.2006.869826}{IEEE
  Trans.\ Nucl.\ Sci.\  \textbf{53} (2006) 270}\relax
\mciteBstWouldAddEndPuncttrue
\mciteSetBstMidEndSepPunct{\mcitedefaultmidpunct}
{\mcitedefaultendpunct}{\mcitedefaultseppunct}\relax
\EndOfBibitem
\bibitem{Agostinelli:2002hh}
Geant4 collaboration, S.~Agostinelli {\em et~al.},
  \ifthenelse{\boolean{articletitles}}{\emph{{\geant~--~a simulation toolkit}},
  }{}\href{https://doi.org/10.1016/S0168-9002(03)01368-8}{Nucl.\ Instrum.\
  Meth.\  \textbf{A506} (2003) 250}\relax
\mciteBstWouldAddEndPuncttrue
\mciteSetBstMidEndSepPunct{\mcitedefaultmidpunct}
{\mcitedefaultendpunct}{\mcitedefaultseppunct}\relax
\EndOfBibitem
\bibitem{LHCb-PROC-2011-006}
M.~Clemencic {\em et~al.}, \ifthenelse{\boolean{articletitles}}{\emph{{The
  \lhcb simulation application, \gauss: design, evolution and experience}},
  }{}\href{https://doi.org/10.1088/1742-6596/331/3/032023}{J.\ Phys.\ Conf.\
  Ser.\  \textbf{331} (2011) 032023}\relax
\mciteBstWouldAddEndPuncttrue
\mciteSetBstMidEndSepPunct{\mcitedefaultmidpunct}
{\mcitedefaultendpunct}{\mcitedefaultseppunct}\relax
\EndOfBibitem
\bibitem{LHCb-DP-2013-002}
LHCb collaboration, R.~Aaij {\em et~al.},
  \ifthenelse{\boolean{articletitles}}{\emph{{Measurement of the track
  reconstruction efficiency at LHCb}},
  }{}\href{https://doi.org/10.1088/1748-0221/10/02/P02007}{JINST \textbf{10}
  (2015) P02007}, \href{http://arxiv.org/abs/1408.1251}{{\normalfont\ttfamily
  arXiv:1408.1251}}\relax
\mciteBstWouldAddEndPuncttrue
\mciteSetBstMidEndSepPunct{\mcitedefaultmidpunct}
{\mcitedefaultendpunct}{\mcitedefaultseppunct}\relax
\EndOfBibitem
\bibitem{LHCb-PAPER-2016-040}
LHCb collaboration, R.~Aaij {\em et~al.},
  \ifthenelse{\boolean{articletitles}}{\emph{{Observation of
  \mbox{\decay{\Bp}{\jpsi 3\pip 2\pim}} and \mbox{\decay{\Bp}{\psitwos
  \pip\pip\pim}} decays}},
  }{}\href{https://doi.org/10.1140/epjc/s10052-017-4610-6}{Eur.\ Phys.\ J.\
  \textbf{C77} (2017) 72},
  \href{http://arxiv.org/abs/1610.01383}{{\normalfont\ttfamily
  arXiv:1610.01383}}\relax
\mciteBstWouldAddEndPuncttrue
\mciteSetBstMidEndSepPunct{\mcitedefaultmidpunct}
{\mcitedefaultendpunct}{\mcitedefaultseppunct}\relax
\EndOfBibitem
\bibitem{LHCb-PROC-2011-008}
A.~Powell {\em et~al.}, \ifthenelse{\boolean{articletitles}}{\emph{{Particle
  identification at \lhcb}},
  }{}\href{https://doi.org/https://doi.org/10.22323/1.120.0020}{PoS
  \textbf{ICHEP2010} (2011) 020},
  \href{https://cdsweb.cern.ch/record/1322666?ln=en}{LHCb-PROC-2011-008}\relax
\mciteBstWouldAddEndPuncttrue
\mciteSetBstMidEndSepPunct{\mcitedefaultmidpunct}
{\mcitedefaultendpunct}{\mcitedefaultseppunct}\relax
\EndOfBibitem
\bibitem{PDG2021}
Particle Data Group, P.~A. Zyla {\em et~al.},
  \ifthenelse{\boolean{articletitles}}{\emph{{\href{http://pdg.lbl.gov/}{Review
  of particle physics}}}, }{}\href{https://doi.org/10.1093/ptep/ptaa104}{Prog.\
  Theor.\ Exp.\ Phys.\  \textbf{2020} (2020) 083C01}, and
  {\href{http://pdglive.lbl.gov/}{2022 update}}\relax
\mciteBstWouldAddEndPuncttrue
\mciteSetBstMidEndSepPunct{\mcitedefaultmidpunct}
{\mcitedefaultendpunct}{\mcitedefaultseppunct}\relax
\EndOfBibitem
\bibitem{Hulsbergen:2005pu}
W.~D. Hulsbergen, \ifthenelse{\boolean{articletitles}}{\emph{{Decay chain
  fitting with a Kalman filter}},
  }{}\href{https://doi.org/10.1016/j.nima.2005.06.078}{Nucl.\ Instrum.\ Meth.\
  \textbf{A552} (2005) 566},
  \href{http://arxiv.org/abs/physics/0503191}{{\normalfont\ttfamily
  arXiv:physics/0503191}}\relax
\mciteBstWouldAddEndPuncttrue
\mciteSetBstMidEndSepPunct{\mcitedefaultmidpunct}
{\mcitedefaultendpunct}{\mcitedefaultseppunct}\relax
\EndOfBibitem
\bibitem{LHCb-PAPER-2020-009}
LHCb collaboration, R.~Aaij {\em et~al.},
  \ifthenelse{\boolean{articletitles}}{\emph{{Study of the $\psitwod$ and
  $\chiconex$ states in $\decay{\Bp}{(\jpsi \pip\pim)\Kp}$ decays}},
  }{}\href{https://doi.org/10.1007/JHEP08(2020)123}{JHEP \textbf{08} (2020)
  123}, \href{http://arxiv.org/abs/2005.13422}{{\normalfont\ttfamily
  arXiv:2005.13422}}\relax
\mciteBstWouldAddEndPuncttrue
\mciteSetBstMidEndSepPunct{\mcitedefaultmidpunct}
{\mcitedefaultendpunct}{\mcitedefaultseppunct}\relax
\EndOfBibitem
\bibitem{LHCb-PAPER-2020-035}
LHCb collaboration, R.~Aaij {\em et~al.},
  \ifthenelse{\boolean{articletitles}}{\emph{{Study of $\Bs\to \jpsi \pip \pim
  \Kp \Km$ decays}}, }{}\href{https://doi.org/10.1007/JHEP02(2021)024}{JHEP
  \textbf{02} (2021) 024},
  \href{http://arxiv.org/abs/2011.01867}{{\normalfont\ttfamily
  arXiv:2011.01867}}\relax
\mciteBstWouldAddEndPuncttrue
\mciteSetBstMidEndSepPunct{\mcitedefaultmidpunct}
{\mcitedefaultendpunct}{\mcitedefaultseppunct}\relax
\EndOfBibitem
\bibitem{LHCb-PAPER-2011-013}
LHCb collaboration, R.~Aaij {\em et~al.},
  \ifthenelse{\boolean{articletitles}}{\emph{{Observation of \jpsi-pair
  production in \proton\proton~collisions at \mbox{$\sqs=7\tev$}}},
  }{}\href{https://doi.org/10.1016/j.physletb.2011.12.015}{Phys.\ Lett.\
  \textbf{B707} (2012) 52},
  \href{http://arxiv.org/abs/1109.0963}{{\normalfont\ttfamily
  arXiv:1109.0963}}\relax
\mciteBstWouldAddEndPuncttrue
\mciteSetBstMidEndSepPunct{\mcitedefaultmidpunct}
{\mcitedefaultendpunct}{\mcitedefaultseppunct}\relax
\EndOfBibitem
\bibitem{Skwarnicki:1986xj}
T.~Skwarnicki, {\em {A~study of the~radiative cascade transitions between
  the~$\PUpsilon^{\prime}$ and $\PUpsilon$~resonances}}, PhD thesis, Institute
  of Nuclear Physics, Krakow, 1986,
  {\href{http://inspirehep.net/record/230779/}{\mbox{DESY-F31-86-02}}}\relax
\mciteBstWouldAddEndPuncttrue
\mciteSetBstMidEndSepPunct{\mcitedefaultmidpunct}
{\mcitedefaultendpunct}{\mcitedefaultseppunct}\relax
\EndOfBibitem
\bibitem{Byckling}
E.~Byckling and K.~Kajantie, {\em Particle kinematics}, John Wiley \& Sons
  Inc., New York, 1973\relax
\mciteBstWouldAddEndPuncttrue
\mciteSetBstMidEndSepPunct{\mcitedefaultmidpunct}
{\mcitedefaultendpunct}{\mcitedefaultseppunct}\relax
\EndOfBibitem
\bibitem{LHCb-PAPER-2020-008}
LHCb collaboration, R.~Aaij {\em et~al.},
  \ifthenelse{\boolean{articletitles}}{\emph{{Study of the lineshape of the
  $\chiconex$ state}},
  }{}\href{https://doi.org/10.1103/PhysRevD.102.092005}{Phys.\ Rev.\
  \textbf{D102} (2020) 092005},
  \href{http://arxiv.org/abs/2005.13419}{{\normalfont\ttfamily
  arXiv:2005.13419}}\relax
\mciteBstWouldAddEndPuncttrue
\mciteSetBstMidEndSepPunct{\mcitedefaultmidpunct}
{\mcitedefaultendpunct}{\mcitedefaultseppunct}\relax
\EndOfBibitem
\bibitem{Wilks:1938dza}
S.~S. Wilks, \ifthenelse{\boolean{articletitles}}{\emph{{The large-sample
  distribution of the likelihood ratio for testing composite hypotheses}},
  }{}\href{https://doi.org/10.1214/aoms/1177732360}{Ann.\ Math.\ Stat.\
  \textbf{9} (1938) 60}\relax
\mciteBstWouldAddEndPuncttrue
\mciteSetBstMidEndSepPunct{\mcitedefaultmidpunct}
{\mcitedefaultendpunct}{\mcitedefaultseppunct}\relax
\EndOfBibitem
\bibitem{Pivk:2004ty}
M.~Pivk and F.~R. Le~Diberder,
  \ifthenelse{\boolean{articletitles}}{\emph{{sPlot: a statistical tool to
  unfold data distributions}},
  }{}\href{https://doi.org/10.1016/j.nima.2005.08.106}{Nucl.\ Instrum.\ Meth.\
  \textbf{A555} (2005) 356},
  \href{http://arxiv.org/abs/physics/0402083}{{\normalfont\ttfamily
  arXiv:physics/0402083}}\relax
\mciteBstWouldAddEndPuncttrue
\mciteSetBstMidEndSepPunct{\mcitedefaultmidpunct}
{\mcitedefaultendpunct}{\mcitedefaultseppunct}\relax
\EndOfBibitem
\bibitem{Blatt:1952ije}
J.~M. Blatt and V.~F. Weisskopf, {\em {Theoretical nuclear physics}},
  \href{https://doi.org/10.1007/978-1-4612-9959-2}{ Springer, New York,
  1952}\relax
\mciteBstWouldAddEndPuncttrue
\mciteSetBstMidEndSepPunct{\mcitedefaultmidpunct}
{\mcitedefaultendpunct}{\mcitedefaultseppunct}\relax
\EndOfBibitem
\bibitem{Okubo:1963fa}
S.~Okubo, \ifthenelse{\boolean{articletitles}}{\emph{{$\Pphi$\nobreakdash-meson
  and unitary symmetry model}},
  }{}\href{https://doi.org/10.1016/S0375-9601(63)92548-9}{Phys.\ Lett.\
  \textbf{5} (1963) 165}\relax
\mciteBstWouldAddEndPuncttrue
\mciteSetBstMidEndSepPunct{\mcitedefaultmidpunct}
{\mcitedefaultendpunct}{\mcitedefaultseppunct}\relax
\EndOfBibitem
\bibitem{Zweig2}
G.~Zweig, \ifthenelse{\boolean{articletitles}}{\emph{{An SU$_3$ model for
  strong interaction symmetry and its breaking; Version 2}}}{}
  \href{http://cds.cern.ch/record/570209}{CERN-TH-412}, CERN, Geneva,
  1964\relax
\mciteBstWouldAddEndPuncttrue
\mciteSetBstMidEndSepPunct{\mcitedefaultmidpunct}
{\mcitedefaultendpunct}{\mcitedefaultseppunct}\relax
\EndOfBibitem
\bibitem{Iizuka:1966fk}
J.~Iizuka, \ifthenelse{\boolean{articletitles}}{\emph{{A systematics and
  phenomenology of meson family}},
  }{}\href{https://doi.org/10.1143/PTPS.37.21}{Suppl.\ Prog.\ Theor.\ Phys.\
  \textbf{37} (1966) 21}\relax
\mciteBstWouldAddEndPuncttrue
\mciteSetBstMidEndSepPunct{\mcitedefaultmidpunct}
{\mcitedefaultendpunct}{\mcitedefaultseppunct}\relax
\EndOfBibitem
\bibitem{Singh:1977sa}
C.~P. Singh and C.~Singh,
  \ifthenelse{\boolean{articletitles}}{\emph{{Phenomenology for
  Iizuka\nobreakdash--Okubo\nobreakdash--Zweig rule breaking}},
  }{}\href{https://doi.org/10.1016/0370-2693(77)90492-0}{Phys.\ Lett.\
  \textbf{B68} (1977) 350}\relax
\mciteBstWouldAddEndPuncttrue
\mciteSetBstMidEndSepPunct{\mcitedefaultmidpunct}
{\mcitedefaultendpunct}{\mcitedefaultseppunct}\relax
\EndOfBibitem
\bibitem{Singh:1978mi}
C.~A. Singh, \ifthenelse{\boolean{articletitles}}{\emph{{Violation of
  the~Okubo\nobreakdash--Zweig\nobreakdash--Iizuka rule in the~decay mode
  \mbox{$\decay{\B(1235)}{\Pphi(1020)+\pion}$} }},
  }{}\href{https://doi.org/10.1007/BF02848150}{Pramana -- J.\ Phys.\
  \textbf{9} (1977) 629}\relax
\mciteBstWouldAddEndPuncttrue
\mciteSetBstMidEndSepPunct{\mcitedefaultmidpunct}
{\mcitedefaultendpunct}{\mcitedefaultseppunct}\relax
\EndOfBibitem
\bibitem{LHCb-DP-2018-001}
R.~Aaij {\em et~al.}, \ifthenelse{\boolean{articletitles}}{\emph{{Selection and
  processing of calibration samples to measure the particle identification
  performance of the LHCb experiment in Run 2}},
  }{}\href{https://doi.org/10.1140/epjti/s40485-019-0050-z}{EPJ Tech.\
  Instrum.\  \textbf{6} (2019) 1},
  \href{http://arxiv.org/abs/1803.00824}{{\normalfont\ttfamily
  arXiv:1803.00824}}\relax
\mciteBstWouldAddEndPuncttrue
\mciteSetBstMidEndSepPunct{\mcitedefaultmidpunct}
{\mcitedefaultendpunct}{\mcitedefaultseppunct}\relax
\EndOfBibitem
\bibitem{Student}
{Student~(W.\ \,S.\ ~Gosset)},
  \ifthenelse{\boolean{articletitles}}{\emph{{The~probable error of a~mean}},
  }{}\href{https://doi.org/10.1093/biomet/6.1.1}{Biometrika \textbf{6} (1908)
  1}\relax
\mciteBstWouldAddEndPuncttrue
\mciteSetBstMidEndSepPunct{\mcitedefaultmidpunct}
{\mcitedefaultendpunct}{\mcitedefaultseppunct}\relax
\EndOfBibitem
\bibitem{Jackman}
S.~Jackman, {\em Bayesian analysis for the social sciences}, John Wiley \&
  Sons, Inc., Hoboken, New Jersey, USA, 2009\relax
\mciteBstWouldAddEndPuncttrue
\mciteSetBstMidEndSepPunct{\mcitedefaultmidpunct}
{\mcitedefaultendpunct}{\mcitedefaultseppunct}\relax
\EndOfBibitem
\bibitem{Santos:2013gra}
D.~Mart{\'\i}nez~Santos and F.~Dupertuis,
  \ifthenelse{\boolean{articletitles}}{\emph{{Mass distributions marginalized
  over per-event errors}},
  }{}\href{https://doi.org/10.1016/j.nima.2014.06.081}{Nucl.\ Instrum.\ Meth.\
  \textbf{A764} (2014) 150},
  \href{http://arxiv.org/abs/1312.5000}{{\normalfont\ttfamily
  arXiv:1312.5000}}\relax
\mciteBstWouldAddEndPuncttrue
\mciteSetBstMidEndSepPunct{\mcitedefaultmidpunct}
{\mcitedefaultendpunct}{\mcitedefaultseppunct}\relax
\EndOfBibitem
\bibitem{LHCb-PAPER-2010-001}
LHCb collaboration, R.~Aaij {\em et~al.},
  \ifthenelse{\boolean{articletitles}}{\emph{{Prompt \KS~production in
  $\proton\proton$~collisions at~\mbox{$\sqs=0.9\tev$}}},
  }{}\href{https://doi.org/10.1016/j.physletb.2010.08.055}{Phys.\ Lett.\
  \textbf{B693} (2010) 69},
  \href{http://arxiv.org/abs/1008.3105}{{\normalfont\ttfamily
  arXiv:1008.3105}}\relax
\mciteBstWouldAddEndPuncttrue
\mciteSetBstMidEndSepPunct{\mcitedefaultmidpunct}
{\mcitedefaultendpunct}{\mcitedefaultseppunct}\relax
\EndOfBibitem
\bibitem{LHCb-PAPER-2015-041}
LHCb collaboration, R.~Aaij {\em et~al.},
  \ifthenelse{\boolean{articletitles}}{\emph{{Measurements of prompt charm
  production cross\nobreakdash-sections in \proton\proton~collisions
  at~\mbox{$\sqs = 13\tev$}}},
  }{}\href{https://doi.org/10.1007/JHEP03(2016)159}{JHEP \textbf{03} (2016)
  159}, Erratum \href{https://doi.org/10.1007/JHEP09(2016)013}{ibid.\
  \textbf{09} (2016) 013}, Erratum
  \href{https://doi.org/10.1007/JHEP05(2017)074}{ibid.\   \textbf{05} (2017)
  074}, \href{http://arxiv.org/abs/1510.01707}{{\normalfont\ttfamily
  arXiv:1510.01707}}\relax
\mciteBstWouldAddEndPuncttrue
\mciteSetBstMidEndSepPunct{\mcitedefaultmidpunct}
{\mcitedefaultendpunct}{\mcitedefaultseppunct}\relax
\EndOfBibitem
\bibitem{LHCb-PAPER-2012-010}
LHCb collaboration, R.~Aaij {\em et~al.},
  \ifthenelse{\boolean{articletitles}}{\emph{{Measurement of relative branching
  fractions of \B~decays to \psitwos and \jpsi mesons}},
  }{}\href{https://doi.org/10.1140/epjc/s10052-012-2118-7}{Eur.\ Phys.\ J.\
  \textbf{C72} (2012) 2118},
  \href{http://arxiv.org/abs/1205.0918}{{\normalfont\ttfamily
  arXiv:1205.0918}}\relax
\mciteBstWouldAddEndPuncttrue
\mciteSetBstMidEndSepPunct{\mcitedefaultmidpunct}
{\mcitedefaultendpunct}{\mcitedefaultseppunct}\relax
\EndOfBibitem
\end{mcitethebibliography}
 
\clearpage
\centerline
{\large\bf LHCb collaboration}
\begin
{flushleft}
\small
R.~Aaij$^{32}$\lhcborcid{0000-0003-0533-1952},
A.S.W.~Abdelmotteleb$^{50}$\lhcborcid{0000-0001-7905-0542},
C.~Abellan~Beteta$^{44}$,
F.~Abudin{\'e}n$^{50}$\lhcborcid{0000-0002-6737-3528},
T.~Ackernley$^{54}$\lhcborcid{0000-0002-5951-3498},
B.~Adeva$^{40}$\lhcborcid{0000-0001-9756-3712},
M.~Adinolfi$^{48}$\lhcborcid{0000-0002-1326-1264},
P.~Adlarson$^{77}$\lhcborcid{0000-0001-6280-3851},
H.~Afsharnia$^{9}$,
C.~Agapopoulou$^{13}$\lhcborcid{0000-0002-2368-0147}
% ...

{\footnotesize \it

% ...
$^{9}$Universit{\'e} Clermont Auvergne, CNRS/IN2P3, LPC, Clermont-Ferrand, France\\
% ...
$^{13}$LPNHE, Sorbonne Universit{\'e}, Paris Diderot Sorbonne Paris Cit{\'e}, CNRS/IN2P3, Paris, France\\
% ...
$^{32}$Nikhef National Institute for Subatomic Physics, Amsterdam, Netherlands\\
% ...
$^{40}$Instituto Galego de F{\'\i}sica de Altas Enerx{\'\i}as (IGFAE), Universidade de Santiago de Compostela, Santiago de Compostela, Spain\\
% ...
$^{44}$Physik-Institut, Universit{\"a}t Z{\"u}rich, Z{\"u}rich, Switzerland\\
% ...
$^{48}$H.H. Wills Physics Laboratory, University of Bristol, Bristol, United Kingdom\\
% ...
$^{50}$Department of Physics, University of Warwick, Coventry, United Kingdom\\
% ...
$^{53}$School of Physics and Astronomy, University of Glasgow, Glasgow, United Kingdom\\
$^{54}$Oliver Lodge Laboratory, University of Liverpool, Liverpool, United Kingdom\\
% ...
$^{77}$Department of Physics and Astronomy, Uppsala University, Uppsala, Sweden, associated to $^{53}$\\
% ...
}
\end{flushleft}

\end{document}